\documentclass[12pt]{article}
\usepackage{amsfonts,amsmath,epstopdf}
\usepackage{graphicx}
\newcommand\encadremath[1]{\vbox{\hrule\hbox{\vrule\kern8pt
\vbox{\kern8pt \hbox{$\displaystyle #1$}\kern8pt}
\kern8pt\vrule}\hrule}}
\def\enca#1{\vbox{\hrule\hbox{
\vrule\kern8pt\vbox{\kern8pt \hbox{$\displaystyle #1$}
\kern8pt} \kern8pt\vrule}\hrule}}

\newcommand\figureframex[3]{
\begin{figure}[bth]
\hrule\hbox{\vrule\kern8pt
\vbox{\kern8pt \vbox{
\begin{center}
{\mbox{\epsfxsize=#1.truecm\epsfbox{#2}}}
\end{center}
\caption{#3}
}\kern8pt}
\kern8pt\vrule}\hrule
\end{figure}
}
\newcommand\figureframey[3]{
\begin{figure}[bth]
\hrule\hbox{\vrule\kern8pt
\vbox{\kern8pt \vbox{
\begin{center}
{\mbox{\epsfysize=#1.truecm\epsfbox{#2}}}
\end{center}
\caption{#3}
}\kern8pt}
\kern8pt\vrule}\hrule
\end{figure}
}

\makeatletter
\@addtoreset{equation}{section}
\makeatother
\newtheorem{theorem}{Theorem}[section]

\newtheorem{remark}{Remark}[section]
\newtheorem{proposition}{Proposition}[section]
\newtheorem{lemma}{Lemma}[section]
\newtheorem{corollary}{Corollary}[section]
\newtheorem{definition}{Definition}[section]
\def\br{\begin{remark}\rm\small}
\def\er{\end{remark}}
\def\bt{\begin{theorem}}
\def\et{\end{theorem}}
\def\bd{\begin{definition}}
\def\ed{\end{definition}}
\def\bp{\begin{proposition}}
\def\ep{\end{proposition}}
\def\bl{\begin{lemma}}
\def\el{\end{lemma}}
\def\bc{\begin{corollary}}
\def\ec{\end{corollary}}
\def\beaq{\begin{eqnarray}}
\def\eeaq{\end{eqnarray}}

\newcommand{\beq}{\begin{equation}}
\newcommand{\eeq}{\end{equation}}
\newcommand{\bea}{\begin{eqnarray}}
\newcommand{\eea}{\end{eqnarray}}

\renewcommand{\and}{{\qquad {\rm and} \qquad}}

\newcommand{\Res}{\mathop{\,\rm Res\,}}

\newcommand{\ee}[1]{{{\rm e}^{#1}}}

\newcommand{\Pint}{{\int\kern -1.em -\kern-.25em}}

\usepackage{hyperref}
\hypersetup{colorlinks,urlcolor=magenta,citecolor=red,linkcolor=blue,filecolor=black}

\begin{document}

\sloppy

\pagestyle{empty}
\begin{flushright}
IPhT-T10/107 \\ CERN-PH-TH/2010-227
\end{flushright}
\addtolength{\baselineskip}{0.20\baselineskip}
\begin{center}
\vspace{26pt}
{\large \bf {Large deviations of the maximal eigenvalue \\ of random matrices}}
\end{center}

\vspace{26pt}

\begin{center}
{\sl G.\ Borot}\hspace*{0.05cm}\footnote{gaetan.borot@cea.fr},
{\sl B.\ Eynard}\hspace*{0.05cm}\footnote{bertrand.eynard@cea.fr },
{\sl S.\ N.\ Majumdar}\hspace*{0.05cm}\footnote{satya.majumdar@u-psud.fr},
{\sl C.\ Nadal}\hspace*{0.05cm}\footnote{celine.nadal@u-psud.fr}

\vspace{6pt}
$\,^{1,2}$ Institut de Physique Th\'{e}orique de Saclay,\\
F-91191 Gif-sur-Yvette Cedex, France.\\
$\,^{3,4}$ Univ. Paris Sud, CNRS, LPTMS,\\
UMR 8626, F-91405 Orsay , France
\end{center}

\vspace{20pt}
\begin{center}
{\bf Abstract}
\end{center}

%

\vspace{0.5cm}

We present detailed computations of the nondecaying terms (three dominant orders) of the free energy in a one cut matrix model with a hard edge $a$, in $\beta$-ensembles, with any polynomial potential. $\beta > 0$ is not restricted to the standard values $\beta = 1$ (hermitian matrices), $\beta = 1/2$ (symmetric matrices), $\beta = 2$ (quaternionic self-dual matrices). This model allows to study the statistic of the maximum eigenvalue of random matrices. We compute the large deviation function to the left of the expected maximum. We specialize our results to the gaussian $\beta$-ensembles and check them numerically. Our method is based on general results and procedures already developed in the literature to solve the Pastur equations (also called "loop equations"). It allows to compute the left tail of the analog of Tracy-Widom laws for any $\beta$, including the constant term.

\vspace{0.5cm}

\newpage


\vspace{26pt}
\pagestyle{plain}
\setcounter{page}{1}
\setcounter{footnote}{0}


\section{Introduction}

Extreme value statistics, in particular the statistics of the maximum
of a set of random variables, play a very important role
as they have a large variety of applications
in various fields such as finance, hydrology (rainfall, flood...) and engineering.
On the other hand, random matrix theory
has raised a huge interest in both physics and mathematics
communities. The distribution of the maximal eigenvalue of a random matrix
is thus of great importance.

For usual Gaussian ensembles of random matrices
(Gaussian Orthogonal Ensemble, Gaussian Unitary Ensemble and Gaussian Symplectic Ensemble),
 the distribution of the maximal eigenvalue close to its mean value is known
to converge to the Tracy-Widom distribution~\cite{TW93,TW95} for large $N$
(where  the matrix is of size $N\times N$). This describes small typical
fluctuations around the mean value. We are interested here in the large deviation
of the distribution, which describes large and rare events (well-spaced from the mean value).
We actually consider as starting point a distribution on $\mathbb{R}^N$
that generalizes the distribution of the eigenvalues of a random matrix for a general
potential $V(x)$ (instead of a quadratic potential for Gaussian ensembles)
and a general $\beta$ (Dyson index).

We consider the following joint distribution:
\beq
\label{eq:ProbDist}
P(\lambda_1,...,\lambda_N) = \frac{1}{Z(\infty)}\: |\Delta(\lambda_i)|^{2\beta}\,\, \prod_{i = 1}^N e^{-\frac{N\beta}{t}\,V(\lambda_i)}
\;\;\;\;
{\rm with}\;\;\; \Delta(\lambda_i)=\prod_{1\leq i<j \leq N}\left(\lambda_i-\lambda_j\right)
\eeq
$Z(\infty)=\int_{\mathbb{R}^N}\, \mathrm{d}\lambda_1\dots\, \mathrm{d}\lambda_N\,\, |\Delta(\lambda_i)|^{2\beta}\,\, \prod_{i = 1}^N e^{- \frac{N\beta}{t}\,V(\lambda_i)}$ is the (unconstrained) partition function.

We define:
\beq
\label{eq:matint}Z(a) = \int_{]-\infty,a]^N}\, \mathrm{d}\lambda_1\dots\, \mathrm{d}\lambda_N\,\, |\Delta(\lambda_i)|^{2\beta}\,\, \prod_{i = 1}^N e^{-\frac{N\beta}{t}\,V(\lambda_i)}
\eeq
$Z(a)$ is the partition function for the weight given by Eqn.~\ref{eq:ProbDist}, multiplied by $\prod_{i = 1}^N\Theta(a - \lambda_i)$. The probability that the maximal eigenvalue $\lambda_{\rm max}$ is smaller or equal to $a$ (i.e. the probability that all eigenvalues $\lambda_i$ are smaller or equal to $a$) is:
\beq
\mathcal{P}(a) =\int_{]-\infty,a]^N}\, \mathrm{d}\lambda_1\dots\, \mathrm{d}\lambda_N\,\,P(\lambda_1,...\lambda_N)
= \frac{Z(a)}{Z(\infty)}
\eeq
The probability that the largest eigenvalue lies between $a$ and $a + \mathrm{d}a$ is:
\beq
\partial_a \mathcal{P}\,\mathrm{d}a = \frac{\mathrm{d}_a Z(a)}{Z(\infty)}
\eeq
For $\beta = 1/2,1,2$, the probability distribution on $\mathbb{R}^N$ (the space of eigenvalues) given in Eqn.~\ref{eq:ProbDist} derives from an invariant probability distribution on a set of matrices, whereas no such derivation is known for other values of $\beta$. We still call $Z(a)$ a "matrix model", and talk about "$\beta$-ensembles". This is only a convention of language to refer to a statistical weight of the form Eqn.~\ref{eq:matint}.
In the case of a quadratic potential $V(x)=\frac{x^2}{2}$, one recovers the usual Gaussian ensembles:
Gaussian Orthogonal Ensemble ($\beta=1/2$), Gaussian Unitary Ensemble ($\beta=1$)
and Gaussian Symplectic Ensemble ($\beta=2$). And in general, one defines the Gaussian $\beta$ Ensemble. Note that $\beta$ differs from the usual Dyson index $\tilde{\beta}$ by a factor $2$: $\tilde{\beta}=2 \beta$. The normalization we choose in the definition of G$\beta$E is also the one used in \cite{RRV}. Let us mention Dimitriu and Edelman \cite{DE} have found a tridiagonal matrix model, whose distribution of eigenvalues reproduces the G$\beta$E ensemble. As far as our methods are concerned, these sparse matrix models are very different in nature from those we consider.

We want to compute the large $N$ expansion of $Z(a)$. For a one cut solution, we assume\footnote{The proof of this assumption under some hypothesis is a work in progress \cite{BG11}} that $\ln Z(a)$ has an expansion in powers of $1/N$. It takes the form:
\beq
\ln Z(a) = D_{N,\beta} + \sum_{g, k \geq 0} \left(\frac{N\sqrt{\beta}}{t}\right)^{2 - 2g - k}\,\left(\sqrt{\beta} - \frac{1}{\sqrt{\beta}}\right)^{k}\,F^{g,k}(a)
\eeq
In the following, we will write $F(a) =  \sum_{g, k \geq 0} \left(\frac{N\sqrt{\beta}}{t}\right)^{2 - 2g - k}\,\left(\sqrt{\beta} - \frac{1}{\sqrt{\beta}}\right)^{k}\,F^{g,k}(a)$.
The coefficients $F^{g,k}$ are independent of $N$ and $\beta$. They may depend on $a$, on $t$, and on the potential $V(x)$. They are given by geometric quantities associated to a plane curve, which were introduced in \cite{CE06}. In this case, the plane curve is related to the equilibrium density of eigenvalues. $D_{N,\beta}$ is a normalization constant, it depends only on $N$ and $\beta$, and not of $a,t$ and $V$.
We tackle the question of this constant in Section~\ref{sec:constant}.

For instance, limiting ourselves to the non-negative powers of $N$
(we will call them "nondecaying terms" or "stable" terms), we have:
\beq
\ln Z(a) =  D_{N,\beta} + \frac{N^2}{t^2}\beta F^{0,0}(a) + \frac{N}{t}(\beta - 1)F^{0,1}(a) + F^{1,0}(a) + (\beta + \beta^{-1} - 2)F^{0,2}(a)
\eeq
$F^{0,0}$ is called the prepotential, $F^{0,1}$ is the entropy, $F^{1,0}$ is related to a Laplacian determinant, and $F^{0,2}$ is the Polyakov anomaly. All the higher $F^{g,k}$ are also given by geometric quantities (their interpretation is less known), and there exists an algorithm to compute them recursively.

For the statistics of the largest eigenvalue, $F^{0,0}(a)$ is the function of large deviation, and higher $F^{g,k}$ are its subleading corrections. This expansion of $F(a)$ as a series in $1/N$
for large $N$ is actually valid only on the left of
the mean value. On the right of the mean value, $\ln\big(Z(a)/Z(\infty)\big)$
 is expected to be exponentially small, i.e. much smaller than any power of $1/N$.
Note that, for the Gaussian case (quadratic potential $V(x)=\frac{x^2}{2}$),
 the leading term $F^{0,0}(a)$ of the left-tail large deviation
has already been computed a few years ago~\cite{DM06,DM08} using a Coulomb gas technique~\cite{For}.
Here, we explain how to compute recursively all the $F^{g,k}$ for a general potential and a general
$\beta$. We also give an explicit expression for the nondecaying terms
$F^{0,0}$, $F^{0,1}$, $F^{1,0}$ and $F^{0,2}$ for the Gaussian case (quadratic potential)
and compare our results with Monte Carlo numerical simulations.

\subsection*{Outline of the article}

We first present (Section~\ref{sec:coco}) the topological expansion of matrix models and the method of loop equations. We describe in details the one-cut solution of the loop equations in Section~\ref{sec:cacq}. We specialize it for the Gaussian-$\beta$ ensembles in Section~\ref{sec:Gogo}, with comparison to numerics in Section~\ref{sec:num}. We discuss in general the scaling limit at the edge of the spectrum, and make the link with Tracy-Widom law in Section~\ref{sec:scalingG}. The appendices give the proof of the results used in the text to settle the "constant problem".

\newpage

\section{Correlators and loop equations}
\label{sec:coco}
\subsection{Definitions}

We shall consider the expectation value of the resolvent
\beq
W_1(x) = \left< \sum_{i = 1}^N \frac{1}{x-\lambda_{i}}\right>
\eeq
and more generally, the non-connected correlation of $n$ resolvents:
\beq
\overline{W}_n(x_1,\dots,x_n) = \,\,\left< \prod_{j=1}^n\, \sum_{i_j = 1}^N \frac{1}{x_j-\lambda_{i_j}}\right>
\eeq
Their cumulants (the connected correlation functions) are called for short "correlators":
\beq
\label{defWn}
W_n(x_1,\dots,x_n) = \,\,\left< \prod_{j=1}^n\, \sum_{i_j = 1}^N \frac{1}{x_j-\lambda_{i_j}}\right>_C
\eeq
For instance: $\overline{W}_2(x_1,x_2)=W_2(x_1,x_2)+W_1(x_1) W_1(x_2)$.

Let us write the potential:
\beq
V(x) = t_0 + \sum_{j \geq 1} \frac{t_j}{j}\,x^j
\eeq
and introduce the loop insertion operator:
\beq
\frac{\partial}{\partial V(x)} = - x^{-1}\,\frac{\partial}{\partial t_0} - \sum_{j \geq 1} x^{-(j + 1)}\,\frac{1}{j}\,\frac{\partial}{\partial t_j}
\eeq
Applying successively this operator to $\ln Z(a)$ gives the $n$-point correlation functions:
\beq
W_n(x_1,\ldots,x_n) = \left(\frac{t}{N\beta}\right)^{n}\,\frac{\partial}{\partial V(x_1)}\cdots\frac{\partial}{\partial V(x_n)}\,\ln Z(a)
\eeq

The correlators satisfy loop equations. These are named Schwinger-Dyson equations (in physics literature), or Pastur equations (in maths literature). They follow from the invariance of an integral under a change of variable. Alternatively, they can be obtained by integration by parts. One must just pay attention to boundary terms at $\lambda_i=a$. We derive them by two equivalent methods, giving complementary information.

\subsection{Loop equation - 1st method}

\medskip
Consider an infinitesimal change of variables $\lambda_i\to \lambda_i+\epsilon\, \frac{\lambda_i-a}{x-\lambda_i}$. This change of variable preserve the range of integration $]-\infty,a]$. We have, to first order in $\epsilon$:
\bea
\prod_{i = 1}^N d\lambda_i & \to & \prod_{i = 1}^N d\lambda_i\,\, \left(1+\epsilon\sum_i \frac{x-a}{(x-\lambda_i)^2}+ O(\epsilon^2) \right)
 \nonumber \\
|\Delta(\lambda_i)|^{2\beta}& \to & |\Delta(\lambda_i)|^{2\beta}\,\left[ 1 + \beta\epsilon \sum_{1 \leq j\neq i \leq N}\left(\frac{1}{x-\lambda_i}-\frac{1}{x-\lambda_j}\right)\,\frac{x-a}{\lambda_i-\lambda_j} + O(\epsilon^2) \right] \nonumber \\
& \to & |\Delta(\lambda_i)|^{2\beta}\,\left[ 1 + \beta\epsilon \sum_{1 \leq j\neq i \leq N}\frac{x-a}{(x-\lambda_i)(x-\lambda_j)}\, + O(\epsilon^2) \right]
\eea
\beq
\prod_i \ee{-\frac{N\beta}{t}\,V(\lambda_i)} \to \prod_i \ee{-\frac{N\beta}{t}\,V(\lambda_i)}\,\Big(1-\epsilon \frac{N\beta}{t}\sum_{i = 1}^N \frac{V'(\lambda_i)\,(\lambda_i-a)}{x-\lambda_i} + O(\epsilon^2) \Big)
\eeq
Writing that $\delta_{\epsilon} \ln Z(a) = 0$, we get to order 1 in $\epsilon$:
\bea
& & \left<\frac{N\beta}{t}\sum_{i = 1}^N \frac{V'(\lambda_i)\,(\lambda_i-a)}{x-\lambda_i}\right> \nonumber \\
& = & \Big< \sum_{i = 1}^N \frac{x-a}{(x-\lambda_i)^2} + \beta \sum_{1 \leq j\neq i \leq N}\frac{x-a}{(x-\lambda_i)(x-\lambda_j)} \Big>  \nonumber \\
&=& (x-a)\,\,\Big< \sum_{i = 1}^N \frac{1-\beta}{(x-\lambda_i)^2} + \beta \sum_{i = 1}^N \frac{1}{(x-\lambda_i)}\sum_{j = 1}^{N}\frac{1}{(x-\lambda_j)} \Big>  \nonumber \\
&=& (x-a)\,\left( (\beta-1)\,W_1'(x) + \beta W_1(x)^2 + \beta W_2(x,x)\right)
\eea
where we have used the notations of Eqn.~\ref{defWn} above.

The left hand side can also be written:
\bea
\left<\sum_{i = 1}^N \frac{V'(\lambda_i)\,(\lambda_i-a)}{x-\lambda_i}\right>
&=& (x-a)\,\left<\sum_{i = 1}^N \frac{V'(\lambda_i)}{x-\lambda_i}\right> - \left<\sum_{i = 1}^N V'(\lambda_i)\right> \nonumber \\
&=& (x-a)\,\left<\sum_{i = 1}^N \frac{V'(\lambda_i)-V'(x)+V'(x)}{x-\lambda_i}\right> - \left<\sum_{i = 1}^N V'(\lambda_i)\right> \nonumber \\
&=& (x-a)V'(x)W_1(x)-(x-a)\,\left<\sum_{i = 1}^N \frac{V'(x)-V'(\lambda_i)}{x-\lambda_i}\right> \nonumber \\
&& \quad - \left<\sum_{i = 1}^N V'(\lambda_i)\right>
\eea
Let us define:
\beq
P_1(x) = \left<\sum_{i = 1}^N \frac{V'(x)-V'(\lambda_i)}{x-\lambda_i}\right> = \left(V'(x)W_1(x)\right)_+
\eeq
which is a polynomial of $x$, of degree $\deg V'-1$, and
\beq
c = \left<\sum_{i = 1}^N V'(\lambda_i)\right>
\eeq

\medskip

The loop equation can thus be written:
\beq
\label{eq:master} \beta W_1(x)^2 + \beta W_2(x,x) + (\beta-1)W_1'(x) = \frac{N\beta}{t}\,\left(V'(x)W_1(x)-P_1(x) - \frac{c}{x-a} \right)
\eeq

\subsection{Loop equation - 2nd method}
We can also perform a infinitesimal change of variable changing the bound of integration:
\beq
\lambda_i\to \lambda_i+\epsilon\,\frac{1}{x - \lambda_i} + O(\epsilon^2)
\eeq
We have to first order in $\epsilon$:
\bea
\prod_{i = 1}^N \mathrm{d}\lambda_i & \rightarrow &\prod_{i = 1}^N \mathrm{d}\lambda_i\cdot\left(1 + \epsilon\sum_{j = 1}^{N} \frac{1}{(x - \lambda_i)^2} + O(\epsilon^2)\right) \nonumber \\
\Delta(\mathbf{\lambda})^{2\beta} & \rightarrow & \Delta(\mathbf{\lambda})^{2\beta}\left[1 + \epsilon\,\beta \sum_{1 \leq i \neq j \leq N} \left(\frac{1}{x - \lambda_i} - \frac{1}{x - \lambda_j}\right)\frac{1}{\lambda_i - \lambda_j} + O(\epsilon^2)\right] \nonumber \\
& \rightarrow & \Delta(\mathbf{\lambda})^{2\beta}\left[1 + \epsilon\,\beta  \left(\sum_{1 \leq i,j \leq N}\frac{1}{x - \lambda_i}\frac{1}{x - \lambda_j}  - \sum_{i = 1}^N \frac{1}{(x - \lambda_i)^2}\right) + O(\epsilon^2)\right] \nonumber \\
& & \\
\prod_{i = 1}^{N} e^{-\frac{N\beta}{t}V(\lambda_i)} & \rightarrow & \prod_{i = 1}^N e^{-\frac{N\beta}{t}V(\lambda_i)}\left[1 + \epsilon\sum_{j = 1}^{N} \frac{\beta N}{t} \frac{V'(\lambda_i)}{x - \lambda_i} + O(\epsilon^2)\right]
\eea

Under this change of variable, the integral changes by $\epsilon \frac{\partial_a Z(a)}{x - a} + O(\epsilon^2)$. Collecting the first order in $\epsilon$, we obtain:
\bea
\label{eq:loop2}\beta W_2(x,x) + \beta (W_1(x))^2 + (\beta - 1)W_1'(x) - \frac{\beta N}{t}\left(V'(x)W_1(x) - P_1(x)\right) & = & \frac{\partial_a F(a)}{x - a} \nonumber \\
& &
\eea
This is again Eqn.~\ref{eq:master}, with the alternative information that $\frac{\beta N}{t}c = -\partial_a\,F(a)$.

\subsection{Higher order loop equations}

By application of $\left(\frac{t}{N\beta}\right)\frac{\partial}{\partial V(x_1)},\cdots,\left(\frac{t}{N\beta}\right)\frac{\partial}{\partial V(x_{n - 1})}$ to Eqn.~\ref{eq:loop2}, one can obtain loop equations for the $n$-point correlators. We have to introduce:
\beq
P_n(x\:;\:x_1,\ldots,x_{n - 1}) = \Big\langle\mathrm{Tr}\,\frac{V'(x) - V'(M)}{x - M}\,\cdot\,\prod_{i = 1}^{n - 1} \mathrm{Tr}\,\frac{1}{x_i - M}\Big\rangle_c
\eeq
Let us write $I = \{x_1,\ldots,x_{n -1}\}$ for the spectator variables.
The result is:
\bea
&& \sum_{J \subseteq I} \beta W_{|J| + 1}(x,J)W_{n - |J|}(x,I\setminus J) + \beta W_{n + 1}(x,x,I) + (\beta - 1)\frac{\mathrm{d}W_{n}(x,I)}{\mathrm{d}x} \nonumber \\
& = & \frac{\beta N}{t}\left(V'(x)W_n(x,I) - P_n(x\:;\:I)\right) + \sum_{x_i \in I} \frac{\mathrm{d}}{\mathrm{d}x_i}\left(\frac{W_{n - 1}(x,I\setminus\{x_i\}) - W_{n - 1}(I)}{x - x_i}\right) \nonumber \\
\label{eq:loopeqh2} & & + \frac{\partial_a W_{n - 1}(I)}{x - a}
\eea

\subsection{Recovering $F$}

The easiest way to find $F$ is to integrate its derivative with respect to the independent parameters $(a,t,t_j)$. These derivatives can be expressed in terms of the correlators.
\begin{itemize}
\item[$\bullet$] Wrt the hard edge $a$. By comparison of the loop equations obtained by the two change of variables, we have seen that:
\beq
\partial_a F = -\frac{\beta N}{t}\,\Big\langle \mathrm{Tr} V'(M) \Big\rangle = \frac{\beta N}{t}\,\Res_{x \rightarrow \infty}\mathrm{d}x\,V'(x)W_1(x)
\eeq
This could also be obtained by comparing the $1/x$ term (when $x \rightarrow \infty$) in Eqn.~\ref{eq:loop2}.
Another method uses the fact that $P_1$ is a polynomial, thus regular at $x = a$. So, we have:
\beq
\label{eq:derF}\partial_a F = \lim_{x \rightarrow a} (x - a)\left[\beta W_2(x,x) + \beta W_1(x)^2 + (\beta - 1)W_1'(x) - \frac{N\beta}{t}V'(x)W_1(x)\right]
\eeq
We will see later another method based on a residue formula.
\item[$\bullet$] Wrt the times $t$ and $t_j$. We have from the matrix integral:
\bea
\partial_t F & = & \frac{N\beta}{t^2} \Big\langle V(M) \Big\rangle \nonumber \\
& = & - \frac{N\beta}{t^2}\,\Res_{x \rightarrow \infty} \mathrm{d}x\,V(x)W_1(x) \\
\partial_{t_j} F & = & -\frac{N\beta}{t} \Big\langle \frac{M^{j}}{j} \Big\rangle \nonumber \\
& = & \frac{N\beta}{t}\,\Res_{x \rightarrow \infty} \mathrm{d}x\,\frac{x^{j}}{j}W_1(x)
\eea
\end{itemize}

\subsection{Topological and $\hbar$ expansion}

\subsubsection{Definition}

We introduce two parameters:
\beq
\nu = \frac{N\sqrt{\beta}}{t},\qquad \hbar = \frac{t}{N}\left(1 - \frac{1}{\beta}\right)
\eeq
We shall consider the large $N$ expansion of the free energy $F(a)$ and of the correlators $W_n(x_1,\ldots,x_n)$, in the form:
\bea
F & = & \sum_{g,k \geq 0} \nu^{2 - 2g}\hbar^k\,F^{g,k} \\
\label{eq:AAA1} W_n(x_1,\ldots,x_n) & = & \beta^{- n/2} \sum_{g,k \geq 0} \nu^{2 - 2g - n} \hbar^{k}\,W_n^{g,k}(x_1,\ldots,x_n)
\eea
The prefactor in Eqn.~\ref{eq:AAA1} was chosen such that:
\beq
\frac{\partial}{\partial V(x_{n + 1})}\,W_{n}^{g,k}(x_1,\ldots,x_n) = W_{n + 1}^{g,k}(x_1,\ldots,x_n,x_{n+1})
\eeq
For the $\beta$-matrix ensembles, $\hbar$ is of order $1/N$. Those definition are such that $F^{g,k}$ and the functions $W_n^{g,k}$ are independent of $N$ and $\beta$. They depend on $a$, on $t$, and on the potential $V(x)$. For instance, the free energy is:
\bea
F(a)  & = &  \frac{N^2}{t^2}\beta F^{0,0} + \frac{N}{t}(\beta - 1)F^{0,1} + F^{1,0} + (\beta + \beta^{-1} - 2)F^{0,2} \nonumber \\
& & + \frac{t}{N}\left((1 - \beta^{-1})F^{1,1} + (\beta - 3 + 3\beta^{-1} - \beta^{-2})F^{0,3}\right) \nonumber \\
& & + \frac{t^2}{N^2}\left(\beta^{-1}F^{2,0} + (1 - 2\beta^{-1} + \beta^{-2})F^{1,2} + (\beta - 4 + 6\beta^{-1} - 4\beta^{-2} + \beta^{-3})F^{0,4}\right) \nonumber \\
& & + o(1/N^2)
\eea

\subsubsection{Expansion of the loop equations}
All the same, we expand:
\beq
P_n(x,I) = \beta^{-n/2}\,\sum_{g,k \geq 0}^{\infty} \nu^{2 - 2g} \hbar^k\,P_n^{g,k}(x,I) \nonumber
\eeq
\label{sec:loopeq}
To the leading order, the loop equation (Eqn.~\ref{eq:loop2}) reads:
\beq
\left(W_1^{0,0}(x)\right)^2 = V'(x) W_1^{0,0}(x) - P_1^{0,0}(x) - \frac{c^{0,0}}{x-a}
\eeq
This relation is sometimes called \textbf{master loop equation}. And, for higher $g,k$, we have:
\bea
\label{eq:master1}(V'(x)-2W_1^{0,0}(x))\,W_1^{g,k}(x)
&=&  W_2^{g-1,k}(x,x) + \frac{\mathrm{d}}{\mathrm{d}x}W_1^{g,k-1}(x)  \cr
&& + \sum_{0 \leq g'\leq g,\,0 \leq k'\leq k}'\,W_1^{g',k'}(x)W_1^{g-g',k-k'}(x) \cr
&& + P_1^{g,k}(x) + \frac{c^{g,k}}{x-a}
\eea
And, for $n$-point correlators, the expansion of Eqn.~\ref{eq:loopeqh2} is:
\bea
& & (V'(x)-2W_1^{0,0}(x))\,W_{n}^{g,k}(x,I) \nonumber \\
&= &  W_{n+1}^{g-1,k}(x,x,I) + \frac{\mathrm{d}}{\mathrm{d}x} W_{n}^{g,k-1}(x,J)  \nonumber \\
&& + \sum_{0 \leq g'\leq g,\,0 \leq k' \leq k,\,J\subseteq I}'\, W_{|J|+1}^{g',k'}(x,J)W_{n-|J|}^{g-g',k-k'}(x,I\setminus J) \nonumber \\
&& +\sum_{x_i \in I}  \frac{\mathrm{d}}{\mathrm{d}x_i}\left(\frac{W_{n - 1}^{g,k}(x,I\setminus\{x_i\})- \frac{x_i - a}{x - a}\,W_{n - 1}^{g,k}(I)}{x-x_i}\right)  \nonumber \\
&& + P_{n}^{g,k}(x,I) + \frac{c_n^{g,k}(I)}{x-a}
\eea
In those expressions, $\sum'$ means that we exclude of the sum the terms where $W_1^{0,0}$ appear (we put them by hand in the LHS).

The general solution of those equations was found in \cite{CE06,CEM09}. All $W_n^{g,0}$ are "geometric covariants", and all $F^{g,0}$ are "geometric invariants", associated to the plane curve $\mathcal{L}$:
\beq
(x,y) \in \mathcal{L} \subseteq \mathbb{C}^2 \qquad \Leftrightarrow \qquad  y = \frac{V'(x)}{2} - W_1^{0,0}(x)
\eeq
The function $y$ defined on $\mathcal{L}$ is closely related to the density of eigenvalues in the thermodynamic limit ($N \rightarrow \infty$). The triplet $(x,y,\mathcal{L})$ is often called the \textbf{spectral curve}. A deep property of stable $F^{g,0}$ is their invariance under symplectic transformations, i.e. all transformations $(x,y) \rightarrow (x_1,y_1)$ such that $|\mathrm{d}x\wedge \mathrm{d}y| = |\mathrm{d}x_1\wedge \mathrm{d}y_1|$. It is conjectured that stable $F^{g,k}$ for $k \neq 0$ are also invariant in this sense, but there is no proof at the moment. So, we often call $F^{g,0}$ the "symplectic invariants of the curve", but we use the name "geometric quantities associated to the curve" to refer to general $F^{g,k}$.

In practice, the first step is to compute the \textbf{unstable} correlators, that is the correlators $W_n^{g,k}$ such that $2 - 2g - k - n \geq 0$. The most important one is the resolvent to leading order, $W_1^{0,0}$. The other unstable correlators are $W_2^{0,0}$ and $W_1^{0,1}$. Then, all the remaining correlators follow. In a similar way\footnote{We may identify the free energies $F^{g,k}$ to "$0$-point correlators, $W_0^{g,k}$".}, there are general formulas for unstable free energies ($F^{0,0}$, $F^{0,1}$, $F^{0,2}$, $F^{1,0}$) and a uniform algorithm to compute all the other $F^{g,k}$.

Notice that unstable terms come with a nonnegative power of $N$ in the topological expansion, whereas \textbf{stable} terms are coefficients in the $o(1)$. In a sense, this $o(1)$ is easier to compute than the nondecaying terms.

Let us also mention that all $W_n^{g,k}$ (and $W_0^{g,k} \equiv F^{g,k}$) have a combinatorial interpretation, as counting (possibly non-orientable) ribbon graphs of genus $g$, with $k$ M\"{o}bius strips, and $n$ marked vertices \cite{CEM09}.

\subsubsection{First few loop equations}

\begin{itemize}
\item[$\bullet$] For $n = 1$:
\bea
\label{eq:W101L}2y(x)W_1^{0,1}(x) & = & \frac{\mathrm{d}}{\mathrm{d}x}W_1^{0,0}(x) + P_1^{0,1}(x) + \frac{c^{0,1}}{x - a} \\
2y(x)W_1^{1,0}(x) & = & W_2^{0,0}(x,x) + P_1^{1,0}(x) + \frac{c^{1,0}}{x - a} \\
2y(x)W_1^{0,2}(x) & = & \frac{\mathrm{d}}{\mathrm{d}x}W_1^{0,1}(x) + \left(W_1^{0,1}(x)\right)^2 + P_1^{0,2}(x) + \frac{c^{0,2}}{x - a}
\eea
and so on.
\item[$\bullet$] For $n = 2$:
\bea
2y(x)W_2^{0,1}(x,x') & = & \frac{\mathrm{d}}{\mathrm{d}x}W_2^{0,0}(x,x') + W_1^{0,1}(x)\left(2W_2^{0,0}(x,x') + \frac{1}{(x - x')^2}\right) \nonumber \\
& & - \frac{\mathrm{d}}{\mathrm{d}x'}\left(\frac{x' - a}{x - a}\,\frac{W_1^{0,1}(x')}{x - x'}\right) + P_2^{0,1}(x;x') + \frac{c_2^{0,1}(x')}{x - a} \nonumber \\
2y(x)W_2^{1,0}(x,x') & = & W_3^{0,0}(x,x,x') + W_1^{0,1}(x)\left(2W_2^{0,0}(x,x') + \frac{1}{(x - x')^2}\right) \nonumber \\
& & - \frac{\mathrm{d}}{\mathrm{d}x'}\left(\frac{x' - a}{x - a}\,\frac{W_1^{1,0}(x')}{x - x'}\right) + P_2^{1,0}(x;x') + \frac{c_2^{1,0}(x')}{x - a} \nonumber \\
2y(x)W_2^{0,2}(x,x') & = & \frac{\mathrm{d}}{\mathrm{d}x}W_2^{0,1}(x,x') + 2W_1^{0,1}(x)W_2^{0,1}(x,x') \nonumber \\
& & + 2W_1^{0,2}(x)\left(W_2^{0,0}(x,x') + \frac{1}{(x - x')^2}\right) \nonumber \\
 & & - \frac{\mathrm{d}}{\mathrm{d}x'}\left(\frac{x' - a}{x - a}\,\frac{W_1^{0,2}(x')}{x - x'}\right) + P_2^{0,2}(x;x') + \frac{c_2^{0,2}(x')}{x - a}\nonumber
\eea
and so on.

\item[$\bullet$] For $n = 3$:
\bea
2y(x)W_3^{0,0}(x,x_1,x_2) & = & 2W_2^{0,0}(x,x_1)W_2^{0,0}(x,x_2) + \frac{W_2^{0,0}(x,x_2)}{(x - x_1)^2} + \frac{W_2^{0,0}(x,x_1)}{(x - x_2)^2} \nonumber \\
& & + \left(\frac{\mathrm{d}}{\mathrm{d}x_1}\frac{\frac{x_1 - a}{x - a}}{x - x_1} + \frac{\mathrm{d}}{\mathrm{d}x_2}\frac{\frac{x_2 - a}{x - a}}{x - x_2}\right)W_2^{0,0}(x_1,x_2) \nonumber \\
& & + P_3^{0,0}(x;x_1,x_2) + \frac{c_3^{0,0}(x_1,x_2)}{x - a} \nonumber
\eea
and so on.
\end{itemize}

\subsection{Density of eigenvalues}

The expected density of eigenvalues is defined as
\beq
\varrho(x) = \left< \sum_{i = 1}^N \delta(x-\lambda_i)\right>
\eeq
The linear form $f \rightarrow \int \rho f$, defined on an appropriate space of test functions, has a large $N$ expansion (in the weak topology) of the type\footnote{In general, this is not true when $\rho(x)$ is considered as a function. $\rho(x)$ has a pointwise $\rho^{0}$, but exhibits beyond leading order fast oscillations, even in the one cut case. We thank a referee for pointing an erroneous statement in an earlier version of the text.}
\beq
\varrho(x) \mathop{=}^{\mathrm{weak}} \frac{1}{\sqrt\beta}\sum_{g,k \geq 0} \left(\frac{N\sqrt{\beta}}{t}\right)^{1-2g-k}\,\left(\sqrt{\beta} - \frac{1}{\sqrt{\beta}}\right)^k\,\varrho^{g,k}(x)
\eeq

The $1$-point correlator, $W_1(x)$, is the Stieltjes transform of the density. This means that, if $\mathcal{D}$ contains the eigenvalue support,
\beq
W_1(x) = \int_{\mathcal{D}}\,\frac{\mathrm{d}x'\,\varrho(x')}{x-x'}
\eeq
Conversely, one can see the density as the discontinuity of $W_1(x)$:
\beq
2i\pi \varrho(x) = W_1(x-i0^+)-W_1(x+i0^+)
\eeq

In the thermodynamic limit ($N \rightarrow \infty$):
\beq
\frac{t}{N}\,\varrho(x) \mathop{\sim}_{N \rightarrow \infty} \varrho^{0,0}(x) = \frac{-1}{2i\pi}\left(W_1^{0,0}(x + i0^+) - W_1^{0,0}(x - i0^+)\right)
\eeq
The origin of $\mathcal{D}$ will become more precise when we present the analytical structure of $W_1^{0,0}$ derived from the master loop equation.

When $x \rightarrow \infty$, we know that $W_1^{0,0} \sim t/x$, and this implies the normalization:
\bea
\int_{\mathcal{D}} \mathrm{d}x\,\varrho^{0,0}(x) & = & t
\eea

\section{One-cut solution to the loop equations}
\label{sec:cacq}

The spectral curve is defined by the master loop equation, that we rewrite:
\beq
\label{eq:W100s}
W_1^{0,0}(x) = \frac{V'(x)}{2} - \sqrt{\frac{V'(x)^2}{4}- P_1^{0,0}(x) - \frac{c^{0,0}}{x-a}}
\eeq
If eigenvalues accumulate in the large $N$ limit on one segment, of the form $[b,a[$, we know that $W_1^{(0,0)}(x)$ must have only one cut $[b,a]$ in the complex plane, and thus $W_1^{0,0}(x)$ must be of the form
\beq
\label{eq:formrho}W_1^{0,0}(x) = \frac{V'(x)}{2} - \frac{M(x)}{2}\:\sqrt{\frac{x-b}{x-a}}
\eeq
where $M(x)$ is a polynomial to determine.

In principle, knowing that $W_1^{0,0}(x)$ has one cut and matching Eqn.~\ref{eq:formrho} with the large $x$ behavior $W_1^{0,0}(x) \sim {t/x}$, allow to compute explicitly the endpoint $b$, as well as the polynomials $M(x)$ and $P_1^{0,0}(x)$. The easiest way to do this computation is to use Zhukovsky map.

The purpose of this article is to illustrate our methods, so we assume that:
\begin{itemize}
\item[$\bullet$] the support of eigenvalues is of the form $[b,a[$, in particular it is \textbf{connected}
\item[$\bullet$] $[b,a[ \subseteq \mathbb{R}$
\end{itemize}
Then, one can prove that $W_n^{g,k}(x_1,\ldots,x_n)$ has one cut on $[b,a[$ in each variable for $x_i \in [b,a[$, and that it is holomorphic in each variable outside this cut.

We present detailed computations of the unstable correlators ($W_1^{0,0}$, $W_2^{0,0}$, $W_1^{0,1}$) and the unstable free energies ($F^{0,0}$, $F^{0,1}$, $F^{0,2}$, $F^{1,0}$), and the algorithm to compute the other $W_n^{g,k}$ and $F^{g,k}$. General formulas for $W_1^{0,0}$ and the procedure to compute all $W_n^{k,g}$ when the support of eigenvalues is a finite set of paths in the complex plane can be found in the literature \cite{C06,CE06}.

\subsection{Spectral curve, $W_1^{0,0}$ and density}

The Zhukovsky map is defined as:
\beq
x(z) = \alpha + \gamma(z + 1/z)
\eeq
with $\alpha = (a + b)/2$ and $\gamma = (a - b)/4$. We also have:
\beq
\gamma(z - 1/z) = \sqrt{(x - a)(x - b)},\qquad \sqrt{\frac{x - b}{x - a}} = \pm \frac{1 + z}{1 - z}
\eeq
It is then clear that $W_1^{0,0}(x(z))$ and $\rho(x(z))$ are rational function of $z$ (for this reason, $z$ is called a uniformizing variable).

Consider any potential $V$ of finite degree $d$:
\beq
V(x) = t_0 + \sum_{j = 1}^{d} \frac{t_j}{j}\,x^j
\eeq
We have:
\beq
V'(x) \mathop{\sim}_{x \rightarrow \infty} t_d\,x^{d - 1}
\eeq
We assume that $V$ is such that the matrix integral converges. Let us decompose
\beq
\label{eq:Vu}V'(x(z)) = u_0 + \sum_{k = 1}^{d - 1} u_k(z^k + z^{-k})
\eeq
With the Zhukovsky variable, $w_1(z) = W_1^{0,0}(x(z))$ is the solution to the problem:
\begin{itemize}
\item[$(i)$] $w_1$ is a meromorphic function defined on $\mathbb{C}$.
\item[$(ii)$] $w_1$ is meromorphic outside the unit disk, and :
\beq
w_1(z) \mathop{\sim}_{z \rightarrow \infty} t/x(z) \nonumber
\eeq
\item[$(iii)$] $w_1$ is regular at $z = -1$ (i.e $x = b$), and has atmost a simple pole at $z = 1$ (i.e $x = a$).
\item[$(iv)$] $w_1$ satisfies the following equation (deduced from the loop equation for a 1-cut solution):
\beq
\label{eq:W1lin}\forall z \in \mathbb{C},\qquad w_1(x(z)) + w_1(x(1/z)) = V'(x(z)) \nonumber
\eeq
\end{itemize}
The solution is unique, and it is easy to find the answer in the form:
\beq
\label{eq:W10} \encadremath{w_1(z) = \frac{r}{z - 1} + \sum_{k = 1}^{d - 1} u_k\,z^{-k}}
\eeq
The behavior when $z \rightarrow \infty$ determines $b$ through:
\beq
u_0 = -r\,,\qquad \gamma(r + u_1) = t
\eeq
Then, the density in the thermodynamic limit $\rho(x) = \rho^{0,0}(x)$ is:
\bea
\rho(x(z)) & = & \frac{1}{2i\pi}\left(w_1(1/z) - w_1(z)\right) \nonumber \\
\label{eq:rhoz}& = & \frac{1}{2i\pi}\left(r\;\frac{1 + z}{1 - z} + \sum_{k = 1}^{d - 1} u_k(z^k - z^{-k})\right)
\eea

\subsubsection{More on the density}

It is sometimes convenient to use the variable $\theta$ such that $z = e^{i\theta}$, that is $x = \alpha + 2\gamma\cos\theta$. We have:
\beq
\rho(x(e^{i\theta})) = \frac{1}{2\pi} \left(r\,\mathrm{cotan}(\theta/2) + 2 \sum_{k = 1}^{d - 1} u_k\,\sin(k\theta) \right)
\eeq
Eqn.~\ref{eq:Vu} is in fact a decomposition of $V'(x)$ on the basis of Chebyshev polynomials of the first kind $T_n(\cos\theta) = \cos(n\theta)$.
\beq
V'(x) = u_0 + \sum_{k \geq 1} 2u_k\,T_k(\cos \theta)
\eeq
Likewise, it is useful to decompose $V$ on the basis $(T_n)_{n \in \mathbb{N}}$:
\beq
V(x) = v_0 + \sum_{k \geq 1} 2v_k\,T_k(\cos\theta)
\eeq
Then, $V'$ is naturally decomposed on the basis of Chebyshev polynomials of the second kind $U_n(\cos\theta) = \frac{\sin (n+1)\theta}{\sin\theta}$.
\beq
\label{eq:Vv}V'(x) = \frac{1}{\gamma} \sum_{k \geq 1} kv_k\,U_k(\cos\theta)
\eeq

We recall that $T_n$ are orthogonal polynomials for some scalar product $(\cdot|\cdot)_T$:
\bea
\big(T_m\,\big|\,T_n\big)_T  =  \frac{1}{\pi}\int_{0}^{\pi} \mathrm{d}\theta\,T_m(\cos\theta)T_n(\cos\theta) & = & \frac{1}{2}\delta_{n,m}\,\qquad \mathrm{if}\, n \neq 0 \\
& = & \phantom{\frac{1}{2}}\delta_{0,m}\,\qquad \mathrm{if}\, n = 0
\eea
Similarly, $U_n$ are orthogonal polynomials for a related scalar product $\big(\cdot\,\big|\,\cdot\big)_U$:
\bea
\label{eq:orthoU}\big(U_m\,\big|\,U_n\big)_U = \frac{1}{\pi}\int_0^{\pi} \mathrm{d}\theta\,\sin^2(\theta)\,U_n(\theta)U_m(\theta) & = & \frac{1}{2}\delta_{n,m}\,\qquad \mathrm{if}\, n \neq 0 \\
& = & \phantom{\frac{1}{2}}\delta_{0,m}\,\qquad\mathrm{if}\, n = 0
\eea
The expansion of Eqn.~\ref{eq:Vv} is related to Eqn.~\ref{eq:Vu} through:
\bea
\forall k \geq 2 & & \gamma(u_{k - 1} - u_{k + 1}) = \Big(T_{k - 1} - T_{k + 1}\,\Big|\,\sum_{n \geq 1} nv_nU_{n - 1}\Big)_T \nonumber \\
& & \phantom{\gamma(u_{k - 1} - u_{k + 1}) } = \Big(2\sin^2\theta\,U_{k - 1}\,\Big|\,\sum_{n \geq 1} nv_n\,U_{n - 1}\Big)_T \nonumber \\
& & \phantom{\gamma(u_{k - 1} - u_{k + 1}) } = 2\Big(U_{k - 1}\,\Big|\,\sum_{n \geq 1} nv_nU_{n - 1}\Big)_U \nonumber \\
& & \phantom{\gamma(u_{k - 1} - u_{k + 1}) } = kv_{k} \nonumber
\eea
One can check that this formula holds for $k = 1$ as well.
\beq
\forall k \geq 1\,,\qquad \gamma(u_{k - 1} - u_{k + 1}) = kv_k
\eeq

So, we have a compact expression for the density $\mathrm{d}x\,\rho(x)$:
\bea
\mathrm{d}x(z)\,\rho(x(z)) & = & \frac{1}{2i\pi}\,\frac{\mathrm{d}z}{z}\,\gamma\left(z - \frac{1}{z}\right)\left(r \frac{1 + z}{1 - z} + \sum_{k \geq 1} u_k(z^k + z^{-k})\right) \nonumber \\
& = & \frac{1}{2i\pi}\frac{\mathrm{d}z}{z}\,\left(-2\gamma(r + u_1) + \sum_{k \geq 1} \gamma(u_{k - 1} - u_{k + 1})(z^k + z^{-k})\right) \nonumber \\
\label{eq:dens} & = & \frac{\mathrm{d}\theta}{\pi}\left(- t + \sum_{k \geq 1} kv_k\,T_k(\cos\theta)\right)
\eea

The eigenvalue support $\mathcal{D} = [b,a[$ is mapped to the upper unit circle from $-1$ to $1$ (we call it $-\mathcal{C}_+$), and to $[\pi,0[$ in the variable $\theta$. The $-1$ sign in Eqn.~\ref{eq:dens} is consistent with this reverse orientation in the variable $\theta$, since the total weight of $\rho$ is $t$.

\subsubsection{The moment polynomial}

The density can also be written:
\beq
\rho(x) = -\frac{M(x)}{2\pi}\sqrt{\frac{x - b}{a - x}}
\eeq
From Eqn.~\ref{eq:W100s}, one knows that $\rho(x)$ behaves as $\sqrt{x - b}$ when $x \rightarrow b$. So, $M$ must be a polynomial, and it is called the \textbf{moment}. Since the support of eigenvalues stays connected, $b \in ]-\infty,a[$ is determined as the unique zero of the polynomial in bracket such that $\rho(x)$ remains nonnegative for $x \in [b,a[$. Besides, the degree of $M$ is $d - 1$, so it admits $d - 1$ zeroes $x_1,\ldots,x_{d - 1}$. For generic $V$ (i.e a non critical potential), those zeroes are distinct from $a$ and $b$. They must lie outside $]a,b[$, and are either real, or come in pairs of complex conjugate values. Let us call $s_j$, the point such that $x(s_j) = x_j$ and $|s_j| > 1$ (we also have $x(1/s_j) = x_j$). We have:
\beq
M(x) = t_d\prod_{j = 1}^{d - 1} (x - x_j)
\eeq
Also, we may write in a factorized form:
\beq
\label{eq:facto}\rho(x(e^{i\theta})) = \frac{t_d\gamma^{d - 1}}{2\pi}\,\mathrm{cotan}(\theta/2)\,\prod_{j = 1}^{d - 1} \frac{(e^{i\theta} - s_j)(e^{i\theta} - 1/s_j)}{e^{i\theta}}
\eeq
For $p \in \mathbb{N}$, we may also introduce the following quantities:
\beq
m_a^{(p)} = \sum_{k \geq 1} k^p\,u_k,\qquad m_b^{(p)} = \sum_{k \geq 1} (-1)^kk^p\,u_k
\eeq
They allow to relate the values of the $p$-th derivative of $M(x)$ at points $a$ and $b$, to the coefficients $u_k$. By expanding Eqn.~\ref{eq:rhoz2} near $z \rightarrow 1$ ($x = a$), or $z \rightarrow -1$ ($x = b$):
\bea
\left\{\begin{array}{rcl} M(a) & = & r \\ M'(a) & = & m_a^{(1)}/\gamma \\ \ldots & & \end{array}\right. &\qquad & \left\{\begin{array}{rcl} M(b) & = & 4m_b^{(1)} - r \\ M'(b) & = & -(m_b^{(1)} + 2m_b^{(3)})/3\gamma \\ \ldots & &\end{array}\right. \nonumber
\eea

\subsubsection{The spectral curve}

The function $y$ defining the spectral curve is given by:
\bea
y(z) & = & i\pi\rho(z) = \frac{1}{2}\left(w_1(1/z) - w(z)\right) \nonumber \\
 & = & \frac{1}{2}\frac{z + 1}{z - 1}\,M(x(z)) \nonumber \\
\label{eq:rhoz2} & = & \frac{1}{2}\left(r\,\frac{z + 1}{z - 1} + \sum_{k \geq 1} u_k(z^k - z^{-k})\right)
\eea

\subsection{The prepotential, $F^{0,0}$}

\subsubsection{General formula for $F^{0,0}$}

The prepotential may be found by integrating Eqn.~\ref{eq:F00der}, which might be useful for simple examples or numerics, but do give a hint for a general formula. For this purpose, one has from homogeneity (invariance of $F$ with respect to $(t,t_j) \mapsto (\lambda t,\lambda t_j)$ for $\lambda > 0$):
\beq
\label{eq:F00} F^{0,0} = \frac{1}{2}\Res_{x \rightarrow \infty} \mathrm{d}x\,V(x)W_1^{0,0}(x) + \frac{t}{2}\partial_t F^{0,0}
\eeq
We have introduced the chemical potential $\mu = \partial_t F^{0,0}$. We state \cite{E06} that it can be computed as:
\beq
\label{eq:mu}\mu = 2t\ln\gamma + \Res_{x \rightarrow \infty} \frac{\mathrm{d}z(x)}{z(x)}\,V(x)
\eeq
With our notations, we find $\mu = 2t\ln\gamma - v_0$.

To compute the first term in $F^{0,0}$, we can move the contour from $\infty$ to $\mathcal{C}$ surrouding the cut. We have:
\bea
& & \Res_{x \rightarrow \infty}\mathrm{d}x\,V(x)W_1^{0,0}(x) \nonumber \\
& = & \Res_{z \rightarrow \infty} \frac{\mathrm{d}z}{z}\left(v_0 + \sum_{k \geq 1} v_k(z^k + z^{-k})\right)\left(r(1 + 1/z) + \sum_{k \geq 1} kv_k z^{-k}\right) \nonumber \\
& = & -\frac{1}{2}t\,v_0 + \frac{1}{2}\sum_{k \geq 1} k\,v_k^2 \nonumber
\eea
Eventually, we collect the terms
\beq
\encadremath{F^{0,0} = \frac{1}{2}\sum_{k \geq 1} k\,v_k^2 - tv_0 + t^2\ln\gamma}
\eeq

\subsubsection{Computation of $\partial_a F^{0,0}$}
After Eqn.~\ref{eq:derF}, $\partial_a F^{0,0}$ appears as the coefficient of the simple pole $1/(x - a)$ in:
\beq
\left[(W_1^{0,0}(x))^2 - V'(x)W_1^{0,0}(x) - P_1^{0,0}(x)\right] \nonumber
\eeq
Since $P_1^{(0)}$ is regular at $x = a$, and $W_1^{0,0}$ behaves as $(x - a)^{-1/2}$, we have when $x \rightarrow a$ (or $z \rightarrow 1$):
\bea
(W_1^{0,0}(x))^2 & \sim & \frac{\partial_a F^{0,0}}{x - a} \nonumber \\
& \sim & \frac{\partial_a F^{0,0}}{\gamma}\frac{1}{(z - 1)^2} \nonumber
\eea
With Eqn.~\ref{eq:W10}, we obtain:
\beq
\label{eq:F00der}\encadremath{\partial_a F^{0,0} = \gamma\,r^2} \nonumber
\eeq

\subsection{Finite correction to the $1$-point correlator, $W_1^{0,1}$}

The loop equation (Eqn.~\ref{eq:W101L}) is:
\beq
M(x)\sqrt{\frac{x - b}{x - a}}\,W_1^{0,1}(x) = \frac{\mathrm{d}}{\mathrm{d}x} W_1^{0,0}(x)  + P_1^{0,1}(x) + \frac{c^{0,1}}{x-a}
\eeq
Hence:
\beq
W_1^{0,1}(x) = \frac{1}{4(x-a)}-\frac{1}{4(x-b)}-\frac{M'(x)}{2M(x)}  + \frac{Q(x)}{M(x)\sqrt{(x-a)(x-b)}}
\eeq
where $Q(x)$ is a polynomial. In the Zhukovsky variable $z$, this expression is clearly a rational function of $z$.
We define:
\beq
\omega_1^{0,1}(z) \equiv W_1^{0,1}(x(z))\,x'(z)
\eeq

We see that $\omega_1^{0,1}(z)$ has simple poles at $z=1$ (with residue $1/2$), at $z=-1$ (with residue $-1/2$), and at $z = 1/s_j$, $z = s_j$ and $z = 0$. Since $W_1^{0,1}(x)$ is holomorphic outside $[b,a[$, $\omega_1^{0,1}(z)$ must be meromorphic outside the unit disk. Therefore, the polynomial $Q(x)$ is determined such that the poles at $s_j$ cancel.
This implies that the poles at $1/s_j$ have residue $-1$.
Besides, $W_1^{0,1}(x) \in O(1/x^2)$ when $x \rightarrow \infty$ imply that $\omega_1^{0,1}(z) \in O(1/z^2)$ when $z \rightarrow \infty$. This determines the residue of the pole at $z = 0$. Eventually we get:
\beq
\label{eq:W101}
\encadremath{\omega_1^{0,1}(z) = \frac{1}{2(z-1)} - \frac{1}{2(z+1)} - \sum_{j = 1}^{d - 1} \left(\frac{1}{z-1/s_j}-\frac{1}{z}\right)}
\eeq

\subsection{The entropy, $F^{0,1}$}

\subsubsection{General formula for $F^{0,1}$}

The first correction $F^{0,1}$ was computed by Dyson himself \cite{Dyson}. In our notations, it is the opposite of the entropy:
\beq
\label{eq:F01} F^{0,1} =  \int_{b}^a \mathrm{d}x\,\rho(x)\ln(\rho(x)/t) \nonumber
\eeq

Let us insert the expression the density (Eqn.~\ref{eq:dens}), and its factorized form (Eqn.~\mbox{\ref{eq:facto}}):
\bea
\label{eq:fff}F^{0,1} & = & \int_{0}^{\pi} \frac{\mathrm{d}\theta}{\pi}\,\left(t - \sum_{k \geq 1} kv_k\,T_k(\cos\theta)\right)\ln\rho(e^{i\theta}) \nonumber \\
& = &  \int_{0}^{\pi} \frac{\mathrm{d}\theta}{\pi}\,\left(t - \sum_{k \geq 1} kv_k\,T_k(\cos\theta)\right)\left|\frac{t_d\gamma^{d - 1}}{2\pi t}\,\mathrm{cotan}(\theta/2)\,\prod_{j = 1}^{d - 1} \frac{(e^{i\theta} - s_j)(e^{i\theta} - 1/s_j)}{e^{i\theta}}\right| \nonumber \\
& = & t\,\ln\left(\frac{t_d\gamma^{d -1}}{2\pi t}\right) - \mathcal{I} + \sum_{j = 1}^{d - 1} [\mathcal{J}(s_j) + \mathcal{J}(1/s_j)]
\eea
where:
\bea
\mathcal{I} & = & \int_{0}^{\pi} \frac{\mathrm{d}\theta}{\pi}\,\left(t - \sum_{k \geq 1} kv_k\,T_k(\cos\theta)\right)\ln\left|\mathrm{tan}(\theta/2)\right| \nonumber \\
\mathcal{J}(s) & = & \int_{0}^{\pi} \frac{\mathrm{d}\theta}{\pi}\,\left(t - \sum_{k \geq 1} kv_k\,T_k(\cos\theta)\right)\ln\left|e^{i\theta} - s\right| \nonumber
\eea
To compute $\mathcal{I}$, we use the classical integral $\int_{0}^{\pi} \ln|\mathrm{tan}(\theta/2)| = 0$ and do an integration by part:
\bea
\mathcal{I} & = & \int_{0}^{\pi} \frac{\mathrm{d}\theta}{\pi}\,\sum_{k \geq 1} v_k\,\frac{\sin(k\theta)}{\sin\theta} \nonumber \\
& = & \sum_{k\,\mathrm{odd}} v_{k} \nonumber
\eea
We compute $\mathcal{J}(s)$ for $|s| > 1$:
{\small
\bea
\mathcal{J}(s) & = & \int_{0}^{\pi} \frac{\mathrm{d}\theta}{\pi}\,\left(t - \sum_{k \geq 1} kv_k\,T_k(\cos\theta)\right)\left(\ln|s| + \frac{1}{2}\ln(1 - e^{i\theta}/s) + \frac{1}{2}\ln(1 - e^{-i\theta}/\overline{s})\right) \nonumber \\
& = & t\ln|s| + \int_{0}^{\pi}\frac{\mathrm{d}\theta}{\pi}\,\left(t - \sum_{k \geq 1} kv_k\,T_k(\cos\theta)\right)\nonumber \\
& & \phantom{t\ln|s| + \int_{0}^{\pi}\frac{\mathrm{d}\theta}{\pi}\,\left(\right)} \cdot \left(-\sum_{m = 1}^{\infty} \mathrm{Re}(s^{-m})\,\frac{T_m(\cos\theta)}{m} + \mathrm{Im}(s^{-m})\,\frac{\sin(m\theta)}{m}\right) \nonumber \\
& = & t\ln|s| + \frac{1}{2}\sum_{k \geq 1} v_k\,\mathrm{Re}(s^{-k}) - \sum_{k \geq 1} v_k\,\mathrm{Im}\left(\sum_{m = 1}^{\infty} A_{m,k}\frac{1}{s^m}\right) \nonumber
\eea}
$A_{m,k}$ is a real coefficient which vanish when $m$ and $k$ are of the same parity, and else is equal to:
\beq
A_{m,k} = \frac{1}{\pi}\left(\frac{1}{m + k} - \frac{1}{m - k}\right) \nonumber
\eeq
We compute $\mathcal{J}(1/s)$ for $|s| > 1$ in the same way:
{\normalsize \bea
\mathcal{J}(1/s) & = & \int_0^{\pi} \frac{\mathrm{d}\theta}{\pi}\,\left(t - \sum_{k \geq 1} kv_k\,T_k(\cos\theta)\right)\left(\frac{1}{2}\ln(1 - e^{-i\theta}/s) + \frac{1}{2}\ln(1 - e^{i\theta}/\overline{s})\right) \nonumber \\
& = &  \int_0^{\pi} \frac{\mathrm{d}\theta}{\pi}\,\left(t - \sum_{k \geq 1} kv_k\,T_k(\cos\theta)\right)\nonumber \\
& & \phantom{\int_0^{\pi} \frac{\mathrm{d}\theta}{\pi}\,\left(\right)}\cdot \left(\sum_{m = 1}^{\infty} - \mathrm{Re}(s^{-m})\frac{T_m(\cos\theta)}{m} - \mathrm{Im}(s^{-m})\frac{\sin(m\theta)}{m}\right) \nonumber \\
& = & \frac{1}{2}\sum_{k \geq 1} v_k\,\mathrm{Re}(s^{-k}) - \sum_{k \geq 1} v_k\,\mathrm{Im}\left(\sum_{m = 1}^{\infty} A_{m,k}\frac{1}{s^m}\right) \nonumber
\eea}
In the sum $\mathcal{J}(s) + \mathcal{J}(1/s)$, the term containing an imaginary part cancel. Moreover, we know that if $s_j$ is a zero of $\rho$, then $\overline{s_j}$ is also a zero of $\rho$. So, $\sum_j s_j^{-k}$ is already real, and the absolute value in the log only remove the sign of $s_j$. Collecting the terms:
\beq
\label{eq:F01e}\encadremath{F^{0,1} = t\,\ln\left(\frac{t_d}{2\pi t}\prod_{j = 1}^{d - 1} \gamma |s_j|\right) - \sum_{k\,\mathrm{odd}} v_k + \sum_{k \geq 1} v_k\,\left(\sum_{j = 1}^{d - 1} \frac{1}{s_j^k}\right)}
\eeq

\subsubsection{Computation of $\partial_a F^{0,1}$}

One may also compute $\partial_a F^{0,1}$, which requires less algebra. We have when $x \rightarrow a$ (or $z \rightarrow 1$) by expansion of Eqn.~\ref{eq:derF}:
\bea
\left[\left(2W_1^{0,0}(x) - V'(x)\right)W_1^{0,1}(x) + \partial_x\,W_1^{0,0}(x)\right] & \sim & \frac{\partial_a F^{0,1}}{x - a} \nonumber \\
& \sim & \frac{\partial_a F^{0,1}}{\gamma}\,\frac{1}{(z - 1)^2} \nonumber
\eea
We may extract $\partial_a F^{0,1}$ by computing a residue. Since $x'(z) \sim 2\gamma\,(z - 1)$ when $z \rightarrow 1$, we write:
\bea
\partial_a F^{0,1} & = & \frac{1}{2} \Res_{z \rightarrow 1} \mathrm{d}z\,x'(z)\left[\left(2w_1^{0,0}(z) - V'(x(z))\right)w_1^{0,1}(z) + \frac{\partial_z w_1^{0,0}(z)}{x'(z)}\right] \nonumber \\
& = & -\Res_{z \rightarrow 1} \mathrm{d}z\,y(z) \omega_1^{0,1}(z)
\eea
With $\omega_1^{0,1}$ given by Eqn.~\ref{eq:W101}, and $y(z) = i\pi\rho(z)$ given by Eqn.~\ref{eq:rhoz2}, we obtain:
\bea
\partial_a F^{0,1} & = & \Res_{z \rightarrow 1} \left(\frac{r}{z - 1} + \frac{r}{2} + O(z - 1)\right)\left(\frac{1}{2}\frac{1}{z - 1} - \frac{1}{4} - \sum_{j = 1}^{d - 1} \frac{1}{s_j - 1} + O(z - 1)\right) \nonumber
\eea
Hence:
\beq
\encadremath{\partial_a F^{0,1} = r\,\sum_{j = 1}^{d - 1} \frac{1}{s_j - 1}}
\eeq

\subsection{Leading order of the 2-point correlator, $W_2^{0,0}$}

By application of the loop insertion operator to Eqn.~\ref{eq:W1lin}, we find that $w_2(z_1,z_2) = W_2^{0,0}(x(z_1),x(z_2))$ is the unique solution to the following problem:
\begin{itemize}
\item[$(i)$] $w_2(z_1,z_2)$ is a meromorphic function in $z_1$ and $z_2$, which is symmetric in its two variables.
\item[$(ii)$] $w_2(z_1,z_2)$ is meromorphic for $z_1$ and $z_2$ outside the unit disk, and behaves as $\frac{1}{z_1^2}$ when $z_1 \rightarrow \infty$.
\item[$(iii)$] For $z_2 \neq \pm 1$, $w_2(z_1,z_2)$ has atmost simple poles when $z_1 \rightarrow \pm 1$ (i.e when $x_1 = a$ or $x_1 = b$).
\item[$(iv)$] $w_2$ satisfies the following equation (deduced from the loop equation for a 1-cut solution):
\beq
\label{eq:W2lin}\forall z_1 \in \mathbb{C},\;\forall z_2 \in \mathbb{C}\setminus\{\pm 1\},\qquad w_2(z_1,z_2) + w_2(1/z_1,z_2) = - \frac{1}{(x(z_1) - x(z_2))^2} \nonumber \eeq
\end{itemize}
This problem is the same with or without hard edges, and its solution is universal (it only depends on the endpoints $a$ and $b$, through the Zhukovsky map), it is given by:
\beq
w_2(z_1,z_2)  =  \frac{1}{x'(z_1)x'(z_2)}\,\frac{1}{(z_1 - z_2)^2} - \frac{1}{(x(z_1) - x(z_2))^2}
\eeq
or equivalently:
\beq
\encadremath{w_2(z_1,z_2) =  \frac{1}{x'(z_1)\,x'(z_2)}\frac{1}{(z_1z_2 - 1)^2}}
\eeq

\subsection{Unstable $W_n^{k,g}$}

\subsubsection{Analytical structure}

Initially, $W_n^{g,k}(x(z_1),\ldots,x(z_n))$ was defined and meromorphic for $|z_i| > 1$ and $z_i \neq \pm 1$. We shall see that it can be analytically continued (this is true under the 1-cut assumption) as a meromorphic function on the whole complex plane, i.e. as a rational function of its variables. We call this function $w_n^{g,k}(z_1,\ldots,z_n)$.
Let us introduce
\beq
\omega_n^{g,k}(z_1,\ldots,z_n) = x'(z_1)\cdots x'(z_n)\,w_n^{g,k}(z_1,\ldots,z_k)
\eeq

We shall now investigate the analytical properties of $\omega_n^{g,k}$. We already know the important property:
\begin{itemize}
\item[$(ii)$] $\omega_n^{g,k}(z_1,\ldots,z_n)$ has no singularity for $|z_i| \geq 1$ and $z_i \neq \pm 1$.
\end{itemize}
Let us rewrite the loop equation by separating "singular" and "regular" terms. We recall the notation:
\beq
w_1^{0,0}(x(z)) = \frac{V'(x(z))}{2} - y(z) \nonumber
\eeq
and we introduce
\bea
\mathcal{E}_{n}^{g,k}(z,I) & = & \left[\omega_{n + 1}^{g - 1,k}(z,z,I) + \right.\nonumber \\
 & & + \sum_{J \subseteq I,\,0 \leq g' \leq g,\,0 \leq k' \leq k}^{'} \omega_{|J|+1}^{g',k'}(z,J)\,\omega_{n - |J|}^{g - g',k - k'}(z,I\setminus J) \nonumber \\
& & + \partial_z(\omega_n^{g,k-1}(z,I)) - \frac{x''(z)}{x'(z)}\,\omega_n^{g,k - 1}(z,I)  \nonumber \\
& & \left.+ \sum_{z_i \in I} (x'(z))^2\,\partial_{z_i}\left(\frac{\frac{\omega_{n - 1}^{g,k}(z,I\setminus\{z_i\})}{x'(z)} - \frac{x(z_i) - a}{x(z) - a}\,\frac{\omega_{n - 1}^{g,k}(I)}{x'(z_i)}}{x(z) - x(z_i)}\right)\right] \\
& & \nonumber \\
\mathcal{S}_n^{g,k}(z,I) & = & \frac{\mathcal{E}_n^{g,k}(z,I)}{2y(z)x'(z)} \\
& & \nonumber \\
\label{eq:Rdef}\mathcal{R}_n^{g,k}(z,I) & \equiv & \left(P_n^{g,k}(x(z),I) + \frac{c_n^{g,k}}{x(z) - a}\right)\frac{x'(z)}{2y(z)}\,\prod_{z_i \in I} x'(z_i)
\eea
The loop equations may be rewritten:
\beq
\label{eq:z}\omega_n^{g,k}(z,I) = \mathcal{S}_n^{g,k}(z,I) + \mathcal{R}_n^{g,k}(z,I)
\eeq

$\mathcal{S}_n^{g,k}$ has the same polar structure as $\omega_n^{g,k}$, apart from the fact that it may have a pole in the physical sheet, at $z = s_j$. To find $\omega_n^{g,k}$, we should add $\mathcal{R}_n^{g,k}$ to $\mathcal{S}_n^{g,k}$, thus cancelling this pole. The remainder $\mathcal{R}_n^{g,k}(z,I)$ has the following properties:
\begin{itemize}
\item[$\bullet$] $\mathcal{R}_n^{g,k}(z,I)$ is regular when $z \rightarrow \pm 1$.
\item[$\bullet$] $\mathcal{R}_n^{g,k}(z,I)$ may have poles only at $z = s_j$, $z = 1/s_j$, and they are atmost simple.
\end{itemize}

By a recursion of $2g - 2 + n + k$ using the loop equations, It is then possible to complete the description of the analytical structure of $\omega_n^{g,k}$:
\begin{itemize}
\item[$(i)$] $\omega_n^{g,k}(z_1,\ldots,z_n)$ is a rational function of all its variables (thus defined on $\mathbb{C}^n$).
\item[$(iii)$] $\omega_n^{g,k}(z_1,\ldots,z_n)$ may have poles at $z_l = \{\pm 1\}$, $z_l = 1/s_j$ ($1 \leq j \leq d -1$), $z_l = 1/z_i$ ($i \neq l$), and $z_l = 0$.
\item[$(iv)$] $\omega_n^{g,k}(z_1,\ldots,z_n) \in O(1/z_i^2)$ when $z_i \rightarrow \infty$.
\end{itemize}

It is clear that the loop equations, the analytical properties $(i)-(iv)$, and the data of $\omega_1^{0,0}$, $\omega_1^{0,1}$, $\omega_2^{0,0}$ determines uniquely the family of rational functions $\omega_n^{g,k}$.

\begin{itemize}
\item[$(v)$] We know an extra property of the $\omega_n^{g,k}$ we are looking for: they are symmetric functions of all their variables. Although not obvious, the fact that this unique solution is symmetric can be proven from the loop equations.

\item[$(vi)$] It is also possible to prove by recursion that the pole at $z = 0$ is atmost simple (we assume $d \geq 2$). It comes from the leading term of $P_n^{g,k}(x(z),I)$ (in $\mathcal{R}_n^{g,k}(z,I)$) when $z \rightarrow 0$ ($x \rightarrow \infty$), all the other terms in $\mathcal{S}_n^{g,k}(z,I)$ and $\mathcal{R}_n^{g,k}(z,I)$ are regular when $z \rightarrow 0$.

\end{itemize}

\subsubsection{Algorithm}
\label{sec:algo}
Let us recall that $x(z) = x(1/z)$ and $y(z) = -y(1/z)$. Since $P_n^{g,k}(x(z),I)$ is a polynomial in $x(z)$, we know that:
\beq
\label{eq:Rsym}\mathcal{R}_n^{g,k}(1/z,I) = z^2\,\mathcal{R}_n^{g,k}(z,I)
\eeq
Since this rational function in $z$ has only simple poles at $z = s_j$ and $z = 1/s_j$, it must be of the form:
\beq
\label{eq:Req1}\mathcal{R}_n^{g,k}(z,I) = \sum_{j = 1}^{d - 1} \mathcal{R}_{n,j}^{g,k}(I)\,\left(\frac{1}{z - 1/s_j} - \frac{1}{z - s_j}\right)
\eeq
To cancel the pole at $z = s_j$ in $\mathcal{S}_n^{g,k}$, we must have:
\beq
\label{eq:Req2}\mathcal{R}_{n,j}^{g,k}(I) = \Res_{\xi \rightarrow s_j} \mathrm{d}\xi\,\mathcal{S}_n^{g,k}(\xi,I)
\eeq
Since $\mathcal{S}_{n}^{g,k}(z,J)$ has only a simple pole at $z = s_j$, we have:
\beq
\label{eq:resj1}\mathcal{R}_{n,j}^{g,k}(I) = \frac{\mathcal{E}_n^{g,k}(s_j,I)}{2y'(s_j)x'(s_j)}
\eeq

This is enough to compute $\omega_n^{g,k}$ with Eqn.~\ref{eq:z}, by recursion on $2g + k + n$. We can also find a residue formula, based on the following principle. The divergent part of a rational function $f$ at a pole $\lambda$ is given by:
\beq
[f(z_0)]_{-,\lambda} = \Res_{z \rightarrow \lambda} \frac{\mathrm{d}z\,f(z)}{z_0 - z}
\eeq

$\omega_n^{g,k}(z_0,I)$ is the sum of its divergent part near $z_0 = \pm 1$, $z = 1/z_i$, $z = 1/s_j$ and $z = 0$. After $(vi)$, $\omega_n^{g,k}(z_0,I)$ has only a simple pole at $z_0 = 0$, and it must be such that $\omega_n^{g,k}(z_0,I) \in O(1/z_0^2)$ when $z_0 \rightarrow 0$. Thus:
\bea
\omega_n^{g,k}(z_0,I) & = & \Res_{z \rightarrow 1,-1,1/z_i,1/s_j}\,\mathrm{d}z\left(\frac{1}{z_0 - z} - \frac{1}{z_0}\right)\mathcal{S}_n^{g,k}(z,I) \nonumber \\
\label{eq:resfo} & & + \sum_{j = 1}^{d - 1} \Res_{z \rightarrow s_j}\,\mathrm{d}z\left(\frac{1}{z_0 - 1/s_j} - \frac{1}{z_0}\right) \mathcal{S}_n^{g,k}(z,I)
\eea

To compute the residues at $z = \pm 1$, the term $-\partial_{z_i}\left(\omega_n^{g - 1}(I)/\cdots\right)$ do not contribute and we can replace $\mathcal{S}_n^{g,k}$ by:
\bea
\widehat{\mathcal{E}}_n^{g,k}(z,I) & = & \left(\omega_{n + 1}^{g - 1}(z,z,I) + \right.\nonumber \\
& & + \sum_{J \subseteq I,\,0 \leq g' \leq g,\,0 \leq k' \leq k}^{'} \omega_{|J| + 1}^{g',k'}(z,J)\omega_{n - |J|}^{g - g',k - k'}(z,I\setminus J) \nonumber \\
& & + \partial_z(\omega_n^{g,k - 1}(z,I)) - \frac{x''(z)}{x'(z)}\omega_n^{g,k - 1}(z,I) \nonumber \\
& & \left. + \sum_{z_i \in I} \frac{x'(z)\,x'(z_i)}{(x(z) - x(z_i))^2}\omega_{n - 1}^{g,k}(z,I\setminus\{z_i\})\right) \\
\widehat{\mathcal{S}}_n^{g,k}(z,I) & = & \frac{\mathcal{E}_n^{g,k}(z,I)}{2y(z)x'(z)}
\eea

\subsubsection{Improvement for $\omega_n^{g,0}$}

This algorithm can be further simplified in the case $k = 0$, i.e the limit of a hermitian matrix model. Notice first that $\omega_n^{g,0}$ is obtained by the former algorithm without computing any $\omega_n^{g,k}(z,I)$ with $k \neq 0$. Then, one can prove the following "mirror relation" by a recursion using the loop equations:
\beq
\label{eq:312}\omega_n^{g,0}(1/z,I) = z^2\,\omega_n^{g,0}(z,I)
\eeq
There is no similar mirror relation for $\omega_n^{g,k}(z,I)$ with $k \neq 0$ because of the term $\partial_{z}\omega_n^{g,k - 1}(z,I)$ in the loop equations. As a consequence, $\omega_n^{g,0}$ can only have pole at $z = \pm 1$. Indeed, no pole at $z = z_i$ (resp. $z = s_j$) imply no pole at $z = 1/z_i$ (resp. $z = 1/s_j$) by the mirror relation. All the same, $\omega_n^{g,0} \in O(1/z^2)$ when $z \rightarrow \infty$ imply that $\omega_n^{g,0}$ is regular at $z = 0$. Besides:
\bea
\label{eq:r2}\Res_{\xi \rightarrow \pm 1} \omega_n^{g,0}(\xi,I) & = & -\Res_{\xi \rightarrow \pm 1}\mathrm{d}\xi\,\omega_n^{g,0}(\xi,I) = 0
\eea

Therefore, we can compute $\omega_n^{g,0}$ in a simple way if the $\omega_{n'}^{g',0}$'s with $2g' + n' < 2g + n$ are known: $\omega_n^{g,0}$ is the divergent part in the Laurent expansion of $\mathcal{S}_n^{g,k}(z,I)$ at $z \rightarrow \pm 1$. We can again replace $\mathcal{S}_n^{g,0}(z,I)$ by $\widehat{\mathcal{S}}_n^{g,0}(z,I)$ and write:
\bea
\label{eq:lab}\omega_n^{g,0}(z,I) & = & \Res_{z \rightarrow 1,-1}\,\frac{\mathrm{d}z}{z_0 - z} \widehat{\mathcal{S}}_n^{g,0}(z,I)
\eea

\subsubsection{Standard form of the topological recursion}

Formula \ref{eq:lab} is valid for one-cut solution of loop equations: we used the fact that $\omega_n^{g,0}$ is a rational function of its variables. It is false for other cases. However, we are going to transform slightly this formula into the form of the "residue formula of the topological recursion" \cite{EORev}, which still hold in more general settings.

With the mirror relation, one can prove that $\mathcal{S}_n^{g,0}(1/z,I) = z^2$ (or use directly Eqn.~\ref{eq:z} and Eqn.~\ref{eq:Rsym}).
Hence:
\bea
\omega_n^{g,0}(z,I) & = & \frac{1}{2}\Res_{z \rightarrow 1,-1}\,\mathrm{d}z\left(\frac{1}{z_0 - z} - \frac{1}{z_0 - 1/z}\right)\widehat{\mathcal{S}}_n^{g,0}(z,I) \nonumber \\
& = & \Res_{z \rightarrow 1,-1}\,\frac{\mathrm{d}z}{2y(z)x'(z)} \left(\frac{1}{z_0 - z} - \frac{1}{z_0 - 1/z}\right) \nonumber \\
& & \cdot\left[\omega_{n + 1}^{g - 1,0}(z,z,I) + \sum_{J \subseteq I,\,0 \leq h \leq g}^{'} \omega_{|J| + 1}^{h,0}(z,I)\,\omega_{n - |J|}^{g - h,0}(z,I\setminus J) \right. \nonumber \\
& & \left.+ \sum_{z_i \in I} \frac{x'(z)\,x'(z_i)}{(x(z) - x(z_i))^2}\,\omega_{n - 1}^{g,0}(z,I\setminus\{z_i\})\right]
\eea

Let us define:
\bea
\varpi_2^{0,0}(z_1,z_2) & = & \omega_2^{g,0}(z_1,z_2) + \frac{1}{2}\frac{\mathrm{d}x(z_1)\,\mathrm{d}x(z_2)}{(x(z_1) - x(z_2))^2} \\
"\varpi_2^{0,0}(z_1,z_2)" & = & \omega_2^{g,0}(z_1,z_2) + \frac{\mathrm{d}x(z_1)\,\mathrm{d}x(z_2)}{(x(z_1) - x(z_2))^2}
\eea
and if $(n,g) \neq (2,0)$:
\beq
\varpi_n^{g,0}(z_1,\ldots,z_n) = "\varpi_n^{g,0}(z_1,\ldots,z_n)" = \omega_n^{g,0}(z_1,\ldots,z_n)
\eeq
We also define the point $\overline{z} \neq z$ such that $x(\overline{z}) = x(z)$. In general, it is defined at least locally around the simple zeroes of $x'$ (the branchpoints at which we take the residues). Here, we have $\overline{z} = 1/z$, which happens to be defined globally.

Using the mirror property, we can rewrite:
\bea
\omega_n^{g,0}(z_0,I) & = & \Res_{z \rightarrow 1,-1}\mathrm{d}z\,K(z_0,z)\left["\varpi_{n + 1}^{g - 1,0}(z,\overline{z},I)" + \right.\nonumber \\ & & \phantom{\Res_{z \rightarrow 1,-1}\mathrm{d}z\,K(z_0,z)} \left. + \sum_{J \subseteq I,\, 0 \leq h \leq g}^{'} \varpi_{|J| + 1}^{g,0}(z,J)\varpi_{n - |J|}^{g - h,0}(\overline{z},I \setminus J)\right] \nonumber \\
\label{eq:toptop}& &
\eea
This is the standard form of the topological recursion associated to a spectral curve $(\mathcal{L},x,y)$, also called "residue formula". The residues are computed at the simple zeroes of $\mathrm{d}x$ on $\mathcal{L}$. $"\varpi_{2}^{0,0}(z_1,z_2)"$ is the heat kernel of $\mathcal{L}$, also called \textbf{Bergman kernel}. $K(z,z_0)$ is the \textbf{recursion kernel}. Here:
\bea
K(z_0,z) & = & \frac{1}{4y(z)x'(1/z)}\left(\frac{1}{z_0 - z} - \frac{1}{z_0 - 1/z}\right) \nonumber \\
& = & \frac{1}{2(y(z) - y(\overline{z}))x'(\overline{z})}\left(\frac{1}{z_0 - z} - \frac{1}{z_0 - \overline{z}}\right) \nonumber \\
& = & \frac{\int_{\overline{z}}^{z} \mathrm{d}z'\,"\varpi_{2}^{0,0}(z',z_0)"}{2(y(z) - y(\overline{z}))x'(\overline{z})}
\eea
The last line is the expression of the recursion kernel in the general case.

 The residue formula (Eqn.~\ref{eq:toptop}) can be seen as a definition of some quantities $\omega_n^{g,0}$ (for $2 - 2g - n < 0$) from the data of the spectral curve $(\mathcal{L},x,y)$. These quantities satisfy some loop equations. When this formalism is applied to a spectral curve coming from a matrix model, it gives the topological expansion of the correlators of the matrix model. As a matter of fact, the residue formula allow to express all the solutions of loop equations (non necessarily with 1-cut). The quantities $\omega_n^{g,0}$ defined by Eqn.~\ref{eq:toptop} enjoy interesting properties, and can be computed with a diagrammatic technique (see the review \cite{EORev}).

 Let us mention that another residue formula and a diagrammatic technique also exists in general for all $\omega_n^{g,k}$ (the case of hyperelliptic spectral curves was treated in \cite{CE06}).

\subsubsection{Example}

We apply the algorithm described in Section~\ref{sec:algo} to find the $1/N$ corrections to the $1$-point correlator.

\begin{itemize}
\item[$\bullet$] $\underline{(n,g,k) = (1,1,0)}$. $\mathcal{E}_1^{1,0}(z) = \omega_2^{0,0}(z,z)$ has a symmetry when $z \mapsto 1/z$. So, killing the pole at $z = s_j$ also kills the pole at $z = 1/s_j$. Thus, $\omega_1^{1,0}(z)$ is the singular part at $z = \pm 1$ of $\mathcal{S}_1^{(1,0)}(z)$. We find:
\bea
\omega_1^{1,0}(z) & = & \frac{1}{16\gamma M(a)}\frac{1}{(z - 1)^2} + \nonumber \\
& & \frac{1}{4\gamma M(b)} \frac{1}{(z + 1)^4} - \frac{1}{4\gamma M(b)}\frac{1}{(z + 1)^3} + \frac{-M(b) + 4\gamma M'(b)}{16\gamma M(b)^2}\frac{1}{(z + 1)^2} \nonumber
\eea
\item[$\bullet$] $\underline{(n,g,k) = (1,0,2)}$. This is the first case for which there is no short cut.
{\footnotesize
\bea
& & \omega_1^{0,2}(z) \nonumber \\
& = &  -\frac{3}{16\gamma M(a)}\frac{1}{(z - 1)^2} + \frac{5}{4\gamma M(b)}\frac{1}{(z + 1)^4} + \frac{1}{4\gamma M(b)}\left(-13 + \sum_{j} \frac{1}{s_j + 1}\right)\,\frac{1}{(z + 1)^3} \nonumber \\
& & + \left[\frac{1}{\gamma M(b)}\left(\frac{3}{16} + \sum_{j \neq l} \frac{1}{(s_j + 1)(s_l + 1)}\right) + \frac{5M'(b)}{4M(b)^2}\right]\,\frac{1}{(z + 1)^2} \nonumber \\
& & + \frac{1}{\gamma M(b)}\left(\sum_{j} \frac{2}{s_j + 1}\right)\left[\left(\sum_j \frac{1}{(s_j + 1)^2} - \frac{1}{s_j + 1}\right) - \frac{\gamma M'(b)}{M(b)}\right]\,\left(\frac{1}{z + 1} - \frac{1}{z}\right) \nonumber \\
& & + \sum_{j} \frac{-2}{\gamma^2M'(x_j)}\frac{1}{(s_j - 1)(s_j + 1)^3}\,\frac{1}{(z - 1/s_j)^3} + \left[- \frac{M''(x_j)}{\gamma (M'(x_j))^2}\frac{1}{(s_j + 1)^3} \right.\nonumber \\
& & \left. + \frac{2}{\gamma M'(x_j)}\frac{2s_j}{(s_j - 1)(s_j + 1)^3}\left(\frac{s_j^2 - s_j + 1}{(s_j - 1)(s_j + 1)} + \sum_{l \neq j} \frac{1}{s_l - s_j}\right) \right]\,\frac{1}{(z - 1/s_j)^2} \nonumber \\
& & + \left[\frac{1}{6}\,\frac{1}{\gamma M'(x_j)^3}\left(2M'''(x_j)M'(x_j) - 3(M''(x_j))^2\right)\frac{s_j - 1}{s_j + 1} \right.\nonumber \\
& & + \frac{M''(x_j)}{\gamma (M'(x_j))^2}\frac{s_j}{(s_j + 1)^3}\left(1 - \sum_{l \neq j} \frac{s_j}{s_l - s_j}\right) \nonumber \\
& & + \frac{1}{\gamma^2M'(x_j)}\frac{s_j^2}{(s_j - 1)(s_j + 1)^3}\left(\sum_{l \neq j} \frac{2s_j^2}{(s_l - s_j)^2} + \frac{s_j}{s_l - s_j}\frac{-s_j^3 + s_j^2 + 2s_j - 3}{(s_j - 1)(s_j + 1)}\right. \nonumber \\
& & + \sum_{l \neq j} \frac{2}{(s_ls_j - 1)^2} + \frac{2(-2s_j + 3)}{s_ls_j - 1} \nonumber \\
& & + \left.\left. \sum_{l \neq j,\,l'\neq j,\,/\,l \neq l'} \frac{1}{(s_ls_j - 1)(s_{l'}s_j - 1)}\right)\right]\cdot\left(\frac{1}{z - 1/s_j} - \frac{1}{z}\right) \nonumber
\eea}
\end{itemize}

\subsubsection{The Bergman-tau function, $F^{1,0}$}

Finding $F_{1,0}$ requires the integration of $\omega_1^{1,0}(x)$ with respect to the variations of the potential, and we have:
\beq
\omega_1^{1,0}(z_0) = \Res_{z \rightarrow 1,-1} \frac{\mathrm{d}z}{z_0 - z}\,\frac{\omega_2^{0,0}(z,z)}{2y(z)x'(z)}
\eeq
A closed formula (when there is no hard edges) was first given by Ambj{\o}rn, Chekhov and Makeenko \cite{ACM}, and it was generalized to hyperelliptic curves by Akemann \cite{Ake96}. As we have seen, $\omega_2^{0,0}(z,z)$ is closely related to the Bergman kernel of the spectral curve: the general method of integration of $\omega_1^{1,0}$ was studied in \cite{EKoKo04}, and the solution was expressed in term of the "Bergman tau-function". It depends on the positions of the edges (hard or not), and is defined as the result of integration of a closed 1-form. When there exist some hard edges, the result is slightly modified and was studied by Chekhov\footnote{Beware, the free energy is defined as $-\ln Z$ in Chekhov, while we use the opposite convention.} \cite{C06}. We apply it in the one-cut case:
\beq
\label{eq:F10} \encadremath{F^{1,0} = -\frac{1}{24}\,\ln\left[M(a)^3\,M(b)\,\left(\frac{a - b}{t}\right)^4\right]}
\eeq

\subsubsection{The Polyakov anomaly, $F^{0,2}$}
\label{sec:F02}
This term was first considered by Wiegmann and Zabrodin (Eqn. 5.20 of \cite{WZ06}). We adapt the expression given in \cite{CE06} when there are some hard edges, in the one-cut case. By comparing with $\partial_a F^{0,2}$ computed for the gaussian potential, we found that the result of \cite{CE06} must be multiplied by $1/4$ to be correct. Namely, the correct formula in our case is:
\beq
F^{0,2} = F^{0,2}_{(I)} + \frac{1}{12}\ln\left[\frac{M(a)^3}{M(b)}\,\left(\frac{a - b}{t}\right)^2\right]
\eeq
where:
\beq
F^{0,2}_{(I)} = \frac{1}{16\pi^2} \int_{\mathcal{C}_{\mathrm{ext}}} \frac{\mathrm{d}\rho(z_1)}{\rho(z_1)} \int_{\mathcal{C}_+} \mathrm{d}z_2\left(\frac{1}{z_2 - z_1} - \frac{1}{z_2 - 1/z_1}\right)\ln(\rho(z_2)/t) \nonumber
\eeq
$\mathcal{C}_{\mathrm{ext}}$ is a contour outside the unit disk, such that $\mathrm{ind}_{\mathcal{C}_{\mathrm{ext}}}(z) = 1$ whenever $|z| \leq 1$, and $\mathrm{ind}_{\mathcal{C}_{\mathrm{ext}}}(s_j) = -1$ for all $1 \leq j \leq d - 1$.

Let us compute, for $|z_1| > 1$:
\bea
A(z_1) & = & \frac{1}{\pi}\int_{\mathcal{C}_+}\mathrm{d}z_2\,\left(\frac{1}{z_2 - z_1} - \frac{1}{z_2 - 1/z_1}\right)\ln(\rho(z_2)/t) \nonumber \\
& = & \int_{0}^{\pi} \frac{\mathrm{d}\theta_2}{\pi}\,\left( - 1 - \sum_{k \geq 1} \frac{2}{k z_1^k}T_k(\cos\theta_2)\right)\ln(\rho(e^{i\theta_2})/t) \nonumber
\eea
We have already computed this kind of integral in Eqn.~\ref{eq:fff}. We just have to substitute:
\beq
t \longrightarrow -1\,,\qquad kv_k \longrightarrow 2/z_1^k \nonumber
\eeq
in the result Eqn.~\ref{eq:F01e}. Therefore:
\beq
A(z_1) = -\ln\left(\frac{t_d}{2\pi t}\prod_{j = 1}^{d - 1} \gamma |s_j|\right) - \sum_{k\,\mathrm{odd}} \frac{2}{k}\frac{1}{z_1^k} + \sum_{k \geq 1} \frac{2}{k}\frac{1}{z_1^k}\left(\sum_{l = 1}^{d - 1} \frac{1}{s_l^k}\right) \nonumber
\eeq
Then:
\bea
F_{(I)}^{0,2} & = & -\frac{1}{8i\pi}\int_{\mathcal{C}_{\mathrm{ext}}}\frac{\mathrm{d}\rho(z)}{\rho(z)} A(z) \nonumber \\
& = & -\frac{1}{8i\pi}\int_{\mathcal{C}_{\mathrm{ext}}}\mathrm{d}z\,\left[\frac{1}{z + 1} - \frac{1}{z - 1} + \sum_{j = 1}^{d - 1} \left(\frac{1}{z - s_j} + \frac{1}{z - 1/s_j} - \frac{1}{z}\right)\right]A(z) \nonumber
\eea
The contour integral can be computed for each term of $A(z)$. Individually, each term gives a meromorphic function to integrate. So, for each term, we can move the contour to surround the poles outside the unit disk, that is $z = \infty$, and $z = s_j$ ($1 \leq s_j \leq d - 1$). We obtain:
\bea
F_{(I)}^{0,2} & = & -\frac{1}{4}\left[-(d - 1)\ln\left(\frac{t_d}{2\pi t}\prod_{j = 1}^{d - 1} \gamma |s_j|\right) + \sum_{j = 1}^{d - 1} \ln\left(\frac{t_d}{2\pi t}\prod_{l = 1}^{d - 1} \gamma |s_l|\right) \right. \nonumber \\
& & \left.+ \sum_{j = 1}^{d - 1} \sum_{k\,\mathrm{odd}} \frac{2}{k}\frac{1}{s_j^k} - \sum_{k \geq 1} \frac{2}{k}\frac{1}{s_j^k}\left(\sum_{l = 1}^{d - 1} \frac{1}{s_l^k}\right)\right] \nonumber \\
& = & -\frac{1}{4}\sum_{j = 1}^{d - 1}\left[\ln\left(\frac{s_j + 1}{s_j - 1}\right) + 2\sum_{l = 1}^{d - 1}\ln\left(1 - \frac{1}{s_js_l}\right)\right] \nonumber \\
& = & \frac{1}{4}\left[\sum_{j = 1}^{d - 1} \ln \left(\frac{s_j^4}{(s_j - 1)(s_j + 1)^3}\right) - \sum_{1 \leq j < l \leq d - 1} \ln\left(1 - \frac{1}{s_js_l}\right)\right] \nonumber
\eea
We also write the second term of $F^{0,2}$ in terms of $s_j$'s:
\beq
\frac{1}{12}\ln\left(16\frac{\gamma^2}{t^2}\,\frac{M(a)^3}{M(b)}\right) = \frac{1}{12}\ln\left(\frac{16t_d^2\gamma^{2d}}{t^2}\right) + \frac{1}{12}\sum_{j = 1}^{d - 1} \ln\left(\frac{1}{s_j^2}\,\frac{(s_j - 1)^6}{(s_j + 1)^2}\right) \nonumber
\eeq
Therefore:
\beq
\label{eq:F02} \encadremath{F^{0,2} = \frac{1}{12}\left[\ln\left(\frac{16t_d^2\gamma^{2d}}{t^2}\right) + \sum_{j = 1}^{d - 1} \ln\left(\frac{s_j^{10}(s_j - 1)^3}{(s_j + 1)^{11}}\right)\right] - \sum_{1 \leq j < l \leq d - 1} \ln \left(1 - \frac{1}{s_js_l}\right)}
\eeq

\subsection{Unstable $F^{g,k}$}

\subsubsection{Computation of $\partial_a F^{g,k}$}

We can repeat the argument which led to the residue representation for $\partial_a F^{0,1}$. It shows that the derivative term $\partial_z w_1^{g,k}(z)$ has a vanishing contribution to the residue, and we have:
\beq
\partial_a F^{g,k} = \frac{1}{2}\Res_{z \rightarrow 1} \left( -2y(z)\omega_1^{g,k}(z) + \sum_{0 \leq g' \leq g,\,0 \leq k' \leq k}^{'}
\frac{\omega_{1}^{g',k'}(z)\,\omega_{1}^{g - g',k - k'}(z)}{x'(z)} +  \frac{\omega_2^{g - 1,k}(z,z)}{x'(z)} \right)
\eeq

Another method is to extract directly $c_1^{g,k} = -\partial_a F^{g,k}$ from $\mathcal{R}_1^{g,k}$, which is a by-product of the algorithm. Indeed, after Eqn.~\ref{eq:Rdef} and recalling $y(z) \sim -r/(z - 1)$ when $z \rightarrow 1$:
\beq
\mathcal{R}_1^{g,k}(z) =  \frac{\partial_a F^{g,k}}{r} + O(z - 1)
\eeq
Then, after Eqns.~\ref{eq:Req1}-\ref{eq:Req2}:
\beq
\encadremath{\partial_a F^{g,k} = -r\sum_{j = 1}^{d - 1} \frac{s_j + 1}{s_j - 1}\,\Res_{\xi \rightarrow s_j} \mathrm{d}\xi\,\widehat{S}_1^{g,k}(\xi)}
\eeq
With this second method, we obtain:
{\small
\bea
\partial_a F^{1,0} & = & \frac{r}{\gamma^2} \sum_{j = 1}^{d - 1} \frac{1}{M'(x_j)}\frac{s_j^4}{(s_j - 1)^4(s_j + 1)^4} \nonumber \\
\partial_a F^{0,2} & = & \sum_{j = 1}^{d - 1} \frac{r}{\gamma^2 M'(x_j)}\frac{s_j^2}{(s_j - 1)^2(s_j + 1)^2}\left(- 2\sum_{l \neq j} \frac{1}{(s_ls_j - 1)^2} - \frac{2}{s_ls_j - 1}\frac{s_j^2 - s_j + 1}{(s_j - 1)(s_j + 1)} \right. \nonumber \\
& & \left.- \sum_{l \neq j,\,l' \neq j\,/\,l\neq l'} \frac{1}{(s_ls_j - 1)(s_{l'}s_j - 1)}  + \frac{2s_j^3 - 3s_j^2 + 4s_j - 2}{(s_j - 1)^2(s_j + 1)^2}\right) \nonumber
\eea}

\subsubsection{Algorithm for $F^{g,k}$}

We state the so-called integration formula \cite{CE06} for the stable free energies:
\beq
\label{eq:Fg} F^{g,k} = \frac{1}{2 - 2g - k} \Res_{z \rightarrow 1,-1,1/s_j} \mathrm{d}z\,\phi(z)\omega_1^{g,k}(z)
\eeq
where $\phi$ is a primitive of $y(z)x'(z)$:
\beq
\phi(z) = \int_1^z y\mathrm{d}x
\eeq
For $k = 0$ and $g \geq 2$, this formula simplifies into:
\beq
F^{g,0} = \frac{1}{2 - 2g} \Res_{z \rightarrow 1,-1} \mathrm{d}z\,\phi(z)\omega_1^{g,k}(z)
\eeq

This formula is not valid (or simply do not make sense) for unstable free energies. The correct formula were Eqns.~\ref{eq:F00}-\ref{eq:mu} for $F^{0,0}$, Eqn.~\ref{eq:F01} for $F^{0,1}$, Eqn.~\ref{eq:F10} for $F^{1,0}$, and Eqn.~\ref{eq:F02} for $F^{0,2}$.

\subsection{Normalization of the geometric quantities}
\label{sec:constant}
\subsubsection{Position of the problem}

Let us define the partition function of a $\beta$-matrix model by:
\beq
\mathcal{Z}_{N,\beta}(a,V) = \int_{]-\infty,a]^{\infty}} \mathrm{d}\lambda_1\cdots\mathrm{d}\lambda_N\,|\Delta(\lambda)|^{2\beta}\,\exp\left(- \frac{N\beta}{t}\sum_{i = 1}^{N} V(\lambda_i)\right)
\eeq
It is known that, when the leading order ($N \rightarrow \infty$) of $W_1(x)$ has only one cut in the complex plane, there exists an asymptotic expansion ($N \rightarrow \infty$) of the form:
\beq
\ln \mathcal{Z}_{N,\beta}(a,\beta,V) = D_{\beta}\ln N + \sum_{g,k \geq 0}^{\infty} \left(\frac{N\sqrt{\beta}}{t}\right)^{2 - 2g - k}\left(\sqrt{\beta} - \frac{1}{\sqrt{\beta}}\right)^k\,\left(f^{g,k}(a,V) + d^{g,k}\right)
\eeq
$D_{\beta}$ depends only on $\beta$, and $d^{g,k}$ is a constant. They do not depend on $a$, $V$ and $N$.

\medskip

To this matrix model, we associate a plane curve $\mathcal{S}_{a,V}$, which do not depend on $N$ and $\beta$:
\beq
\mathcal{S}_{a,V}\; : \quad y = W_1^{0,0}(x) - \frac{V'(x)}{2}
\eeq
Out of the curve were construct some geometric quantities $\mathcal{F}^{g,k}(\mathcal{S}_{a,V})$. We can form a formal asymptotic series in $N$:
\beq
\mathcal{F}_{N,\beta}(\mathcal{S}_{a,V}) = \sum_{g,k \geq 0} \left(\frac{N\sqrt{\beta}}{t}\right)^{2 - 2g - k} \left(\sqrt{\beta} - \frac{1}{\sqrt{\beta}}\right)^k\,\mathcal{F}^{g,k}(\mathcal{S}_{a,V})
\eeq
The theorem completely proved in \cite{CE06} (and only sketched here) is that there exists constants $\widehat{d}^{g,k}$ such that
\beq
\forall g,k\geq 0 \qquad f^{g,k}(a,V) = \mathcal{F}^{g,k}(\mathcal{S}_{a,V}) + \widehat{d}^{g,k}
\eeq
In other words, there exists a constant $C_{N,\beta}$ depending only on $N$ and $\beta$, such that:
\beq
\label{eq:norm}\mathcal{Z}_{N,\beta}(a,V) = C_{N,\beta}\,\exp\left(\mathcal{F}_{N,\beta}(\mathcal{S}_{a,V})\right)
\eeq
and $\ln C_{N,\beta}$ has a topological expansion, modulo an extra $\ln(N)$ term.

If one wishes to test Eqn.~\ref{eq:norm} with Monte-Carlo computations of $\mathcal{Z}$, or to compare the predictions for the statistics of the maximal eigenvalue to Tracy-Widom laws, it is important to know the normalization constant $C_{N,\beta}$.

\subsubsection{Value of $C_{N,\beta}$}

It can be computed explicitly by specializing to some $a$ and $V$ such that:
\begin{itemize}
\item[$\bullet$] one has a close formula for $F^{g,k}$ for any $g,k$.
\item[$\bullet$] one has independently a close formula for the eigenvalue integral.
\end{itemize}
Comparing the two gives the constant $C_{N,\beta}$. Such expressions are available for the Gaussian potential ($V(x) = x^2/2$) in the limit $a \rightarrow -\infty$. In this case indeed, the eigenvalue integral can be reduced to a Selberg integral on one hand, and unstable $F^{g,k}$ vanish on the other hand. We do the computations in Appendix~\ref{app:C}, and the result is:
\bea
C_{N,\beta} & = & \exp\left[-\beta N^2\ln N + \beta N^2\left(\frac{3}{2} - \ln \beta\right) + (\beta - 1)N\ln N + \right. \nonumber \\
& & \left. (\beta - 1)N\left(-1 + \ln(2\pi) + \ln \beta\right)\right]\;\cdot\;2^{\frac{5}{6} - \frac{\beta + \beta^{-1}}{3}}\,\frac{\left(\prod_{j = 1}^N \Gamma(1 + j\beta)\right)^2}{\Gamma(1 + \beta)^N\Gamma(1 + \beta N)} \nonumber \\
\label{eq:CNB} & & \eea

We shall see from this formula how to find the constant prefactor in the asymptotic of the $\beta$ analog of Tracy-Widom laws for arbitrary $\beta$.

\section{Example: Gaussian $\beta$ ensemble}
\label{sec:Gogo}
We consider the Gaussian model, i.e with a quadratic potential
$V(x) = \frac{x^2}{2}$:
\beq
\mathcal{I}_{N,\beta}(a) = \int_{]-\infty,a]^N} \mathrm{d}\lambda_1\cdots\mathrm{d}\lambda_N\,|\Delta(\lambda)|^{2\beta}\,e^{-\frac{\beta N}{2t} \sum_{i = 1}^N \lambda_i^2}
\eeq
We also write $\mathcal{I}_{N,\beta}$ for the unbounded eigenvalue integral $\mathcal{I}_{N,\beta}(\infty)$. Let us note $F^{g,k}(a)$ the geometric quantities of this model, and write:
\beq
F(a) = \sum_{g,k} \nu^{2 - 2g}\,\hbar^{k}\,F^{g,k}(a)
\eeq
We are interested in:
\beq
\mathcal{P}(a) \equiv \mathbb{P}[\lambda_{\mathrm{max}} \leq a] = \frac{\mathcal{I}_{N,\beta}(a)}{\mathcal{I}_{N,\beta}(\infty)} = \frac{C_{N,\beta}}{\mathcal{I}_{N,\beta}}\,e^{F(a)}
\eeq
$C_{N,\beta}=e^{D_{N,\beta}}$ is the normalization constant relating the series constructed out of the geometric quantities to the genuine eigenvalue integral. Its value is given in Eqn.~\ref{eq:CNB}.

\medskip

Notice that $\mathcal{I}_{N,\beta}(a)$ is a function of $a$, $t$, but we have a priori a relation of homogeneity:
\beq
\mathcal{I}_{N,\beta}(a) = |a|^{\beta N^2 + (1- \beta)N}\,I(t/|a|^2)
\eeq
where:
\beq
I(\tau) = \int_{]-\infty,\pm 1]^N} \mathrm{d}\lambda_1\cdots\mathrm{d}\lambda_N\,|\Delta(\lambda)|^{2\beta}\,e^{-\frac{\beta N}{2\tau} \sum_{i = 1}^N \lambda_i^2}
\eeq
Thus:
\bea
F^{0,0}(a) & = & t^2\ln|a| + \mathrm{function}\;\mathrm{of}\;t/|a|^2 \\
F^{0,1}(a) & = & -t\ln|a| + \mathrm{function}\;\mathrm{of}\;t/|a|^2 \\
\mathrm{else}\;F^{g,k} & = & \mathrm{function}\;\mathrm{of}\;t/|a|^2
\eea

\subsection{Theoretical results for arbitrary $a$}

\subsubsection{Density of eigenvalues}

The eigenvalue support is a single interval\footnote{We will not analyze the case where $a > 2\sqrt{t}$, which cannot be reached naively as below} if $a \leq 2\sqrt{t}$. In this case, it is of the form
$[b,a]$ where $b=\frac{1}{3}\left(a-2 \sqrt{a^2+12 t}\right)$.
The parameters $\alpha=\frac{a+b}{2}$ and $\gamma=\frac{a-b}{4}$
are thus given by:
\bea
\alpha & = & \frac{1}{3}(2a - \sqrt{a^2 + 12t})\\
\gamma & = & \frac{1}{6}(a + \sqrt{a^2 + 12t})
\eea
and we have:
\bea
v_0 = \frac{\alpha^2}{2} + \gamma^2\;\;&\;\; v_1 = \alpha\gamma\;\;&\;\; v_2 = \frac{\gamma^2}{2} \\
r = -\alpha\;\;&\;\; u_0 = \alpha\;\; & \;\; u_1 = \gamma
\eea

\medskip

The density of eigenvalue is:
\bea
\rho(x) & = & -\frac{1}{2\pi}M(x)\sqrt{\frac{x - b}{a - x}}\;\;
{\rm for}\;\; x\in [b,a] \nonumber \\
M(x) & = & x - 2\gamma \nonumber
\eea
When $a = 2\sqrt{t}$, the eigenvalue support is $[b,a] =
[-2\sqrt{t},2\sqrt{t}]$ and we recover the Wigner semi-circle law
(with $t=1/2$ and $a=\sqrt{2}$ usually):
\beq
\rho(x) = \frac{1}{2\pi}\sqrt{4t - x^2} \nonumber
\eeq
For $a < 2\sqrt{t}$, $\rho(x)$ has another zero on the real line
at $x=x_1=2\gamma$,
which corresponds to $s=s_1$:
\beq
 s_1 = \frac{1}{2\gamma}\left(2\gamma - \alpha + \sqrt{\alpha(\alpha - 4\gamma)}\right) \nonumber
\eeq
defined by $|s_1| \geq 1$ and $x(s_1)=x_1$, where $x(z)=\alpha+\gamma\left(z+\frac{1}{z}\right)$
is the Zhukovsky map. For $a \leq 2\sqrt{t}$, we have $s \in [1,\infty[$.

\subsubsection{First terms in $F(a)$}

The formulas for the nondecaying terms in the free energy yield:
\bea
\label{eq:fofof} F^{0,0}(a) & = & \frac{\alpha^2\gamma^2}{2} + \frac{\gamma^4}{4} - t\left(\gamma^2 + \frac{\alpha^2}{2}\right) + t^2\ln\gamma \\
F^{0,1}(a) & = & t\ln\left(\frac{\gamma s}{2\pi t}\right) - \left(\alpha\gamma + \frac{\gamma^2}{2}\right) + \left(\frac{\alpha\gamma}{2} + \gamma^2\right)\frac{1}{s} \\
F^{1,0}(a) & = & -\frac{1}{24}\ln\left[16t^{-4}\gamma^4\alpha^3(\alpha - 4\gamma)\right] \\
F^{0,2}(a) & = & \frac{1}{12}\ln\left(16t^{-2}\gamma^4\,\frac{s^{10}(s - 1)^3}{(s + 1)^{11}}\right)
\eea
Indeed, the dominant term (Eqn.~\ref{eq:fofof}) was obtained in \cite{DM06,DM08} for all $\beta$ by Coulomb gas techniques.

\medskip

Their derivatives with respect to $a$ are maybe easier to handle:
\bea
\partial_a F^{0,0}(a) & = & \gamma\alpha^2 \\
\partial_a F^{0,1}(a) & = & -\frac{1}{2}\left(\alpha + \sqrt{\alpha(\alpha - 4\gamma)}\right) \\
\partial_a F^{1,0}(a) & = & -\frac{\gamma^2}{\alpha(\alpha - 4\gamma)^2} \\
\partial_a F^{0,2}(a) & = & \frac{\alpha}{\gamma^2} \frac{s^2(2s^3 - 3s^2 + 4s - 2)}{(s - 1)^4(s + 1)^4}
\eea
For example, in terms of $a$ and $t$ only:
\bea
F^{0,0}(2\sqrt{t})- F^{0,0}(a) & = & \frac{1}{3}ta^2 -
\frac{1}{216}a^4 -\frac{a}{216}(a^2 + 30t)\sqrt{a^2 + 12t} \nonumber \\
& & - t^2\ln\left(\frac{a + \sqrt{a^2 + 12t}}{6 \sqrt{t}}\right) \nonumber \\
& & \\
F^{1,0}(a) & = & \frac{7\ln 3}{24} -\frac{\ln 2}{8} + \frac{\ln t}{6} - \frac{1}{48}\ln(a^2 + 12t) \nonumber \\
 & & - \frac{1}{24}\ln\left[a\left(-a^6 + 6a^4t + 36a^2t^2 - 432t^3\right) + \right. \nonumber \\
& & \phantom{\frac{1}{24}\ln} \left.\left(-a^6 + 12a^4t - 54a^2t^2 + 216t^3\right)\sqrt{a^2 + 12t}\right] \nonumber \\
& &
\eea

\subsubsection{Example: $a = 0$}

$\mathcal{P}(0) = \frac{C_{N,\beta}}{\mathcal{I}_{N,\beta}}e^{F(0)}$ is the probability that all the eigenvalues are negative. The first terms in $F(0)$ are:
\bea
F^{0,0}(0) & = & t^2\left(-\frac{3}{4} - \frac{\ln 3}{2} + \frac{\ln t}{2}\right) \\
F^{0,1}(0) & = & t\left[\frac{1}{2} + \ln\left(1 + \frac{2}{\sqrt{3}}\right)
- \ln(2\pi) - \frac{\ln t}{2}\right]\\
F^{1,0}(0) & = & -\frac{\ln 2}{3} + \frac{\ln 3}{8}\\
F^{0,2}(0) & = & -\frac{3 \ln 3}{8} + \frac{1}{2}\ln\left(1 + \frac{2}{\sqrt{3}}\right)
\eea

\subsection{Numerical simulations: principles}
\label{sec:num}
To verify the analytical predictions derived in the previous
sections, we simulated the joint distribution of eigenvalues in
Eqn.~\ref{eq:matint} in the case of the Gaussian ensemble:
\bea\label{jpdfEV}
P(\lambda_1,\ldots,\lambda_N)&=& B_{M,N}\, \prod_i e^{-\frac{N\beta}{t}\,\frac{\lambda_i^2}{2}}\,
\prod_{i<j} |\lambda_i -\lambda_j|^{2 \beta}\nonumber\\
&=&B_{M,N}\:
e^{-\beta E\left[\{\lambda_i\} \right]}\,,
\eea
where the effective energy $E\left[{\lambda_i}\right]$ is given by
\beq
E\left[\{\lambda_i\}\right]=\frac{ N}{t} \sum_{i=1}^N \frac{\lambda_i^2}{2}
-2 \sum_{i<j} \ln |\lambda_i-\lambda_j|
\eeq
In the simulations, we fix $t=1/2$.
  We sample this probability distribution using
a Monte Carlo Metropolis algorithm (see \cite{Krauth}).

\subsubsection{Standard Metropolis algorithm}

We start with an initial configuration of the $\lambda_i$'s taken at random
within the range $[-2 \sqrt{t},2 \sqrt{t}]=[-\sqrt{2},\sqrt{2}]$
(the average density support).  At each
step, a small modification $\{\lambda_i\} \longrightarrow
\{\lambda_i'\}$ is proposed in the configuration space.  In our
algorithm, the proposed move consists of picking at random an eigenvalue
$\lambda_j$ and proposing to modify it
as $\lambda_j \longrightarrow \lambda_j + \varepsilon$
where $\varepsilon$ is a real number drawn from a Gaussian
distribution with mean zero and with a variance that is set to achieve
an average rejection rate $1/2$
(the proposed move is indeed accepted or rejected as explained below).

The move is accepted with the standard probability
\begin{equation}
\label{DetailedBalance}
p=\min\left(\frac{
P(\lambda_1',\ldots,\lambda_N')}{
P(\lambda_1,\ldots,\lambda_N)},1 \right)
=\min\left( e^{-\beta \left(E\left[ \{\lambda_i' \}\right]-
E\left[\{\lambda_i \}\right]\right)},1\right)\,,
 \end{equation}
 and rejected with probability $1-p$.  This dynamics enforces detailed
 balance and ensures that at long times the algorithm reaches thermal
 equilibrium (at inverse "temperature'' $\beta$) with the correct
 Boltzmann weight $e^{-\beta E\left[\{\lambda_i\}\right]}$.

 At long times (from about $10^6$ steps in our case), the Metropolis
 algorithm thus generates samples of $\{\lambda_i\}$ drawn from the
 joint distribution in Eqn. \ref{jpdfEV}. We can then start to
  construct a histogram for the
 maximal eigenvalue (by keeping the value of $\lambda_{\rm max}$
about every $100$ steps). This histogram describes the probability density function
of $\lambda_{\rm max}$, which can be compared to our analytical predictions:
$\mathbb{P}(\lambda_{\rm max}=a)=\partial_a \mathcal{P}(a)$
with $\mathcal{P}(a)$ given in Eqn. \ref{eq:PbLargeN} for large $N$.
More precisely, we plot here the rate function $\Phi(a)= - \ln \mathbb{P}(\lambda_{\rm max}=a)$.
We compare the numerical data $\Phi_{\rm num}(a)$ with analytical results $\Phi(a)$.

 However, as the distribution of the maximal eigenvalue
 is highly peaked around its average, a standard
 Metropolis algorithm does not allow to explore in a "reasonable'' time
 a wide range of values of $\lambda_{\rm max}$.  The probability to reach a
 value $\lambda_{\rm max}=a$ decreases rapidly with $N$ as $e^{-
   N^2 \Phi_2(a)+O(N)}$ where $\Phi_2(a)$ is  positive  (for $a$
 different from the mean value).  Therefore, we
 modified the algorithm in order to explore the full distribution of
 $\lambda_{\rm max}$ and to compare it with our analytical predictions
for the (left) large deviation of the distribution of $\lambda_{\rm max}$.

\subsubsection{Modified algorithm: conditional probabilities}

It is difficult to reach small (or large) values of $\lambda_{\rm max}$
with a standard Metropolis algorithm.  The idea is thus to force the algorithm to explore a
region $\lambda_{\rm max} \leq a_c$ for different values of $a_c$. We thus add in the
algorithm the constraint $\lambda_{\rm max} \leq a_c$.  More precisely, we start with
an initial configuration that satisfies the constraint $\lambda_{\rm max} \leq a_c$.
 At each step, the proposed move $\{\lambda_i\} \rightarrow \{ \lambda_i'\}$
 is rejected if $\lambda_{\rm max}' > a_c$.
If $\lambda_{\rm max}'\leq a_c$, then the
move is accepted or rejected exactly with the same Metropolis rules as
before.  Because of the new constraint $\lambda_{\rm max}\leq a_c$, the moves are
rejected much more often than before. Therefore the variance of the
Gaussian distribution $P(\varepsilon)$ has to be taken smaller to achieve
a rejection rate $1/2$.

We run the program for several values of $a_c$ (about $60$ different
values on the left of the mean value) and we construct a histogram of $\lambda_{\rm max}$ for each value
$a_c$.  This gives the conditional probability distribution
$\mathbb{P}\left(\lambda_{\rm max}=a \big| \lambda_{\rm max} \leq a_c
\right)$.  Again, as the distribution of the maximal eigenvalue is highly peaked,
the algorithm can only explore a very small range of values of $a$ -
even for a large running time (about $10^9$ steps).  The difference
with the previous algorithm is that we can now explore small regions
of the form $a_c -\eta \leq a \leq a_c $ for every $a_c$, whereas before
we could only explore the neighborhood of the mean value.

The distribution of the maximal eigenvalue is given by
\beq
\mathbb{P}\left(\lambda_{\rm max}=a \right)=
\mathbb{P}\left(\lambda_{\rm max}=a \big| \lambda_{\rm max} \leq a_c \right)\cdot
\mathbb{P}\left(\lambda_{\rm max} \leq a_c\right)\;\;\textrm{(for $a<a_c$)}\,.
\eeq
Therefore the rate function reads:
\bea
  -\ln \mathbb{P}(\lambda_{\rm max}=a)
  &=&-
  \left[ \ln \mathbb{P}\left(\lambda_{\rm max}=a \big| \lambda_{\rm max} \leq a_c \right)
    + \ln \mathbb{P}\left(\lambda_{\rm max} \leq a_c \right) \right] \,.
\eea
The histogram constructed by the algorithm with the constraint $\lambda_{\rm max} \leq
a_c$ is the rate function $\Phi_{a_c}(a)\equiv - \ln
\mathbb{P}\left(\lambda_{\rm max}=a \big| \lambda_{\rm max} \leq a_c
\right)$.  It differs from the exact rate function
$\Phi(a)=- \ln
\mathbb{P}\left(\lambda_{\rm max}=a\right)$ by an additive constant that depends on $a_c$.  In order to
get rid of this constant, we construct from the histogram giving
$\Phi_{a_c}(a)$ the derivative $\frac{\mathrm{d}\Phi_{a_c}(a)}{\mathrm{d}a}$.  This
derivative is equal to $\frac{\mathrm{d}\Phi(a)}{\mathrm{d}a}$ and the constants disappear.

\subsection{Numerical simulations: results}

We now come back to $\Phi(a)$ from its derivative using an
interpolation of the data for the derivative and a numerical
integration of the interpolation.  This allows to compare directly the
numerical results $\Phi_{\rm num}(a)$ with the theoretical rate function $\Phi(a)=- \ln
\mathbb{P}\left(\lambda_{\rm max}=a\right)=-\ln \left[\partial_a \mathcal{P}(a)\right]$.
The interpolation and numerical integration also help to get rid of the fluctuations in numerical data
(by smoothening the plot) and thus increase the precision of the results.

We can follow the same steps to explore the region on the right of the
mean value  by adding in the simulations the condition
$\lambda_{\rm max} \geq a_c$ (instead of $\lambda_{\rm max} \leq a_c$)
for several values of $a_c$. At the end, we can reconstruct the full plot of
the probability density function (pdf) of $\lambda_{\rm max}$. There remains only an additive constant
in the rate function that is not yet determined (as we perform a  numerical integration of the derivative).
This constant can be numerically determined by imposing that the integral (numerical integration
of the data) of the pdf of $\lambda_{\rm max}$ is one (normalization of the probability).
We can now compare the simulations with analytical predictions.

Our analytical predictions give (see Eqn. \ref{eq:PbLargeN}):
\beq
\mathcal{P}(a) = \frac{\mathcal{I}_{N,\beta}(a)}{\mathcal{I}_{N,\beta}(\infty)} = \frac{C_{N,\beta}}{\mathcal{I}_{N,\beta}}\,e^{F(a)}=\exp\left\{F(a)
+\ln C_{N,\beta}-\ln \mathcal{I}_{N,\beta}
\right\}
\eeq
with
\bea
\ln C_{N,\beta}-\ln \mathcal{I}_{N,\beta} & = & \beta N^2 \left(\frac{3}{4}-\frac{\ln t}{2} \right)
+N (\beta -1)\left( -\frac{1}{2}+\ln (2 \pi) +\frac{\ln t}{2} \right) \nonumber \\
&&+\ln N \left( \frac{\beta +\beta^{-1} -3}{12} \right)
+\left( \frac{5-2(\beta +\beta^{-1})}{6} \right) \ln 2 \nonumber \\
&&-\frac{\ln (2\pi)}{2}-\frac{\ln \beta}{2} +\kappa_{\beta}
\eea
and $F(a)$ is given up to order one (for large $N$) by
\beq
F(a)=
\frac{N^2}{t^2}\beta F^{0,0}(a) + \frac{N}{t}(\beta - 1)F^{0,1}(a) + F^{1,0}(a) + (\beta + \beta^{-1} - 2)F^{0,2}(a)
\eeq
Thus
\bea
\Phi(a)&\equiv& -\ln \mathbb{P}(\lambda_{\rm max}=a)=
-\ln \left[\partial_a \mathcal{P}(a) \right]\nonumber
\\
&=&-\left\{F(a)
+\ln C_{N,\beta}-\ln \mathcal{I}_{N,\beta}+\ln\left[\partial_a F(a)\right]\right\} \nonumber\\
&=& N^2 \Phi_2(a) +N \Phi_1(a) + \ln N \: \phi +\Phi_0(a)+ \cdots
\eea
where
\bea
\Phi_2(a)&=&-\frac{1}{t^2}\beta F^{0,0}(a)-\beta \left(\frac{3}{4}-\frac{\ln t}{2} \right)
\nonumber\\
\Phi_1(a)&=& -\frac{1}{t}(\beta - 1)F^{0,1}(a)- (\beta -1)\left( -\frac{1}{2}+\ln (2 \pi) +\frac{\ln t}{2} \right)
\nonumber\\
\phi&=&-\left( \frac{\beta +\beta^{-1} -3}{12} \right)-2  \nonumber\\
\Phi_0(a)&=& - F^{1,0}(a) - (\beta + \beta^{-1} - 2)F^{0,2}(a)
-\ln\left(\partial_a F^{0,0}(a)\right) \nonumber\\
&&-\left( \frac{5-2(\beta +\beta^{-1})}{6} \right) \ln 2
+\frac{\ln (2\pi)}{2}+\frac{\ln \beta}{2} -\kappa_{\beta}-\ln \left(\frac{\beta}{t^2}\right) \nonumber
\eea

\begin{figure}
\begin{center}
\includegraphics[width=0.8\textwidth]{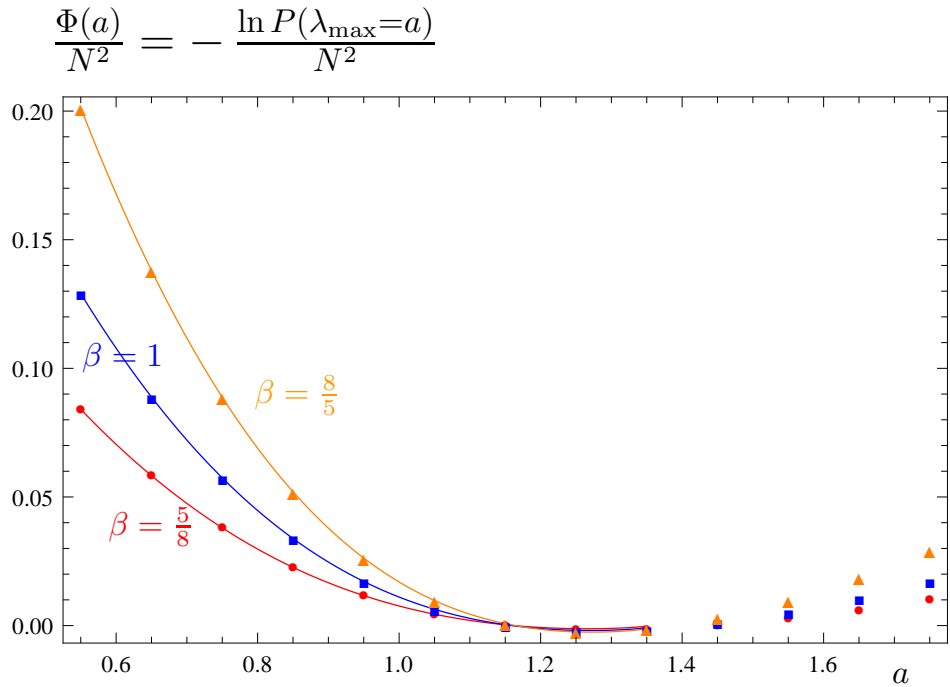}
\caption{Case of a  quadratic potential
(Gaussian ensemble): rate function  $\Phi(a)/N^2=- \ln
\mathbb{P}\left(\lambda_{\rm max}=a\right)/N^2=-\ln \left[\partial_a \mathcal{P}(a)\right]/N^2$
plotted against $a$ for different values of $\beta$:
$\beta=5/8=0.625$ (red disks and red line),
$\beta=1$ (blue squares and blue line)
and $\beta=8/5=1.6$ (orange triangles and orange line).
The points (disks, squares, triangles)
are numerical data obtained with the modified Metropolis algorithm.
The  solid lines are the analytical predictions for the left tail of the large deviation
of the pdf of $\lambda_{\rm max}$: $\mathcal{P}(a)$ is given in Eqn. \ref{eq:PbLargeN}.}
\label{fig:totN2}
\end{center}
\end{figure}
\begin{figure}
\begin{center}
\includegraphics[width=0.8\textwidth]{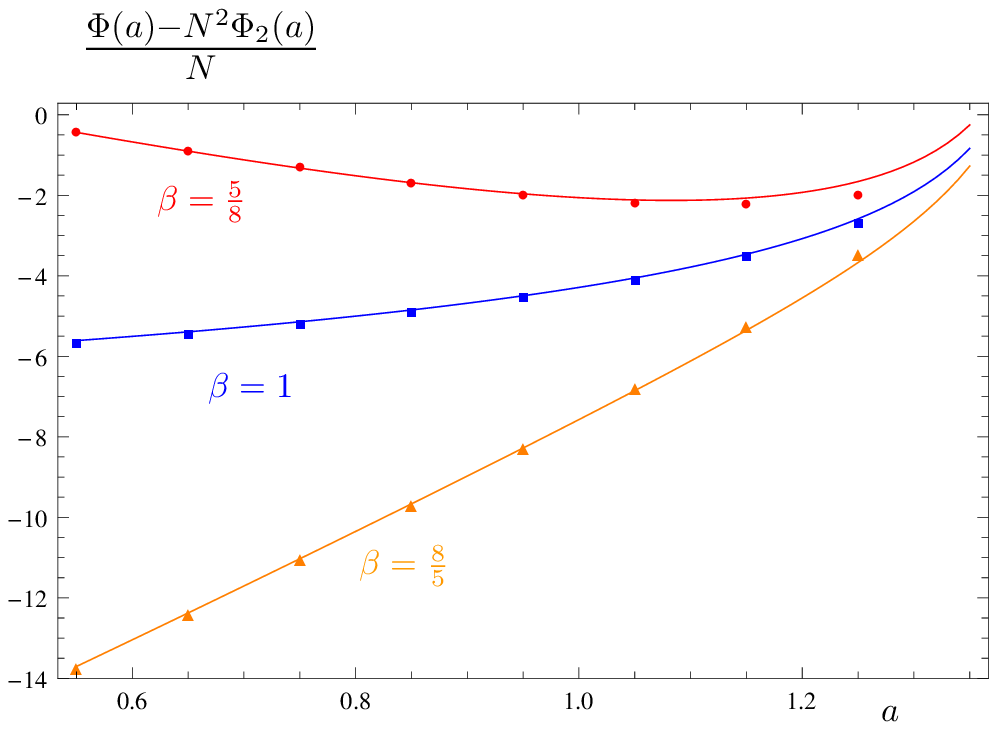}
\caption{Case of a quadratic potential
(Gaussian ensemble): order $N$ of the rate function,  $(\Phi_{\rm num}(a)-N^2 \Phi_2(a))/N$,
plotted against $a$ for different values of $\beta$:
$\beta=5/8=0.625$ (red),
$\beta=1$ (blue)
and $\beta=8/5=1.6$ (orange).
The points (disks, squares, triangles)
are numerical data obtained with the modified Metropolis algorithm.
The  solid lines are the analytical predictions:  $(N \Phi_1(a) +(\ln N) \: \phi +\Phi_0(a))/N$.}
\label{fig:totN1}
\end{center}
\end{figure}
\begin{figure}
\begin{center}
\includegraphics[width=0.8\textwidth]{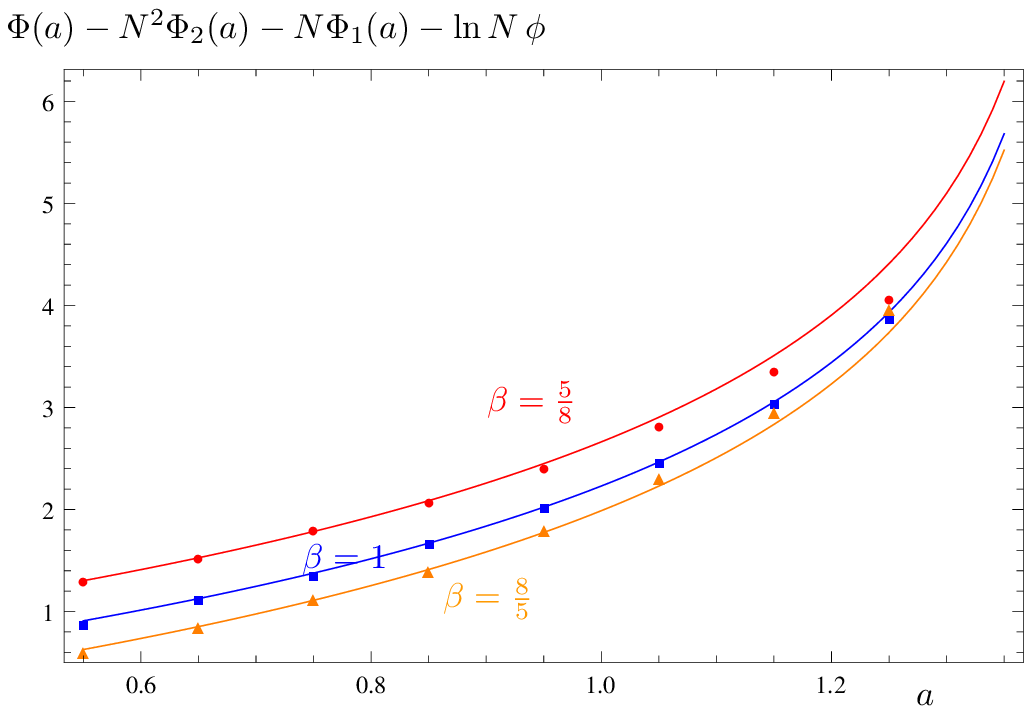}
\caption{Case of a quadratic potential
(Gaussian ensemble): order one of the rate function,
$\Phi_{\rm num}(a)-N^2 \Phi_2(a)-N \Phi_1(a)-(\ln N)\: \phi$,
plotted against $a$ for different values of $\beta$:
$\beta=5/8=0.625$ (red),
$\beta=1$ (blue)
and $\beta=8/5=1.6$ (orange).
The points (disks, squares, triangles)
are numerical data obtained with the modified Metropolis algorithm.
The  solid lines are the analytical predictions:  $\Phi_0(a)$.}
\label{fig:totN0}
\end{center}
\end{figure}

We typically run the simulations for $N = 30$ with $10^9$ iterations.

\medskip

Fig.~\ref{fig:totN2} shows a plot of the rate function
$\Phi_{\rm num}(a)/N^2\equiv -\ln P(\lambda_{\rm max}=a)/N^2$ for different values of $\beta$:
in red, $\beta=\frac{5}{8}=0.625$ ; in blue, $\beta=1$ (Gaussian Unitary Ensemble) ; and in orange, $\beta=\frac{8}{5}=1.6$.
Numerical data $\Phi_{\rm num}(a)$ (disks, squares and triangles) are compared with
analytical predictions for the left tail
$\Phi(a)= N^2 \Phi_2(a) +N \Phi_1(a) + (\ln N) \: \phi +\Phi_0(a)$ (solid lines).
In this figure, we see that the agreement between numerics and theory is very good, at
least to dominant order for large $N$ (order $N^2$).

Fig.~\ref{fig:totN1} shows the next order in the large $N$ expansion (order $N$) and figure
\ref{fig:totN0} shows the order one for large $N$.
Up to order $O(N)$, the agreement between numerics and theory is very good
(see Fig.~\ref{fig:totN2} and Fig.~\ref{fig:totN1}).
To order $O(1)$, the theoretical lines do not fit numerical data as well as
to order $O(N)$ or $O(N^2)$ but the agreement is still quite good.
 At this order, the number of significant digits that is required is very high.
A better precision would require a longer running-time (to have more samples) and
a refined histogram. Also, the determination of the constant depends
on the reconstruction of the full histogram (both left and right tail).
However, we can not go numerically toward infinity, we have to choose a finite
interval.
The precision could thus be increased by exploring a wider range of values (by
going deeper in the left and right tails).

\section{Scaling limit at the edge of the spectrum}
\label{sec:scalingG}

Now, we wish to take the scaling limit $N \rightarrow \infty$, $a \rightarrow \widehat{a}$ the edge of the spectrum, in a one-cut model with arbitrary polynomial potential.
Let us mention that using an asymptotic expansion of eigenvalue statistics to study limiting distribution is not a new idea. For instance, within GUE, Gustavsson \cite{Gustav} has found limiting distributions for the $k$-th eigenvalue in various regimes, using an asymptotic expansion for the first few terms of eigenvalue correlation densities.

\subsection{Blow up of the spectral curve and unstable correlators}
\label{sec:TWcc}

Formula \ref{eq:Fg} computes the stable $F^{g,k}(a)$ only in terms of the spectral curve $y$, $\omega_2^{0,0}$ and $\omega_1^{0,1}$:
\bea
\label{eq:spc} (\Sigma_a)\; :  \; & & \left\{\begin{array}{l} x(z) = a + \gamma\frac{(z - 1)^2}{z} \\ y(z) = \frac{1}{2}\frac{z + 1}{z - 1}\,\prod_{j = 1}^{d - 1} \frac{(z - s_j)(z - 1/s_j)}{z} \end{array}\right. \\
&& \phantom{w}\omega_1^{0,1}(z) = \frac{1}{2(z - 1)} - \frac{1}{2(z + 1)} - \sum_{j = 1}^{d - 1} \left(\frac{1}{z - 1/s_j} - \frac{1}{z}\right) \\
&& \phantom{w}\omega_2^{0,0}(z_1,z_2) = \frac{1}{(z_1z_2 - 1)^2}
\eea
We shall use the notation $F^{g,k}(a) = \mathcal{F}_{g,k}(\Sigma_a)$ for the geometric quantities associated to the curve $\Sigma_a$.

Let us define $\epsilon$ such that $s_{1} - 1 \sim \sqrt{\epsilon}$, and $\epsilon \rightarrow 0$ when $a \rightarrow a^*$. Let us also assume that $\gamma \rightarrow \widehat{\gamma} \neq 0$ and $s_j \rightarrow \widehat{s}_j \neq 1$ for $2 \leq j \leq d - 1$. With the rescaling $z = 1 + \sqrt{\epsilon}\sigma$, we have:
\bea
(\Sigma_a) \; \sim \; & & \left\{\begin{array}{l} x(z) = (a - 2\epsilon\widehat{\gamma}) + \widehat{\gamma}\epsilon\;\widehat{x}(\sigma) \\
y(z) = \widehat{M}_+\:\epsilon^{1/2}\;\widehat{y}(\sigma) \end{array}\right. \\
& & \phantom{w}\omega_1^{0,1}(z) \sim \epsilon^{-1/2}\;\widehat{\omega}_1^{0,1}(\sigma) \\
& & \phantom{w}\omega_2^{0,0}(z_1,z_2) \sim \epsilon^{-1}\;\widehat{\omega}_2^{0,0}(\sigma_1,\sigma_2)
\eea
with the constant:
\beq
\widehat{M}_{\pm} = t_d\,\widehat{\gamma}^{d - 1}\,\prod_{j = 2}^{d - 1} (\pm 1 - \widehat{s}_j)(\pm 1 - 1/\widehat{s}_j) \nonumber
\eeq
We have:
\bea
(\widehat{\Sigma})\; : \; & & \left\{\begin{array}{l} \widehat{x}(\sigma) = \sigma^2  \\ \widehat{y}(\sigma) = \sigma - \frac{1}{\sigma} \end{array}\right. \\
& & \phantom{w}\widehat{\omega}_1^{0,1}(\sigma) = \frac{1}{2\sigma} - \frac{1}{\sigma + 1} \\
& & \phantom{w}\widehat{\omega}_2^{0,0}(\sigma_1,\sigma_2) = \frac{1}{(\sigma_1 + \sigma_2)^2}
\eea
In other words, $\widehat{\Sigma}$ is the plane curve of equation $y^2 = x + 1/x - 2$.

\begin{figure}
\begin{center}
\label{fig:FIA}
\includegraphics[width=0.98\textwidth]{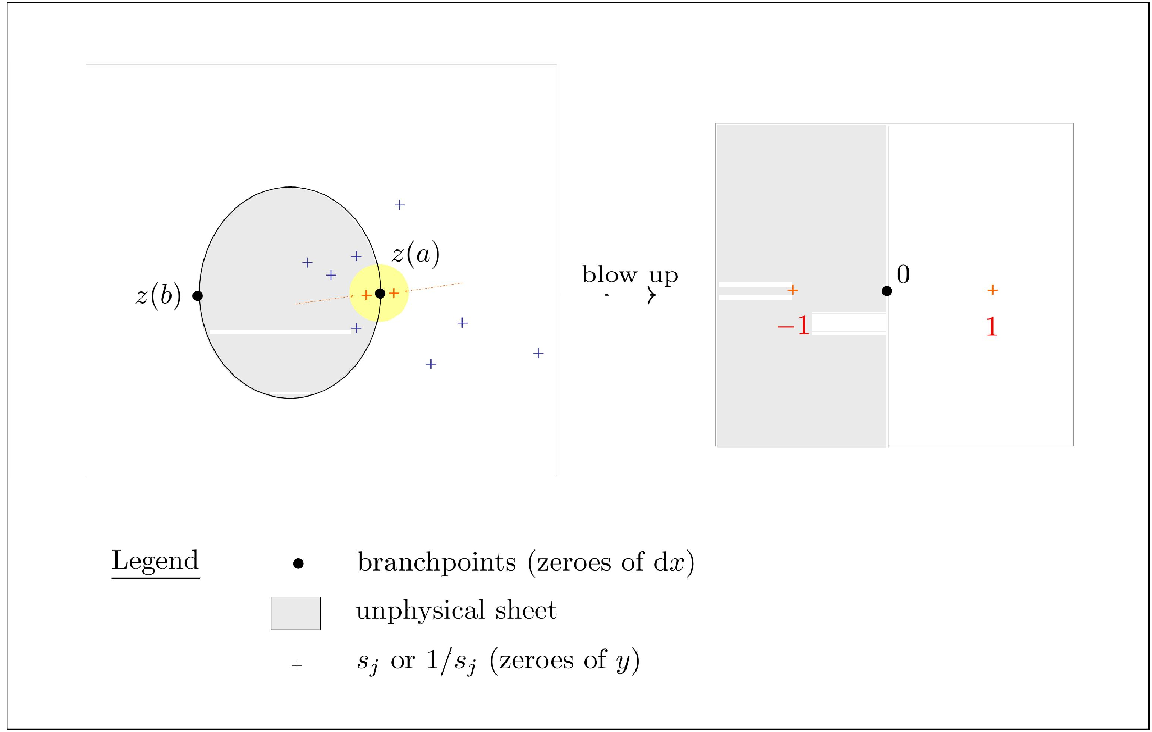}
\caption{When a moving singularity ($x = x(s_i)$ here) merges to a fixed singularity (the hard edge at $x = a$), the blow up around $a$ of the spectral curve (here, of the form Eqn.~\ref{eq:spc}) can be generically described by the situation on the left. Indeed, $(\widehat{\Sigma}) = (\mathbb{C},\widehat{x},\widehat{y})$ is a universal object. We may call it the Tracy-Widom curve. In our parametrization, the physical sheet correspond to $\mathrm{Re}\;\sigma \geq 0$. There is only one simple branchpoint, at $\sigma = 0$, and we have $\overline{\sigma} = - \sigma$ (which is globally defined). $y$ has a pair of zeroes at $\sigma = -1,1$.}
\end{center}
\end{figure}

\subsection{Scaling limit of unstable $F^{g,k}$}

It is easy to obtain the scaling limit of unstable free energies from their expressions. We find:
\bea
F^{0,0}(a) & = & F^{0,0}(\widehat{a}) + \big((\widehat{\gamma}\widehat{M}_+)^{2/3}\,\epsilon\big)^3\,\widehat{F}^{0,0} + o(\epsilon^3) \\
F^{0,1}(a) & = & F^{0,1}(\widehat{a}) + \big((\widehat{\gamma}\widehat{M}_+)^{2/3}\,\epsilon\big)^{3/2}\,\widehat{F}^{0,1} + o(\epsilon^{3/2}) \\
F^{1,0}(a) & = & \ln\big((\widehat{\gamma}\widehat{M}_+)^{2/3}\,\epsilon\big)\,\widehat{F}^{1,0} + \delta^{1,0} + o(1) \\
F^{0,2}(a) & = & \ln\big((\widehat{\gamma}\widehat{M}_+)^{2/3}\,\epsilon\big)\,\widehat{F}^{0,2} + \delta^{0,2} + o(1)
\eea
with:
\bea
\widehat{F}^{0,0} & = & -\frac{2}{3}, \\
\widehat{F}^{0,1} & = & \frac{4}{3}, \\
\widehat{F}^{1,0} & = & -\frac{1}{8},\quad \widehat{F}^{0,2} = \frac{1}{8}
\eea
and:
\bea
\delta^{1,0} & = & - \frac{1}{24}\ln\big(2^{10}\,\widehat{M}_+\widehat{M}_-\,\widehat{\gamma}^2 t^{-4}\big) \\
\delta^{0,2} & = & \frac{1}{12}\ln\left[2^{-7}\,t_d\widehat{\gamma}^{d}t^{-2}\,\prod_{j = 2}^{d - 1} \frac{\widehat{s}_j^{23}(\widehat{s}_j - 1)^{11}}{(\widehat{s}_j + 1)^{11}}\right] - \sum_{2 \leq j < l \leq d - 1} \ln\left(1 - \frac{1}{\widehat{s}_j\widehat{s}_l}\right)
\eea
$\widehat{F}^{0,0}$, $\widehat{F}^{0,1}$, $\widehat{F}^{1,0}$ and $\widehat{F}^{0,2}$ are universal and can be considered as the unstable free energies associated to the curve $\widehat{\Sigma}$. On the contrary, $\delta_{1,0}$ and $\delta^{0,2}$ are non universal. They are relevant to compute the constant prefactor in the scaling limit of $F(a)$.

\subsection{Scaling limit of stable $F^{g,k}$}

The scaling of the correlators is very simple to obtain from the residue formula (Eqn.~\ref{eq:resfo}). The scaling of unstable correlators are computed at hand, and the result for stable correlators is proved by recursion. $\omega_1^{0,0}$ is special:
\beq
\omega_1^{0,0}(z) = \frac{\widehat{\gamma}}{2}V'(\widehat{a})\:\epsilon - \widehat{\gamma}\widehat{M}_+\:\epsilon^{3/2}\;\widehat{y}(\sigma) + O(\epsilon^2) \nonumber
\eeq
And, if $(n,g,k) \neq (1,0,0)$:
\beq
\omega_n^{g,k}(z_1,\ldots,z_n) = (\widehat{\gamma}\widehat{M}_+\:\epsilon^{3/2})^{2 - 2g - k - n}\,\widehat{\omega}_n^{g,k}(\sigma_1,\ldots,\sigma_n) + o(\epsilon^{2 - 2g - k - n})
\eeq
The $\widehat{\omega}_n^{g,k}$ are the result of the residue formula applied to $\widehat{\Sigma}$ and the data of $\widehat{\omega}_2^{0,0}$ and $\widehat{\omega}_1^{0,1}$. The standard form of the topological recursion (Formula~\ref{eq:toptop}) computes $\widehat{\omega}_n^{g,0}$ with the limit of the recursion kernel:
 \beq
 \widehat{K}(\sigma,\sigma_0) = \frac{1}{4}\frac{1}{\sigma(\sigma^2 - 1)(\sigma^2 - \sigma_0)}
 \eeq
In the stable $\omega_n^{g,k}$, the leading behavior correspond to diverging power of $\epsilon$. Thus, only the poles at infinitesimal distance of the hard edge are relevant to find $\widehat{\omega}_n^{g,k}$. In other words, the poles at $z_i = 1$ are replaced by poles at $\sigma_i = 0$, the poles at $z_i = 1/s_1$ are replaced by poles at $\sigma_i = -1$, the poles at $z_i = 1/z_l$ are replaced by poles at $\sigma_i = -\sigma_l$, whereas the poles at $z = 0$ and $z = 1/s_j$ do not give contributions to $\widehat{\omega}_n^{g,k}$.

The integration formula for stable $F^{g,k}$ (Eqn.~\ref{eq:Fg}) yields the scaling:
\beq
\label{eq:hufr} F^{g,k}(a) = (\widehat{\gamma}\widehat{M}_+\:\epsilon^{3/2})^{2 - 2g - k}\;\widehat{F}^{g,k}
\eeq
where $\widehat{F}^{g,k}$ are the geometric quantities associated to the curve $\widehat{\Sigma}$:
\beq
\widehat{F}^{g,k} = \frac{1}{2 - 2g - k}\Res_{\sigma \rightarrow 0,-1}\mathrm{d}\sigma\,\widehat{\phi}(\sigma)\widehat{\omega}_1^{g,k}(\sigma)
\eeq
$\widehat{\phi}$ is the following primitive of $\widehat{y}\,\mathrm{d}\widehat{x}$:
\bea
\widehat{\phi}(\sigma) & = & \int_{0}^{\sigma} \widehat{y}\,\mathrm{d}\widehat{x} \\
& = & 2\left(\frac{\sigma^3}{3} - \sigma\right)
\eea

\subsection{Summary: double scaling limit}
\label{sec:summ}
To find the scaling limit of $F(a)$ when $a$ goes to the edge of the spectrum, a scaling variable appears naturally:
\beq
\encadremath{\eta = \left(\frac{\widehat{\gamma}\widehat{M}_+}{t}\right)^{2/3}N^{2/3}\,\epsilon}
\eeq
Notice that the $N^{2/3}$ scaling is the same for all $\beta$, as pointed out long ago, and proved recently in \cite{RRV}. When $\eta$ is of order $1$, each term in the topological expansion of $F(a(\eta))$ is also of order $1$. It is the so-called "double scaling limit":
\bea
F(a) & = & \sum_{g,k \geq 0}^{\infty} \left(\frac{N}{t}\right)^{2 - 2g - k}\,\beta^{1 - g}\big(1 - \beta^{-1}\big)^{k} \,F^{g,k}(\Sigma_a) \nonumber \\
 & \sim & \frac{N^2 \beta}{t^2} F^{0,0}(\widehat{a}) + \frac{N(\beta - 1)}{t} F^{0,1}(\widehat{a}) + \frac{3 - \beta - \beta^{-1}}{12}\ln\left(\frac{N\widehat{\gamma}\widehat{M}_+}{t}\right) \nonumber \\
  & & + \delta^{1,0} + (\beta + \beta^{-1} - 2)\delta^{0,2} + \sum_{g,k \geq 0}^{\infty} \beta^{1 - g}\big(1 - \beta^{-1}\big)^k\,\eta^{\frac{3}{2}(2 - 2g - k)}\,\widehat{F}^{g,k}
\eea
where it is implied that $\eta^{0}$ should be replaced by $\ln(\eta)$. In particular, for hermitian matrices ($\beta = 1$):
\bea
F(a) & \sim & \frac{N^2}{t^2}F^{0,0} + \frac{1}{12}\ln\left(\frac{N\widehat{\gamma}\widehat{M}_+}{t}\right) \nonumber \\
 & & + \delta_{1,0} + \sum_{g \geq 0}^{\infty} \eta^{3(1 - g)}\,\widehat{F}^{g,0}
\eea
The curve $\widehat{\Sigma}$ was described in Section~\ref{sec:TWcc}. We call it the curve the Tracy-Widom curve, because its geometric quantities $\widehat{F}^{g,k}$ are related to the coefficients of the asymptotic expansion of the left tails of the Tracy-Widom laws (see Section~\ref{sec:sca}).

\medskip

Eventually, the complete answer for $\mathcal{P}(a) = \mathbb{P}(\lambda_{\textrm{max}} \leq a)$ is:
\bea
\ln \mathcal{P}(a) & = & \ln(C_{N,\beta}) - \ln(I_{N,\beta}) + \frac{N^2\beta}{t^2} F^{0,0}(\widehat{a}) + \frac{N(\beta - 1)}{t} F^{0,1}(\widehat{a}) \nonumber \\
& & + \frac{3 - \beta - \beta^{-1}}{12}\ln\left(\frac{N\widehat{\gamma}\widehat{M}_+}{t}\right) + \delta^{1,0} + (\beta + \beta^{-1} - 2)\delta^{0,2} \nonumber \\
\label{eq:hud} & & \sum_{g,k \geq 0}^{\infty} \beta^{1 - g}(1 - \beta^{-1})^{k}\,\eta^{\frac{3}{2}(2 - 2g - k)}\,\widehat{F}^{g,k}
\eea
Notice a posteriori that $\widehat{a}$ must be such that the divergent terms when $N \rightarrow \infty$ cancel in this expression.

\subsection{First few $\widehat{F}^{g,k}$}

We give a few of these geometric quantities associated to $\widehat{\Sigma}$. We have already seen:
\bea
& & \widehat{F}^{0,0} = -\frac{2}{3} \nonumber \\
& & \widehat{F}^{0,1} = \frac{4}{3} \nonumber \\
& & \widehat{F}^{1,0} = -\frac{1}{12},\qquad \widehat{F}^{0,2} = \frac{1}{12} \nonumber
\eea
We can also compute:
\bea
&& \widehat{F}^{1,1} = \frac{31}{2^7\cdot 3},\qquad \widehat{F}^{0,3} = -\frac{9}{2^6} \nonumber \\
&& \widehat{F}^{2,0} = \frac{3}{2^9},\qquad \widehat{F}^{1,2} = \frac{487}{2^{12}\cdot 3},\qquad \widehat{F}^{0,4} = -\frac{595}{2^{11}\cdot 3} \nonumber \\
&& \widehat{F}^{2,1} = \frac{8831}{2^{17}\cdot 3},\qquad \widehat{F}^{1,3} = \frac{9281}{2^{17}\cdot 3^2},\qquad \widehat{F}^{0,5} = -\frac{19977}{2^{16}\cdot 3} \nonumber \\
&& \widehat{F}^{3,0} = \frac{63}{2^{14}},\qquad \widehat{F}^{2,2} = \frac{62761}{2^{20}},\qquad \widehat{F}^{1,4} = -\frac{539193}{2^{23}},\qquad \widehat{F}^{0,6} = -\frac{577879}{2^{22}} \nonumber 
\eea

\subsection{Example : Gaussian $\beta$ ensemble}

To illustrate the general framework explained above, we carry on direct computation in the Gaussian case from the results of Section~\ref{sec:Gogo}. The edge of the spectrum is located at $\widehat{a} = 2\sqrt{t}$. This value is actually the mean value of the largest eigenvalue of the Gaussian model:
\beq
\lim_{N \rightarrow \infty} \mathbb{E}(\lambda_{\mathrm{max}}) = 2\sqrt{t}
\eeq
$F(a)$ gives the large deviation of the distribution of $\lambda_{\rm max}$ on the left of the mean value, which describes atypical large fluctuations of order $1$. On the other hand, it is known that the typical small fluctuations of order $N^{-2/3}$
around the mean value are described by the Tracy-Widom distributions. We expect that the two regimes match smoothly
and thus that a scaling limit of $F(a)$ when $a\rightarrow \left(2\sqrt{t}\right)^{-}$ coincides with the limit $\xi \rightarrow -\infty$
of the Tracy-Widom distribution $\mathrm{F}_{2\beta}(\xi)$. More precisely, we write:
\beq
\label{eq:esp}a = 2\sqrt{t}(1 - \epsilon)
\eeq
and the correspondence
\beq
\epsilon = - c_{\beta}\,N^{-2/3}\,\xi
\eeq
is expected for some constant $c_{\beta} > 0$.

\subsubsection{First terms in $F(a)$}

For bookkeeping, let us write several useful expansion when $\epsilon \rightarrow 0$:
\bea
\alpha & = & -\sqrt{t}\,\epsilon\left(1 + \frac{\epsilon}{8} + \frac{\epsilon^2}{32} + \frac{\epsilon^3}{512}\right) + O(\epsilon^5) \\
\gamma & = & \sqrt{t}\left(1 - \frac{\epsilon}{2} + \frac{\epsilon^2}{16} + \frac{\epsilon^3}{64} + \frac{\epsilon^4}{1024}\right) + O(\epsilon^5)
\eea
The other zero of $\rho(z)$ outside the unit disk is:
\bea
s_1 & = & 1 + \sqrt{\epsilon} + \frac{\epsilon}{2} + \frac{7\epsilon^{3/2}}{16} + \frac{5\epsilon^2}{16} + \frac{103\epsilon^{5/2}}{512} + \frac{9\epsilon^3}{64} \nonumber \\
 & & + \frac{615\epsilon^{7/2}}{8192} + \frac{45\epsilon^4}{1024} + \frac{10187\epsilon^{9/2}}{524288} + O(\epsilon^5)
\eea
We see that the definition of $\epsilon$ through Eqn.~\ref{eq:esp} is compatible with the notation in Section~\ref{sec:TWcc}. Moreover, the natural scaling variable is $\eta = t^{-1/2}\epsilon$.
We have for the nondecaying terms of $F(a)$:
\bea
F^{0,0}(2\sqrt{t}) & = & t^2(-\frac{3}{4} + \frac{\ln t}{2})\\
F^{0,1}(2\sqrt{t}) & = & t\left(\frac{1}{2} - \ln(2\pi) - \frac{\ln t}{2}\right)\\
F^{1,0}(2\sqrt{t}) &  & \mathrm{diverges}\,\mathrm{logarithmicallly} \nonumber \\
F^{0,2}(2\sqrt{t}) & & \mathrm{diverges}\,\mathrm{logarithmically} \nonumber
\eea
And:
\bea
\label{eq:F000}F^{0,0}(a) - F^{0,0}(2\sqrt{t}) & = & t^2\left(-\frac{2\epsilon^3}{3} + \frac{\epsilon^4}{8}\right) + O(\epsilon^5) \\
F^{0,1}(a) - F^{0,1}(2\sqrt{t}) & = & t\left(\frac{4\epsilon^{3/2}}{3} - \frac{\epsilon^2}{2} - \frac{\epsilon^{5/2}}{20} - \frac{\epsilon^3}{24} \right. \nonumber \\
& & \left.+ \frac{23\epsilon^{7/2}}{896} - \frac{\epsilon^4}{128} + \frac{143\epsilon^{9/2}}{18432}\right) + O(\epsilon^5) \\
F^{1,0}(a) & = & - \frac{\ln(\epsilon)}{8} - \frac{\ln 2}{4} + \frac{5\epsilon}{64} + \frac{5\epsilon^2}{1024} - \frac{5\epsilon^3}{2048} - \frac{147\epsilon^4}{131072} + O(\epsilon^5) \nonumber \\
& & \\
F^{0,2}(a) & = & \frac{\ln \epsilon}{8} - \frac{7 \ln 2}{12} + \frac{\sqrt{\epsilon}}{2} - \frac{13\epsilon}{64} \nonumber \\
& & + \frac{13\epsilon^{3/2}}{96} - \frac{69\epsilon^2}{1024} + \frac{147\epsilon^{5/2}}{5120} - \frac{97\epsilon^3}{6144} \nonumber \\
\label{eq:F020}& & - \frac{103\epsilon^{7/2}}{114688} + \frac{83\epsilon^4}{131072} - \frac{35005\epsilon^{9/2}}{9437184} + O(\epsilon^{5})
\eea

\subsubsection{Scaling regime of $F(a)$}
\label{sec:sca}
We have argued in Section~\ref{sec:scalingG} that, when $\epsilon$ is of order $N^{-2/3}$, all the terms in $F(a)$ become of order $1$. This can be checked explicitly on Eqns.~\ref{eq:F000}-\ref{eq:F020}. $N^{-2/3}$ is indeed the typical scale expected of small fluctuations around $\mathbb{E}(\lambda_{\mathrm{max}})$. Let us write $\mathcal{P}(a)$ in the scaling variable $\eta$ such that
\beq
a = \sqrt{t}(2 - 2N^{-2/3}\eta)
\eeq
and taking into account the first terms $F^{0,0}$, $F^{1,0}$, $F^{0,1}$ and $F^{0,2}$. When we use the value in the large $N$ limit of $C_{N,\beta}$ (Appendix~\ref{app:CNBApp}) and $\mathcal{I}_{N,\beta}$ (Appendix~\ref{app:INBApp}), we find as expected that the following limit exists:
\beq
\lim_{N \rightarrow \infty} \mathcal{P}[a = 2\sqrt{t}(1 - N^{-2/3}\eta)] = \mathcal{P}^*_{\mathrm{G}\beta\mathrm{E}}(\eta)
\eeq
For a better comparison to the literature, we set $\xi = -2\eta$, and we find:
\bea
\mathcal{P}_{\mathrm{G}\beta\mathrm{E}}^*(\eta) & = & \tau_{\beta}\,\exp\left[-\frac{\beta|\xi|^3}{12} + \frac{\sqrt{2}(\beta -1)}{3}|\xi|^{3/2} + \frac{\beta + \beta^{-1} - 3}{8}\ln|\xi| + O(|\xi|^{-3/2})\right] \nonumber \\
\label{eq:expa}& &
\eea
$\tau_{\beta}$ is a constant given by:
\beq
\ln\tau_{\beta} = \left(\frac{17}{8} - \frac{25}{24}(\beta + \beta^{-1})\right)\ln 2 - \frac{\ln(2\pi)}{2} - \frac{\ln \beta}{2} + \kappa_{\beta}
\eeq
$\kappa_{\beta}$ is the finite part in the large $N$ asymptotic expansion of  $\sum_{j = 1}^N \ln\Gamma(1 + j\beta)$. It is studied in Appendix~\ref{app:kbeta}, and computed in a somewhat explicit way for rational values of $\beta$ in Appendix~\ref{app:valuek}.

\medskip
\medskip

\begin{tabular}{r|l|l}
Value of $\beta$ & $\ln(\tau_{\beta})$ & Numerical value of $\tau_{\beta}$ \\
\hline & & \\
$1$ & $-\frac{\ln 2}{12} + \zeta'(-1)$ & $0.8723714$ \\
$2$ & $-\frac{37\ln 2}{48} + \frac{\zeta'(-1)}{2}$ & $0.5395545$ \\
$3$ & $-\frac{97\ln 2}{72} - \frac{7\ln 3}{36} - \frac{\ln(2\pi)}{6} + \frac{\ln\Gamma(1/3)}{3} + \frac{\zeta'(-1)}{3}$ & $0.3071491$\\
$4$ & $-\frac{87\ln 2}{32} - \frac{\ln(2\pi)}{4} + \frac{\ln\Gamma(1/4)}{2} + \frac{\zeta'(-1)}{4} $ & $0.1752911$ \\
& & \\
\hline & & \\
$1/2$ & $-\frac{11\ln 2}{48} + \frac{\zeta'(-1)}{2}$ & $0.7854042$ \\
$1/3$ & $-\frac{97\ln 2}{72} + \frac{5\ln 3}{18} - \frac{\ln(2\pi)}{6} + \frac{\ln\Gamma(1/3)}{3} + \frac{\zeta'(-1)}{3}$ & $0.5160081$\\
$1/4$ & $-\frac{185\ln 2}{96} - \frac{\ln(2\pi)}{4} + \frac{\ln\Gamma(1/4)}{2} + \frac{\zeta'(-1)}{4}$ & $0.3034417$\\
& & \\
\hline & & \\
$2/3$ & $-\frac{65\ln 2}{144} + \frac{2\ln 3}{9} + \frac{\ln(2\pi)}{6} - \frac{\ln\Gamma(1/3)}{3} + \frac{\zeta'(-1)}{6}$ & $0.8882751$ \\
$3/2$ & $\frac{17\ln 2}{144} - \frac{25\ln 3}{72} + \frac{\ln(2\pi)}{6} - \frac{\ln\Gamma(1/3)}{3} + \frac{\zeta'(-1)}{6}$ & $0.7051367$
\end{tabular}

\medskip

\subsubsection{Comparison to Tracy-Widom law}
\label{sec:compa}
This expansion should match the left tail of Tracy-Widom law \cite{TW93,TW95} for the usual values of $\beta = 1/2, 1, 2$.  In fact, the leading term in $\epsilon^3$ was obtained in \cite{DM06,DM08} from Eqn.~\ref{eq:fofof}. The correspondence between the scaling variable $\xi$ in the large deviation and the Tracy-Widom variable $\chi$ should be the following:
\begin{itemize}
\item[$\bullet$] Gaussian Unitary Ensemble ($\beta = 1$).
\beq
\mathrm{F}_2(\chi) = \mathcal{P}^*_{\mathrm{GUE}}(-\chi)
\eeq
If we specialize Eqn.~\ref{eq:expa}, we obtain:
\beq
\mathrm{F}_2(\chi) \mathop{=}_{\chi \rightarrow - \infty}\; 2^{1/24}\,e^{\zeta'(-1)}\,\frac{e^{- \frac{|\chi|^3}{12}}}{|\chi|^{1/8}}\left(1 + O(|\chi|^{-3})\right)
\eeq
Notice that only $F^{g,0}$ are involved in Eqn.~\ref{eq:hud} when $\beta = 1$, so the $O(|\chi|^{-3/2})$ is in fact a $O(|\chi|^{-3})$ according to Eqn.~\ref{eq:hufr}.
\item[$\bullet$] Gaussian Orthogonal Ensemble ($\beta = 1/2$).
\beq
\mathrm{F}_1(\chi) = \mathcal{P}_{\mathrm{GOE}}^*(-\chi)
\eeq
We obtain:
\beq
\label{eq:F11c}\mathrm{F}_1(\chi) \mathop{=}_{\chi \rightarrow - \infty}\; 2^{-11/48}\,e^{\zeta'(-1)/2}\,\frac{e^{-\frac{|\chi|^3}{24} - \frac{|\chi|^{3/2}}{3\sqrt{2}}}}{|\chi|^{1/16}}\left(1 + O(|\chi|^{-3/2})\right)
\eeq
\item[$\bullet$] Gaussian Symplectic Ensemble ($\beta = 2$).
\beq
\mathrm{F}_4(\chi) = \mathcal{P}_{\mathrm{GSE}}^*(-2^{-2/3}\chi)
\eeq
We obtain:
\beq
\label{eq:F44c}\mathrm{F}_4(\chi) \mathop{=}_{\chi \rightarrow - \infty}\; 2^{-35/48}\,e^{\zeta'(-1)/2}\,\frac{e^{-\frac{|\chi|^3}{24} + \frac{|\chi|^{3/2}}{3\sqrt{2}}}}{|\chi|^{1/16}}\left(1 + O(|\chi|^{-3/2})\right)
\eeq
\end{itemize}

These expansions match earlier results. Usually, the constant term is considered as the difficult part of the asymptotics. For $\beta = 1$, this constant was first obtained by P.~Deift, A.~Its and I.~Krasovsky by Riemann-Hilbert asymptotic analysis \cite{DIK}. For $\beta = 1/2$ and $2$, it was determined by J.~Baik, R.~Buckingham and J.~DiFranco by a technique allowing representation of total integrals of the Hastings-MacLeod solution of Painlev\'{e} II \cite{BBDF07}. In these works as in ours, an important step was the evaluation of the asymptotic of a Selberg-type integral (see Section~\ref{sec:constant} and Appendix~\ref{app:C}). We remark that for $\beta \neq 1/2,1,2$, the choice of the normalization of the scaling variable is somewhat arbitrary, since at present, there is no other definition of a "$\beta$ Tracy-Widom" distribution than:
\beq
"\mathrm{F}_{2\beta} = \mathcal{P}_{\mathrm{G}\beta\mathrm{E}}^*" \nonumber
\eeq
Our $\tau_{\beta}$ gives the constant of $\mathrm{F}_{2\beta}(\chi) \equiv \mathcal{P}_{\mathrm{G}\beta\mathrm{E}}^*(\eta = -\chi/2)$.

\section{Conclusion}

\begin{itemize}
\item[$\bullet$]
We showed how to determine to all orders the left large deviation function of the maximal eigenvalue of a random matrix in the $\beta$ ensemble, for all values of $\beta$. We have presented in detail the case where the large $N$ spectrum is connected (one-cut case), but the method combined with existing literature would give as well the multi-cut case. We performed numerics to check our results on the case of the Gaussian model at orders $O(N^2)$ and $O(N)$ with a very good agreement, and at order $O(1)$ with a less convincing agreement.
\item[$\bullet$]
We described the repartition function $\mathrm{F}_{2\beta}(\chi)$ of the maximal eigenvalue when it reaches the edge of the spectrum from the left (existence of this distribution was proved in \cite{RRV} with probabilistic methods). More precisely, we gave its full asymptotic expansion when $\chi \rightarrow -\infty$ in terms of geometric quantities associated to a spectral curve of equation:
\beq
\label{eq:STW}\widehat{\Sigma}\;:\; y^2 = x + \frac{1}{x} - 2
\eeq
Since it is well known that $\mathrm{F}_{1/2}$, $\mathrm{F}_{1}$, $\mathrm{F}_{2}$ are the Tracy-Widom laws, we have related these Tracy-Widom laws defined from the world of Painlev\'{e} equations and integrable systems, to the topological recursion defined from algebraic geometry. For $\beta = 1$, in a related work \cite{BETW}, two of us proved directly that the resummation of the symplectic invariants of $\mathcal{S}^*$ are solution to an ODE related to Painlev\'{e} II, as it should be according to the work of Tracy and Widom. For general $\beta$, the topological recursion allows to define a $\mathrm{F}_{2\beta}$ for any value of $\beta$. Here, we have obtained the three first terms of its left tail, the constant term, and we may compute recursively the next terms. We notice that the constant term was computed in three parts, similarly to \cite{BBDF07} where the problem was adressed for $\beta = 1/2,1,2$. Eventually, it would be interesting to relate the $F^{g,k}[\mathcal{S}^*]$, to the parabolic differential equation obtained recently in \cite{BluVir} from which the Tracy-Widom  $\beta$ distribution can be extracted. Nevertheless, at present, nothing is known about the relation between integrable systems and $\beta$ matrix models.
\item[$\bullet$] From the mathematical point of view, our work is based on two assumptions : existence of a $1/N$ expansion, and possibility to take a double scaling limit. The first one will be justified in a forecoming work \cite{BG11}, whereas the second one seems hard to justify rigorously at present in absence of integrability. If one accepts the $1/N$ expansion, the loop equations imposes the special $\beta$-dependance of the expansion. From this, one observe the following duality $\beta \leftrightarrow 1/\beta$ at the level of left-tail asymptotics:
    \beq
    \mathcal{P}_{\mathrm{G}\beta\mathrm{E}}(\chi) = \frac{e^{\kappa_{\beta}}}{\beta\kappa_{1/\beta}}\,\widetilde{\mathcal{P}}_{\mathrm{G}\frac{1}{\beta}\mathrm{E}}(\beta^{2/3}\chi)
    \eeq
 where $\widetilde{\cdots}$ is obtained by taking the other branch of the square root (all half-integer powers of $s$ of the expansion appear with a minus sign). Such a duality is readily observed on the known results for $\mathrm{F}_1$ and $\mathrm{F}_4$ (compare Eqn.~\ref{eq:F11c} and Eqn.~\ref{eq:F44c}). Whether this duality has a counterpart for the functions themselves is an open question.
\item[$\bullet$] One could ask for a combinatorial interpretation of the symplectic invariants of the Tracy-Widom curve $\mathcal{F}^{g,0}(\mathcal{S})$ (let alone $\mathcal{F}^{g,k}(\mathcal{S})$) for fixed $g$. A related challenge would be to have a closed formula, at $g$ fixed, for $F^{g,0}$ and $\omega_n^{g,0}$ of the Tracy-Widom curve.
\end{itemize}

\vspace{2cm}

\section*{Acknowledgments}

G.B. and B.E would like to thank M.~Berg\`{e}re, F.~David and G.~Schehr for useful and fruitful discussions, as well as the organizers of StatPhys2010 in Brisbane and Cairns.
G.B. would like to thank the scientific hospitality of the SISSA and of the Department of Maths and Statistics of Melbourne University, where part of this work was completed. He also would like to thank in general the organizers and participants of the Random Matrix Theory semester at the MSRI, and in particular J.~Baik and P.~Deift for their corrections. He acknowledges the funding of the CEA and CFM for travel to StatPhys conferences.
The work of B.E. is partly supported by the ANR project GranMa "Grandes Matrices Al\'{e}atoires" ANR-08-BLAN-0311-01, by the European Science Foundation through the Misgam program, by the Quebec government with the FQRNT. He also would like to thank the CRM (Centre de recherche math\'ematiques de Montr\'eal, QC, Canada) and the CERN for its hospitality.

\newpage

\appendix

\numberwithin{equation}{section}
\section*{Appendix}

\section{Determination of the normalization $C_{N,\beta}$}
\label{app:C}
When $a \rightarrow -\infty$, we have the asymptotics:
\bea
F^{0,0}(a) & = & -\frac{t|a|^2}{2} - t^2\ln|a| - \frac{3t^2}{2} + t^2\ln t + O(|a|^{-2})\nonumber \\
F^{0,1}(a) & = & t\ln |a| + t - t\ln t - t\ln(2\pi) + O(|a|^{-2}) \nonumber \\
F^{1,0}(a) & = & -\frac{\ln 2}{6} + O(|a|^{-4}) \nonumber \\
F^{0,2}(a) & = & \frac{\ln 2}{3} + O(|a|^{-2}) \nonumber
\eea
and one can prove that all other $F^{g,k}$ are $O(|a|^{-2})$. As $F(a) = \sum_{g,k \geq 0} \nu^{2 - 2g}\hbar^k F^{g,k}(a)$, we have the following result:
\bea
& & \lim_{a \rightarrow -\infty} \left[F(a) - \left(\beta N^2\left\{-|a|^2/(2 t) - \ln|a|\right\}
 + (\beta - 1)N \ln|a|\right)\right] \nonumber \\
& = & \beta N^2\left(-\frac{3}{2} + \ln t\right) + (\beta - 1)N\left(1 - \ln(2\pi) - \ln t\right) \nonumber \\
 & & + \left(\frac{\beta + \beta^{-1}}{3} - \frac{5}{6}\right)\ln 2 \nonumber \\
\label{eq:result}& &
\eea

On the other hand, consider the bounded eigenvalue integral
\beq
\mathcal{I}_{N,\beta}(a) = \int_{]-\infty,a]^{N}}\mathrm{d}\lambda_1\cdots\mathrm{d}\lambda_N\,|\Delta(\lambda)|^{2\beta}\,e^{-\frac{N\beta}{2t}\sum_{i = 1}^N \lambda_i^2}
\eeq
One knows that:
\beq
\mathcal{I}_{N,\beta}(a) = C_{N,\beta}\,e^{F(a)}
\eeq
We want to find $C_{N,\beta}$ by comparing the result of Eqn.~\ref{eq:result} to the large $N$ asymptotic of $\mathcal{I}_{N,\beta}(a)$.
Heuristically, when $a \rightarrow -\infty$, we do not see a Gaussian weight, but the tail of it starting from the point $a$, which looks like a decreasing exponential. Precisely, for any finite $N$, with the change of variable:
\beq
\mu_i = \frac{\beta N |a|}{t}(a - \lambda_i)
\eeq
we find
\bea
\mathcal{I}_{N,\beta}(a) & = & \left(\frac{t}{N\beta|a|}\right)^{\beta N(N - 1) + N}\,e^{-\frac{\beta N^2|a|^2}{2t}} \nonumber \\
& & \times \,\int_{\mathbb{R}_+^N} \mathrm{d}\mu_1\cdots\mathrm{d}\mu_N\,|\Delta(\mu)|^{2\beta}\,e^{-\sum_{i = 1}^{N} \mu_i}\,e^{-\frac{t}{2N\beta|a|^2}\sum_{i = 1}^N \mu_i^2} \nonumber \\
& &
\eea
When $a \rightarrow -\infty$, the integral is of order $1$ and converges to a Selberg integral with Laguerre-type weight:
\beq
\int_{\mathbb{R}_+^N} \mathrm{d}\mu_1\cdots\mathrm{d}\mu_N\,|\Delta(\mu)|^{2\beta}\,e^{-\sum_{i = 1}^{N} \mu_i} = \frac{\left(\prod_{j = 1}^{N} \Gamma(1 + j\beta)\right)^2}{\Gamma(1 + \beta)^N\,\Gamma(1 + N\beta)}
\eeq
Thus, we have:
\bea
&& \lim_{a \rightarrow -\infty}\left[\ln\left(\mathcal{I}_{N,\beta}(a)\right) - \left(\beta N^2\left\{-\frac{|a|^2}{2t}
 - \ln|a|\right\} + (\beta - 1)N\ln|a|\right)\right] \nonumber \\
& = & \beta N^2\left(-\frac{3}{2} + \ln t\right) + (\beta - 1)N\left(1 - \ln(2\pi) - \ln t\right) \nonumber \\
& & - \beta N^2\ln N + \beta N^2\left(\frac{3}{2} - \ln \beta\right) + (\beta - 1)N\ln N + (\beta - 1)N\left(-1 + \ln(2\pi) + \ln\beta\right) \nonumber \\
&&  - \ln\Gamma(1 + N\beta) - N\ln\Gamma(1 + \beta) + \sum_{j = 1}^N 2\ln\Gamma(1 + j\beta)  \nonumber \\
& &
\eea
Hence:
\bea
C_{N,\beta} & = & \exp\left[-\beta N^2\ln N + \beta N^2\left(\frac{3}{2} - \ln \beta\right) + (\beta - 1)N\ln N + \right. \nonumber \\
& & \left. + (\beta - 1)N\left(-1 + \ln(2\pi) + \ln \beta\right)\right]\;\cdot\;2^{\frac{5}{6} - \frac{\beta + \beta^{-1}}{3}}\,\frac{\left(\prod_{j = 1}^N \Gamma(1 + j\beta)\right)^2}{\Gamma(1 + \beta)^N\Gamma(1 + N\beta)} \nonumber \\
\label{eq:CNBBBB}& &
\eea
Note that this normalization constant is the correct one when there is one hard edge. The result is different when there is no hard edge.

\section{Reminder on special functions}
\label{reminder}

\subsection{Gamma function}

We recall the definition of the Gamma function. For $\mathrm{Re}\,s > 1$:
\beq
\Gamma(z) = \int_{0}^{\infty}\mathrm{d}\sigma\,\sigma^{z - 1}\,e^{-\sigma}
\eeq
It satisfies the functional equation:
\beq
\Gamma(z + 1) = z\Gamma(z)
\eeq
It can be analytically continued on the complex plane, and $1/\Gamma(z)$ is an entire function, which can be defined by the product:
\beq
\frac{1}{\Gamma(z)} = z\,e^{\gamma_E z}\,\prod_{m = 1}^{\infty} \left(1 + \frac{z}{m}\right)\,e^{-z/m}
\eeq
Its asymptotic expansion is given by Stirling formula:
\beq
\ln\Gamma(z + 1) \mathop{=}_{z \rightarrow \infty} z\ln z - z + \frac{\ln z}{2} + \frac{\ln(2\pi)}{2} + o(1)
\eeq
It satisfies the reflection property:
\beq
\Gamma(1 - z)\Gamma(z) = \frac{\pi}{\sin(\pi z)}
\eeq
and the "addition of angles" formula:
\beq
\label{eq:double}\prod_{m = 0}^{p - 1} \Gamma\left(\frac{m}{p} + z\right) = (2\pi)^{\frac{p - 1}{2}}\,p^{- pz + \frac{1}{2}}\,\Gamma(pz)
\eeq

\subsection{Riemann zeta function}

We recall the definition of the zeta function. For $\mathrm{Re}\,s > 1$:
\bea
\zeta(s) & \equiv & \sum_{j = 1}^{\infty} \frac{1}{j^s}
\eea
It can be extended by analytical continuation on the complex plane, by the reflection relation:
\beq
\Gamma\left(\frac{s}{2}\right)\,\pi^{-s/2}\,\zeta(s) = \Gamma\left(\frac{1 - s}{2}\right)\,\pi^{-\frac{1 - s}{2}}\,\zeta(1 - s)
\eeq

\subsection{Barnes $G$-function}

We recall that the Barnes $G$-function \cite{Barnes} is an entire function satisfying:
\beq
G(z + 1) = \Gamma(z)G(z)
\eeq
At integers, it coincides with the product of factorials:
\beq
\forall k \in \mathbb{N} \qquad  G(k) = \prod_{j = 1}^{k - 2} j!
\eeq
We take the convention that a void product is equal to $1$. Its asymptotic expansion for $z \rightarrow \infty$ is a classical result \cite{Voros}:
\beq
\label{eq:aG} \ln G(z + 1) = z^2\left(\frac{\ln z}{2} - \frac{3}{4}\right) + \frac{\ln(2\pi)z}{2} - \frac{\ln z}{12} + \zeta'(-1) + o(1)
\eeq
We quote some special values and relations \cite{Adam}:
\bea
\label{eq:G12} \ln G(1/2) & = & \frac{7 \ln 2}{24} - \frac{\ln(2\pi)}{4} + \frac{3\zeta'(-1)}{2} \\
\label{eq:G14} \ln G(1/4) + \ln G(3/4) & = & \frac{\ln 2}{8} - \frac{\ln(2\pi)}{4} - \frac{\ln\Gamma(1/4)}{2} + \frac{9\zeta'(-1)}{4} \\
\label{eq:G13} \ln G(1/3) + \ln G(2/3) & = & \frac{7\ln 3}{36} - \frac{\ln(2\pi)}{3} - \frac{\ln\Gamma(1/3)}{3} + \frac{8\zeta'(-1)}{3} \\
\label{eq:G16} \ln G(1/6) + \ln G(5/6) & = & -\frac{5\ln 2}{36} - \frac{25\ln 3}{72} + \frac{\ln(2\pi)}{6} - \frac{4\ln\Gamma(1/3)}{3} + \frac{5\zeta'(-1)}{3} \nonumber \\
& &
\eea
In principle, $G(p/q)$ can be expressed in terms of an increasing number of fundamental constants when $p$ and $q$ increase.
We do not try in this article to express $G(p/q)$ in other ways. We mention an "addition of angles" formula \cite{Donnon}:
\bea
\sum_{m = 0}^{p - 1} \ln G\left(\frac{m}{p} + z + 1\right) & = & \left(p - \frac{1}{p}\right)\zeta'(-1) + \frac{\ln G(pz + 1)}{p} - z\ln\Gamma(pz) \nonumber \\
& & + \left(pz^2 - z + \frac{1}{6p}\right)\frac{\ln p}{2}  + \sum_{m = 0}^{p - 1} \left(\frac{m}{p} + z\right)\ln\Gamma\left(\frac{m}{p} + z\right) \nonumber \\
\label{eq:aaaa} & &
\eea

\section{Large $N$ expansion of some Selberg integrals}
\label{app:D}
We have defined:
\beq
\mathcal{I}_{N,\beta} \equiv \mathcal{I}_{N,\beta}(\infty) = \int_{\mathbb{R}^N}\mathrm{d}\lambda_1\ldots\mathrm{d}\lambda_N\,|\Delta(\lambda)|^{2\beta}\,e^{-\frac{N\beta}{2t}\sum_{i = 1}^N \lambda_i^2}
\eeq
This is a Selberg integral, and it is given by:
\bea
\mathcal{I}_{N,\beta} & = & (2\pi)^{N/2}\,\left(\frac{t}{N\beta}\right)^{\frac{\beta N^2}{2} - \frac{(\beta - 1)N}{2}}\,\frac{\prod_{j = 1}^N \Gamma(1 + j\beta)}{\Gamma(1 + \beta)^N} \nonumber \\
\label{eq:Inbeta}& &
\eea
In the main text, we are interested in:
\beq
\mathcal{P}(a) = \frac{\mathcal{I}_{N,\beta}(a)}{\mathcal{I}_{N,\beta}(\infty)} = \frac{C_{N,\beta}}{\mathcal{I}_{N,\beta}}\,e^{F(a)}
\eeq
up to a factor equivalent to $1$. So, we need the large $N$ asymptotics of $\ln \mathcal{I}_{N,\beta}$ and $\ln C_{N,\beta}$ up to $o(1)$.

\subsection{Asymptotics of $\sum_{j = 1}^N \ln \Gamma(1 + j\beta)$}
\label{app:kbeta}
\label{app:aSSS}

We first determine the large $N$ asymptotics of $\sum_{j = 1}^N \ln \Gamma(1 + j\beta)$ for a general $\beta$. We start with an integral representation of $\ln\Gamma$:
\bea
\ln \Gamma(z) & = & \left(z - \frac{1}{2}\right)\ln z - z + \frac{\ln(2\pi)}{2} + \nonumber \\
& & + \int_0^{\infty}\mathrm{d}\sigma\,\frac{e^{-\sigma z}}{\sigma^2}\left(\frac{\sigma}{e^{\sigma} - 1} - 1 + \frac{\sigma}{2}\right)
\eea
We have:
\bea
& & \sum_{j = 1}^N \ln \Gamma(1 + j\beta) \nonumber \\
& = & \sum_{j = 1}^N \ln(j\beta) + \ln\Gamma(j\beta) \nonumber \\
& = & \sum_{j = 1}^N \ln j + N\ln(\beta) \nonumber \\
& & + \sum_{j = 1}^N \left(j\beta - \frac{1}{2}\right)\ln(j\beta) - \beta\frac{N(N + 1)}{2} + N\frac{\ln(2\pi)}{2} \nonumber \\
& & + \underbrace{\sum_{j = 1}^N \int_{0}^{\infty} \mathrm{d}\sigma\,\frac{e^{-j\beta\sigma}}{\sigma^2}\left(\frac{\sigma}{e^{\sigma} - 1} - 1 + \frac{\sigma}{2}\right)}_K \nonumber \\
& = & \beta\sum_{j = 1}^N j\ln(j) + \frac{1}{2}\sum_{j = 1}^N \ln(j) + \nonumber \\
& & (\beta\ln\beta - \beta)\frac{N(N + 1)}{2} + N\frac{\ln(\beta) + \ln(2\pi)}{2} + K\nonumber \\
& & \nonumber \\
\eea
We recall the classical asymptotic expansions:
\bea
\sum_{j = 1}^{N} \ln j & = & \ln \Gamma(N+1)=N\ln N - N + \frac{\ln N}{2} + \frac{\ln(2\pi)}{2} + o(1) \nonumber \\
\sum_{j = 1}^{N} j\ln j & = & \frac{N^2\ln N}{2} - \frac{N^2}{4} + \frac{N \ln N}{2} + \frac{\ln N}{12} + \frac{1}{12} - \zeta'(-1) + o(1)
\eea
Let us determine the large $N$ asymptotic of the integral:
\bea
K & \equiv & \sum_{j = 1}^N \int_{0}^{\infty} \mathrm{d}\sigma\,\frac{e^{-j\beta\sigma}}{\sigma^2}\left(\frac{\sigma}{e^{\sigma} - 1} - 1 + \frac{\sigma}{2}\right) \nonumber \\
& = & \frac{1}{12}\sum_{j = 1}^N \int_{0}^{\infty}\mathrm{d}\sigma\,e^{-j\beta\sigma} \nonumber \\
& & + \sum_{j = 1}^N \int_{0}^{\infty}\mathrm{d}\sigma\,\frac{e^{-j\beta\sigma}}{\sigma^2}\left(\frac{\sigma}{e^{\sigma} - 1} - 1 + \frac{\sigma}{2} - \frac{\sigma^2}{12}\right) \nonumber \\
& = & \frac{1}{12\beta}\sum_{j = 1}^N \frac{1}{j} + \int_{0}^{\infty} \frac{\mathrm{d}\sigma}{\sigma^2}\,\frac{1}{e^{\beta\sigma} - 1}\left(\frac{\sigma}{e^{\sigma} - 1} - 1 + \frac{\sigma}{2} - \frac{\sigma^2}{12}\right) \nonumber \\
& = & \frac{\ln N}{12\beta} + \frac{\gamma_E}{12\beta} + \int_{0}^{\infty} \frac{\mathrm{d}\sigma}{\sigma^2}\,\frac{1}{e^{\beta\sigma} - 1}\left(\frac{\sigma}{e^{\sigma} - 1} - 1 + \frac{\sigma}{2} - \frac{\sigma^2}{12}\right)
\eea
where $\gamma_E$ is the Euler-Mascheroni constant. Collecting the results, we find:
\bea
& & \sum_{j = 1}^{N} \ln\Gamma(1 + j\beta) \nonumber \\
& = & \frac{\beta N^2 \ln N}{2} + \beta\left(-\frac{3}{4} + \frac{\ln(\beta)}{2}\right)N^2 + \frac{\beta + 1}{2} N\ln N \nonumber \\
& & + \left(-\frac{1}{2} + \frac{\beta\ln(\beta)}{2} - \frac{\beta}{2} + \frac{\ln(\beta)}{2} + \frac{\ln(2\pi)}{2}\right)N \nonumber \\
\label{eq:divS} & & + \frac{\beta + \beta^{-1} + 3}{12}\ln N + \kappa_{\beta} + o(1)
\eea
where the finite part $\kappa_{\beta}$ is given by:
\bea
\kappa_{\beta} & = & \frac{\ln(2\pi)}{4} + \beta\left(\frac{1}{12} - \zeta'(-1)\right) + \frac{\gamma_E}{12\beta} \nonumber \\
& & + \int_{0}^{\infty} \frac{\mathrm{d}\sigma}{\sigma^2}\,\frac{1}{e^{\beta\sigma} - 1}\left(\frac{\sigma}{e^{\sigma} - 1} - 1 + \frac{\sigma}{2} - \frac{\sigma^2}{12}\right) \nonumber \\
& &
\eea
or equivalently
\bea
\kappa_{\beta} & = & \frac{\ln(2\pi)}{4} + \beta\left(\frac{1}{12} - \zeta'(-1)\right) + \frac{\gamma_E}{12\beta} \nonumber \\
& & + \int_{0}^{\infty} \mathrm{d}\sigma \:\left[\frac{6 \sigma \coth\left(\sigma/2 \right) -12 -\sigma^2}{
12 \sigma^2 \left(e^{\beta \sigma}-1\right)} \right] \nonumber \\
& &
\eea

\subsection{Asymptotics of $\mathcal{I}_{N,\beta}$}
\label{app:INBApp}
If we apply Eqn~\ref{eq:divS} to the definition of $\mathcal{I}_{N,\beta}$ (Eqn.~\ref{eq:Inbeta}), we find:
\bea
\ln \mathcal{I}_{N,\beta} & = & \beta\left(-\frac{3}{4} + \frac{\ln t}{2}\right)N^2 + \beta N\ln N  \nonumber \\
& & + \left(\beta\ln(\beta) - \ln\Gamma(1 + \beta) - \frac{\beta + 1}{2} + \ln(2\pi) - \frac{\beta - 1}{2} \ln t\right)N  \nonumber \\
& & + \frac{\beta + \beta^{-1} + 3}{12}\ln N  + \kappa_{\beta} + o(1) \nonumber \\
& &
\eea

\subsection{Asymptotic of $C_{N,\beta}$}
\label{app:CNBApp}
If we apply Eqn.~\ref{eq:divS} to the expression of $C_{N,\beta}$ (Eqn.~\ref{eq:CNBBBB}), we find:
\bea
\ln C_{N,\beta} & = & \beta N \ln N + N\left(-\beta + \beta\ln\beta + \beta\ln(2\pi) - \ln\Gamma(1 + \beta)\right) + \left(\frac{\beta + \beta^{-1}}{6}\right)\ln(N) \nonumber \\
& & + \left(\frac{5 - 2(\beta + \beta^{-1})}{6}\right)\ln 2 - \frac{\ln(2\pi)}{2} - \frac{\ln\beta}{2} + 2\kappa_{\beta} + o(1) \nonumber \\
& &
\eea

Finally we get the normalization constant for the cumulative distribution
of $\lambda_{\rm max}$. We have indeed
\beq \label{eq:PbLargeN}
\mathcal{P}(a) = \frac{\mathcal{I}_{N,\beta}(a)}{\mathcal{I}_{N,\beta}(\infty)} = \frac{C_{N,\beta}}{\mathcal{I}_{N,\beta}}\,e^{F(a)}=\exp\left\{F(a)
+\ln C_{N,\beta}-\ln \mathcal{I}_{N,\beta}
\right\}
\eeq
with
\bea
\ln C_{N,\beta}-\ln \mathcal{I}_{N,\beta} & = & \beta N^2 \left(\frac{3}{4}-\frac{\ln t}{2} \right)
+N (\beta -1)\left( -\frac{1}{2}+\ln (2 \pi) +\frac{\ln t}{2} \right) \nonumber \\
&&+\ln N \left( \frac{\beta +\beta^{-1} -3}{12} \right)
+\left( \frac{5-2(\beta +\beta^{-1})}{6} \right) \ln 2 \nonumber \\
&&-\frac{\ln (2\pi)}{2}-\frac{\ln \beta}{2} +\kappa_{\beta}
\eea

\subsection{Another computation of $\kappa_{\beta}$ for $\beta$ rational}

\subsubsection{$\beta$ integer}

By definition we have $\kappa_{\beta}=\left[\sum_{j = 1}^N \ln \Gamma(1 + j\beta)  \right]_{\rm finite}$.
Let us come back to the asymptotic of $\sum_{j = 1}^N \ln \Gamma(1 + j\beta)$ when $\beta$ is a positive integer. Here, we only focus on the finite term of this expansion, which we called previously $\kappa_{\beta}$. Of course, one could check with this method that one does obtain the divergent term of Eqn.~\ref{eq:divS}.

For $\beta = 1$, this series is the Barnes function itself:
\beq
\sum_{j = 1}^{N} \ln\Gamma(1 + j) = \ln G(N + 2)
\eeq
Using the asymptotic expansion of $G(z + 1)$ given in Eqn.~\ref{eq:aG}, we obtain for any $w$ in the large $z$ limit:
\beq
\left[\ln G\left(z + w + 1\right)\right]_{\mathrm{finite}} = \frac{w\ln(2\pi)}{2} + \zeta'(-1) + o(1)
\eeq
Thus, we have:
\beq
\kappa_1 = \frac{\ln(2\pi)}{2} + \zeta'(-1)
\eeq

As a matter of fact, one may use only the asymptotic expansion of Barnes function (Eqn.~\ref{eq:aG}) to work out the case $\beta$ integer. We start from the "addition of angles" formula (Eqn.~\ref{eq:double}) and write:
\beq
\prod_{j = 1}^N \Gamma(j\beta) = (2\pi)^{\frac{(1 - \beta)N}{2}}\,\beta^{\frac{\beta N(N + 1)}{2} - \frac{N}{2}}\,\prod_{m = 0}^{\beta - 1} \frac{G\left(N + \frac{m}{\beta} + 1\right)}{G\left(\frac{m}{\beta} + 1\right)}
\eeq
Then:
\bea
\sum_{j = 1}^N \ln\Gamma(1 + j\beta) & = & \ln(N!) + N\ln\beta + N \left(\frac{1 - \beta}{2}\right)\ln(2\pi) + \left(\frac{\beta N(N + 1)}{2} - \frac{N}{2}\right)\ln \beta \nonumber \\
& & + \sum_{m = 0}^{\beta - 1} \left[ \ln G\left(N + \frac{m}{\beta} + 1\right) - \ln G\left(\frac{m}{\beta} + 1\right)
\right] \nonumber \\
& &
\eea
Subsequently:
\bea \label{eq:kappaInteger}
\kappa_{\beta} & = & \frac{\ln(2\pi)}{2} + \sum_{m = 0}^{\beta - 1} \left[\frac{m\ln(2\pi)}{2\beta} + \zeta'(-1) - \ln G\left(\frac{m}{\beta} + 1\right)\right] \nonumber \\
& = & \left(\frac{\beta + 1}{4}\right)\ln(2\pi) + \beta \zeta'(-1) - \sum_{m = 0}^{\beta - 1} \ln G\left(1 + \frac{m}{\beta}\right)
\eea
For example:
\bea
\kappa_{2} & = & \frac{3\ln(2\pi)}{4} + 2\zeta'(-1) - \ln G(3/2) \nonumber \\
& = & \frac{5\ln 2}{24} + \frac{\ln(2\pi)}{2} + \frac{\zeta'(-1)}{2}
\eea
where we used the value of $G(1/2)$ given in Eqn.~\ref{eq:G12} and $\Gamma(1/2) = \sqrt{\pi}$.

\medskip

Actually, $\kappa_{\beta}$ can be expressed only in terms of values of $\Gamma$ at rational points, by means of the formula Eqn.~\ref{eq:aaaa}:
\beq
\label{eq:mul2} \sum_{m = 0}^{p - 1} \ln G\left(1 + \frac{m}{p}\right) = \left(p - \frac{1}{p}\right)\zeta'(-1) + \frac{\ln p}{12p} + \sum_{m = 1}^{p - 1} \frac{m}{p}\,\ln\Gamma\left(\frac{m}{p}\right)
\eeq
Thus, we have:
\beq
\kappa_{\beta} = \left(\frac{\beta + 1}{4}\right)\ln(2\pi) + \frac{\zeta'(-1)}{\beta} - \frac{\ln \beta}{12\beta} - \sum_{m = 1}^{\beta - 1} \frac{m}{\beta}\,\ln\Gamma\left(\frac{m}{\beta}\right)
\eeq
For example:
\bea
\kappa_{3} & = & \frac{11\ln 3}{36} + \frac{\ln(2\pi)}{3} + \frac{\zeta'(-1)}{3} + \frac{\ln\Gamma(1/3)}{3} \\
\kappa_{4} & = & \frac{7\ln 2}{12} + \frac{\ln(2\pi)}{4} + \frac{\zeta'(-1)}{4}  + \frac{\ln\Gamma(1/4)}{2}
\eea

\subsubsection{Value of $\kappa_{\beta}$ for rational $\beta$}

\label{app:valuek}

The case $\beta = p/q \in \mathbb{Q}$ can be treated similarly (we assume $\mathrm{gcd}(p,q) = 1$). It is convenient to do the computation for a given congruence of $N$ mod $q$. Let us write:
\beq
N =  qL +R,\qquad 0 \leq R \leq q - 1 \nonumber
\eeq
We have according to the duplication formula:
\bea
\prod_{j = 1}^{N} \Gamma\left(j\frac{p}{q}\right) & = & (2\pi)^{\frac{(1 - p)N}{2}}\,p^{\frac{p N(N + 1)}{2q} - \frac{N}{2}}\,\prod_{m = 0}^{p - 1} \prod_{j = 1}^{N} \Gamma\left(\frac{j}{q} + \frac{m}{p}\right) \nonumber \\
& = & (2\pi)^{\frac{(1 - p)N}{2}}\,p^{\frac{p N(N + 1)}{2q} - \frac{N}{2}} \nonumber \\
& & \phantom{space} \cdot \prod_{m = 0}^{p - 1} \frac{\prod_{r = 1}^{R} G\left(L + \frac{r}{q} + \frac{m}{p} + 1\right)\cdot\prod_{r = R + 1}^{q} G\left(L + \frac{r}{q} + \frac{m}{p}\right)}{\prod_{r = 1}^q G\left(\frac{r}{q} + \frac{m}{p}\right)} \nonumber \\
& &
\eea
Then:
\beq
\prod_{j = 1}^N \Gamma\left(1 + j\frac{p}{q}\right) = N!\,\left(\frac{p}{q}\right)^N\,\prod_{j = 1}^N \Gamma\left(j\frac{p}{q}\right)
\eeq
We need the large $z$ asymptotic expansion of $\ln G(1 + w + z/q)$ for fixed $w$. Starting from Eqn.~\ref{eq:aG}, we obtain:
\beq
\left[\ln G\left(1 + w + \frac{z}{q}\right)\right]_{\mathrm{finite}} = \left(\frac{1}{12} - \frac{w^2}{2}\right)\ln q + \frac{w\ln(2\pi)}{2} + \zeta'(-1)
\eeq
Then, we have:
\bea
\kappa_{p/q} & = & \frac{\ln(2\pi)}{2} - \sum_{m = 0}^{p - 1}\sum_{r = 1}^{q} \ln G\left(\frac{r}{q} + \frac{m}{p}\right) + pq\zeta'(-1) \nonumber \\
& & + \sum_{m = 0}^{p - 1} \left\{ \sum_{r = 1}^{R} \left[\frac{1}{12} - \frac{1}{2}\left(\frac{-R + r}{q} + \frac{m}{p}\right)^2\right] \right.\nonumber \\
& & \phantom{\sum_{m = 0}^{p - 1} sf} \left.+ \sum_{r = R + 1}^q \left[\frac{1}{12} - \frac{1}{2}\left(\frac{-R + r}{q} - 1 + \frac{m}{p}\right)^2\right]\right\}\ln q \nonumber \\
& & + \sum_{m = 0}^{p - 1} \left\{ \sum_{r = 1}^{R} \left(\frac{-R + r}{q} + \frac{m}{p}\right) + \sum_{r = R + 1}^{q} \left(\frac{-R + r}{q} - 1 + \frac{m}{p}\right)\right\}\frac{\ln(2\pi)}{2} \nonumber \\
& = & \frac{3 - p/q - q/p}{12}\ln q + \frac{p - q + 2}{4}\ln(2\pi) + pq\zeta'(-1) - \sum_{m = 0}^{p - 1} \sum_{r = 1}^{q} \ln G\left(\frac{r}{q} + \frac{m}{p}\right) \nonumber \\
& &
\eea
Notice that the result does not depend on the congruence of $N$ modulo $q$.
For $q=1$, we recover the result for $\beta=p$ integer (see \ref{eq:kappaInteger}).

We give below some values of $\kappa_{\beta}$. To complete the table, we have used repeatedly the reflection formula and the "duplication of angles" formula of the $\Gamma$ function, and Eqns.~\ref{eq:G12}-\ref{eq:G16}.

\vspace{0.2cm}

\begin{center}
\begin{tabular}{r|l}
\label{tab1} Value of $\beta$ & $\kappa_{\beta}$ \\
\hline & \\
$1$ & $\frac{\ln(2\pi)}{2} + \zeta'(-1)$ \\
$2$ & $\frac{5\ln 2}{24} + \frac{\ln(2\pi)}{2} + \frac{\zeta'(-1)}{2}$ \\
$3$ & $\frac{11\ln 3}{36} + \frac{\ln(2\pi)}{3} + \frac{\zeta'(-1)}{3} + \frac{\ln\Gamma(1/3)}{3}$ \\
$4$ & $\frac{7\ln 2}{12} + \frac{\ln(2\pi)}{4} + \frac{\zeta'(-1)}{4} + \frac{\ln\Gamma(1/4)}{2}$ \\
& \\
\hline & \\
$1/2$ & $-\frac{\ln 2}{4} + \frac{\ln(2\pi)}{2} + \frac{\zeta'(-1)}{2}$ \\
$1/3$ & $- \frac{2\ln 3}{9} + \frac{\ln(2\pi)}{3} + \frac{\ln\Gamma(1/3)}{3} + \frac{\zeta'(-1)}{3}$ \\
$1/4$ & $-\frac{5\ln 2}{8} + \frac{\ln(2\pi)}{4} + \frac{\zeta'(-1)}{4} + \frac{\ln\Gamma(1/4)}{2}$ \\
& \\
\hline & \\
$2/3$ & $\frac{13\ln 2}{72} - \frac{5\ln 3}{18} + \frac{2\ln(2\pi)}{3} - \frac{\ln\Gamma(1/3)}{3} + \frac{\zeta'(-1)}{6}$ \\
$3/2$ & $-\frac{\ln 2}{4} + \frac{11\ln 3}{72} + \frac{2\ln(2\pi)}{3} - \frac{\ln\Gamma(1/3)}{3} + \frac{\zeta'(-1)}{6}$ \\
&
\end{tabular}
\end{center}


\begin{thebibliography}{99}
\bibitem{Adam} V.S.~Adamchik, \emph{On the Barnes function}, Proceedings of the 2001 international symposium on Symbolic and algebraic computation (2001), available at \href{http://www.cs.cmu.edu/~adamchik/articles/issac01/issac01.pdf}{\texttt{http://www.cs.cmu.edu/\~{}adamchik/articles/issac01/issac01.pdf}
}
\bibitem{Ake96} G.~Akemann, \emph{Higher genus correlators for the hermitian matrix model with multiple cuts}, Nucl. Phys. \textbf{B482}  p403-430 (1996), \href{http://arxiv.org/abs/hep-th/9606004}{\texttt{hep-th/9606004}}
\bibitem{ACM} J.~Ambj{\o}rn, L.~Chekhov, Yu.~Makeenko, \emph{Higher genus correlators from the hermitian one-matrix model}, Phys.Lett. \textbf{B282} p341-348 (1992), \href{http://arxiv.org/abs/hep-th/9203009}{\texttt{hep-th/9203009}}
\bibitem{Barnes}  E.W.~Barnes, \emph{The theory of the G-function}, Quart. J. Math. \textbf{31}, p264-314 (1899)
\bibitem{BBDF07} J.~Baik, R.~Buckingham, J.~DiFranco, \emph{Asymptotics of Tracy-Widom distributions and the total integral of a Painlev\'{e} II function}, Comm. Math. Phys. \textbf{280}, Number 2, pp 463-497 (2008) \href{http://arxiv.org/abs/0704.3636}{\texttt{math/0704.3636}}
\bibitem{BluVir} A.~Bloemendal, B.~Virag, \emph{Limits of spiked random matrices I}, \href{http://arxiv.org/abs/1011.1877}{\texttt{math.PR/1011.1877}} (2010) 
\bibitem{TAS} A.~Borodin, P.L.~Ferrari, T.~Sasamoto, \emph{Transition between Airy$_1$ and Airy$_2$ processes and TASEP fluctuations}, Comm. Pure Appl. Math. \textbf{61} p1603-1629 (2008), \href{http://arxiv.org/abs/math-ph/0703023}{\texttt{math-ph/0703023}} (2007)
\bibitem{BETW} G.~Borot, B.~Eynard, \emph{The asymptotic expansion of Tracy-Widom GUE law and symplectic invariants}, (2010), \href{http://arxiv.org/abs/1012.2752}{\texttt{nlin.SI/1012.2752}}
\bibitem{BG11} G.~Borot, A.~Guionnet, \emph{Asymptotic expansion of $\beta$ matrix models with strictly convex potential}, in progress.
\bibitem{C06} L.~Chekhov, \emph{Matrix models with hard walls: geometry and solutions}, J.Phys A39: p8857-8894 (2006), \href{http://arxiv.org/abs/hep-th/0602013}{\texttt{hep-th/0602013}}
\bibitem{CE06} L.~Chekhov, B.~Eynard, \emph{Matrix eigenvalue model: Feynman graph technique for all genera}, JHEP: 0612:026 (2006), \href{http://arxiv.org/abs/math-ph/0604014}{\texttt{math-ph/0604014}}
\bibitem{CEM09} L.~Chekhov, B.~Eynard, O.~Marchal, \emph{Topological expansion of the Bethe ansatz, and quantum algebraic geometry}, (2009), \href{http://arxiv.org/abs/0911.1664}{\texttt{math-ph/0911.1664}}
\bibitem{DE} I.~Dumitriu, A.~Edelman, \emph{Matrix models for beta ensembles}, J. Math. Phys. \textbf{43}, Number 11, pp 5830–5847 (2002), \href{http://arxiv.org/abs/math-ph/0206043}{\texttt{math-ph/0206043}}
\bibitem{DM06} D.S.~Dean, S.N.~Majumdar, \emph{Large deviations of extreme eigenvalues of random matrices},
Phys. Rev. Lett., \textbf{97}, 160201 (2006), \href{http://arxiv.org/abs/cond-mat/0609651}{\texttt{cond-mat/0609651}}
\bibitem{DM08} D.S.~Dean, S.N.~Majumdar, \emph{Extreme value statistics of
eigenvalues of Gaussian random matrices}, Phys. Rev. E, \textbf{77}, 041108 (2008), \href{http://arxiv.org/abs/0801.1730}{\texttt{cond-mat/0801.1730}}
\bibitem{DIK} P.~Deift, A.~Its, I.~Krasovsky, \emph{Asymptotics of the Airy-kernel determinant}, Communications in Mathematical Physics
Comm. Math. Phys. \textbf{278}, Number 3, pp 643-678 (2008), \href{http://arxiv.org/abs/math/0609451}{\texttt{math.FA/0609451}}
\bibitem{Donnon} D.F.~Donnon, \emph{New proofs of the duplication and multiplication formulae for the gamma and the Barnes double gamma function}, (2009), \href{http://arxiv.org/abs/0903.4539}{\texttt{math/0903.4539}}
\bibitem{Dyson} F.J.~Dyson \emph{Statistical theory of the energy levels of complex systems}, Parts I, II and III, J. Math. Phys. \textbf{3}, 140, 157, 166 (1962)
\bibitem{E06} B.~Eynard, \emph{Formal matrix integrals and combinatorics of maps}, \href{http://arxiv.org/abs/math-ph/0611087}{\texttt{math-ph/0611087}} (2006) 
\bibitem{EORev} B.~Eynard, N.~Orantin, \emph{Algebraic methods in random matrices and enumerative geometry}, \href{http://arxiv.org/abs/math-ph/0811.3531}{\texttt{math-ph/0811.3531}} (2008)
\bibitem{EKoKo04} B.~Eynard, A.~Kokotov, D.~Korotkin, \emph{Genus one contribution to free energy in hermitian two-matrix model}, Nucl.Phys. \textbf{B694}, p443-472 (2004), \href{http://arxiv.org/abs/hep-th/0403072}{\texttt{hep-th/0403072}}
\bibitem{Gustav} J.~Gustavsson, \emph{Gaussian fluctuations of eigenvalues in the GUE}, Ann. I. H. Poincaré, PR \textbf{41}, pp 151–178 (2005), \href{http://arxiv.org/abs/math/0401076}{\texttt{math.PR/0401076}}
\bibitem{For} P.J.~Forrester, \emph{Log-Gases and Random Matrices}, Princeton University Press (2010), \href{http://www.ms.unimelb.edu.au/~matpjf/matpjf.html}{\texttt{http://www.ms.unimelb.edu.au/\~{}matpjf/matpjf.html}}
\bibitem{RRV} J.A.~Ramirez, B.~Rider, B.~Virag, \emph{Beta ensembles, stochastic Airy spectrum, and a diffusion}, (2006), \href{http://arxiv.org/abs/math/0607331}{\texttt{math-ph/0607331}}
\bibitem{TW93} C.~Tracy, H.~Widom, \emph{Level spacing distributions and the Airy kernel}, Commun.Math.Phys. \textbf{159} p151-174 (1994), \href{http://arxiv.org/abs/hep-th/9211141}{\texttt{hep-th/9211141}}
\bibitem{TW95} C.~Tracy, H.~Widom, \emph{On orthogonal and symplectic matrix ensembles}, Commun.Math.Phys. 177 p727-754 (1996), \href{http://arxiv.org/abs/solv-int/9509007}{\texttt{solv-int/9509007}}
\bibitem{Voros} A.~Voros, \emph{Spectral functions, special functions, and the Selberg zeta function}, Commun. Math. Phys. 110, p439-465 (1987)
\bibitem{WZ06} P.~Wiegmann, A.~Zabrodin, \emph{Large N expansion for the 2D Dyson gas}, J. Phys. A \textbf{39} p8933-8964, (2006),\href{http://arxiv.org/abs/hep-th/0601009}{\texttt{hep-th/0601009}}
\bibitem{Krauth} W.~Krauth, \emph{Statistical Mechanics: Algorithms and Computation}, Oxford Univ. Press, Oxford, (2006)
\end{thebibliography}
\end{document}